\documentclass[a4paper,journal,romanappendices]{IEEEtran}
\usepackage{amssymb}
\usepackage[cmex10]{amsmath}
\usepackage{overpic}
\usepackage{algorithm}
\usepackage{algorithmic}
\usepackage{multirow}
\usepackage{array}
\usepackage{graphicx}
\usepackage{bm}
\usepackage[caption=false]{subfig}
\usepackage{color}
\usepackage{tikz}
\usepackage{epstopdf}
\usepackage{stfloats}
\usepackage{cite}
\usepackage{tikz}
\usepackage{amsmath}
\usepackage{psfrag}
\usepackage{acronym}
\usepackage{enumerate}
\usetikzlibrary{arrows}
\usetikzlibrary{decorations}

\newcounter{tempEqCounter}

\definecolor{BLUE}{rgb}{0,0,1}

\newtheorem{corollary}{Corollary}
\newtheorem{proposition}{Proposition}
\newtheorem{remark}{Remark}
\newtheorem{lemma}{Lemma}
\newtheorem{definition}{Definition}
\newtheorem{theorem}{Theorem}

\newtheorem{condition}{Condition}

\newcommand{\tr}[1]{{\rm tr}\left\{#1\right\}}

\newcommand{\subblk}[3]{\left[#1\right]_{#2,#3}}

\acrodef{aoa}[AOA]{angle-of-arrival}
\acrodef{bcrb}[BCRB]{Bayesian Cram\'{e}r-Rao bound}
\acrodef{bp}[BP]{belief propagation}
\acrodef{cdi}[CDI]{cooperative dilution intensity}
\acrodef{cl}[CL]{cooperative localization}
\acrodef{crb}[CRB]{Cram\'{e}r-Rao bound}
\acrodef{crlb}[CRLB]{Cram\'{e}r-Rao lower bound}
\acrodef{dpeb}[DPEB]{directional position error bound}
\acrodef{fim}[FIM]{Fisher information matrix}
\acrodef{efim}[EFIM]{equivalent Fisher information matrix}
\acrodef{ici}[ICI]{information coupling intensity}
\acrodef{mse}[MSE]{mean-squared error}
\acrodef{pdf}[PDF]{probability density function}
\acrodef{peb}[PEB]{position error bound}
\acrodef{speb}[SPEB]{squared position error bound}
\acrodef{pll}[PLL]{phase-locked loop}
\acrodef{rbs}[RBS]{reference broadcast synchronization}
\acrodef{rhs}[RHS]{right hand side}
\acrodef{rii}[RII]{ranging information intensity}
\acrodef{rss}[RSS]{received signal strength}
\acrodef{rc}[RC]{ranging coefficient}
\acrodef{speb}[SPEB]{squared position error bound}
\acrodef{toa}[TOA]{time-of-arrival}
\acrodef{tdoa}[TDOA]{time-difference-of-arrival}
\acrodef{tpsn}[TPSN]{time synchronization protocol for sensor network}
\acrodef{vmp}[VMP]{variational message passing}
\acrodef{wsn}[WSN]{wireless sensor network}
\acrodef{efim}[EFIM]{equivalent Fisher information matrix}
\acrodef{dio}[DIO]{distance-information-only}
\acrodef{aio}[AIO]{angle-information-only}
\acrodef{saaf}[SAAF]{squared array aperture function}
\acrodef{uoa}[UOA]{uniformly oriented array}
\acrodef{rgg}[RGG]{random geometric graph}
\acrodef{snr}[SNR]{signal-to-noise ratio}
\acrodef{eoc}[EoC]{efficiency of cooperation}
\acrodef{npi}[NPI]{nominal position information}
\acrodef{gnss}[GNSS]{global navigation satellite system}
\acrodef{mimo}[MIMO]{multiple-input multiple-output}
\acrodef{mcs}[MCS]{minimally constrained system}
\acrodef{los}[LOS]{line-of-sight}

\acrodefplural{rgg}[RGGs]{random geometric graphs}

\newcommand{\efim}[1]{\M{J}_{\rm e}(#1)}
\newcommand{\invefim}[1]{\M{J}_{\rm e}^{-1}(#1)}
\newcommand{\refim}[1]{\RM{J}_{\rm e}(#1)}

\newcommand{\bd}{\begin{description}}
\newcommand{\ed}{\end{description}}
\newcommand{\be}{\begin{enumerate}}
\newcommand{\ee}{\end{enumerate}}
\newcommand{\bi}{\begin{itemize}}
\newcommand{\ei}{\end{itemize}}
\newcommand{\bl}{\begin{list}}
\newcommand{\el}{\end{list}}
\newcommand{\bt}{\begin{tabbing}}
\newcommand{\et}{\end{tabbing}}

\newcommand{\hnm}{\jmath_\mathrm{\barred{b}{-0.41ex}{10mu}{-1mu}\mathrm{m}}}

\newcommand\T  { ^\text{T} }

\usepackage{bbm}
\usepackage{winsnotation}
\usepackage{rotating}
\usepackage{xr}
\usepackage{array}

\newcommand{\fitsize}{\fontsize{8.8pt}{\baselineskip}\selectfont}
\newcommand{\fitsizesm}{\fontsize{6.6pt}{\baselineskip}\selectfont}

\usepackage{xcolor}
\usepackage{colortbl}

\begin{document}
\title{Cooperative Localization in Massive Networks}

\author{Yifeng Xiong, \IEEEmembership{Student Member, IEEE}, Nan Wu, \IEEEmembership{Member, IEEE}, Yuan Shen, \IEEEmembership{Senior Member, IEEE}, \\and Moe Z. Win, \IEEEmembership{Fellow, IEEE}
\thanks{This fundamental research was supported in part by the National Natural Science Foundation of China under Grant 61871256. This paper was presented in part at 2017 IEEE International Conference on Communications. \textit{(Corresponding author: Nan Wu.)}}

\thanks{Y. Xiong and N. Wu are with the School of Information and Electronics, Beijing Institute of Technology, Beijing 100081, China (e-mail: yfxiong@bit.edu.cn, wunan@bit.edu.cn).}
\thanks{Y. Shen is with Electronic Engineering Department, Tsinghua University and Beijing National Research Center for Information Science and Technology, Tsinghua University, Beijing 100084, China (e-mail: shenyuan\_ee@tsinghua.edu.cn).}
\thanks{M.\ Z.\ Win is with the Laboratory for Information and Decision Systems (LIDS), Massachusetts Institute of Technology, Cambridge, MA 02139 USA (e-mail: {moewin@mit.edu}).}
}

\markboth{Cooperative Localization in Massive Networks}{\paperTitleMarkboth}

\maketitle

\begin{abstract}
Network localization is capable of providing accurate and ubiquitous position information for numerous wireless applications. This paper studies the accuracy of cooperative network localization in large-scale wireless networks. Based on a decomposition of the equivalent Fisher information matrix (EFIM), we develop a random-walk-inspired approach for the analysis of EFIM, and propose a position information routing interpretation of cooperative network localization. Using this approach, we show that in large lattice and stochastic geometric networks, when anchors are uniformly distributed, the average localization error of agents grows logarithmically with the reciprocal of anchor density in an asymptotic regime. The results are further illustrated using numerical examples.
\end{abstract}


\section{Introduction}
\IEEEPARstart{N}{etwork localization} is a key enabler of various wireless applications requiring location-awareness \cite{proc_ieee_win, poor_msp,SheMazWin:J12}, including Internet-of-things \cite{iot1,iot2,wlmp,iot_win}, autonomous vehicles \cite{av1,av2,av3}, and big data \cite{bigdata,sn1,SanMou:J14}. In most outdoor environments, \ac{gnss} can provide meter-level accuracy for various location-based services, however, their performance are severely degraded in harsh environments such as urban canyons, underground and indoor scenarios. Exploiting the cooperation among nodes, network localization can provide satisfactory positioning performance in these \ac{gnss}-denied environments \cite{error_evolution,netOp_PIEEE,mercury}.

Nodes in network localization fall into two categories: agents with unknown positions and anchors with precisely known positions. To elaborate further, anchors are usually used to model base stations in cellular networks, road-side units in vehicular networks, leading nodes in unmanned aerial vehicle swarms, and access points in wireless local area networks. Agents are widely used to model devices requesting positioning services, such as users in cellular networks and vehicles in vehicular networks.

Typically, agents communicate with neighboring nodes, and infer their positions using some signal metrics extracted from received signals including \ac{toa}, \ac{tdoa}, and \ac{aoa} \cite{sun_ho_toa,proc_henk,moura_cayley_menger,geert_ml,yuan_tvt, MazConAllWin:J18, ConMazBarLinWin:J19, zd1,zd2,zd3,BarDaiConWin:J15}. \ac{toa} and \ac{tdoa} can be obtained by measuring the propagation delay of the received signal, which can provide ranging information between nodes. \ac{aoa} is a metric characterizing the incoming direction of the received signal, which can be obtained using directional antennas or antenna arrays. \ac{toa}- and \ac{tdoa}-based network localization algorithms have been well studied in the literature. Recently, driven by the application of massive \ac{mimo} technology in the forthcoming new generation of wireless systems, techniques combining ranging and bearing information have also been studied \cite{TarMupRauSloSveWym:J14,henk_5g,win_5g}.

In the classical non-cooperative scheme, network localization is performed using only anchor-agent communication. By contrast, the emerging scheme of cooperative network localization also incorporates agent-agent communications. Intuitively, the inter-agent measurements provide additional position information, and hence has the potential of improving the localization accuracy and preventing the outage of localization services, which has been verified empirically by many existing works \cite{proc_henk,outage_shen,mb_coop_nlos,xiong_tcom}. However, these results are based on the performance of specific localization algorithms, which cannot reflect the essential relation between the localization performance and network parameters.

In order to gain more insights into the cooperative localization scheme, some performance limits have been studied \cite{WinConMazSheGifDarChi:J11,array_localization,add_ref1,add_ref2,add_ref3,add_ref4,add_ref6,add_ref5,latincom}. These limits have been exploited to address network operation tasks such as power allocation \cite{DaiSheWin:J15a,netOp_TIT}. The tool of \ac{speb}, which gives a lower bound for the position \ac{mse} using the inverse of \ac{efim} \cite{SheWymWin:J10,SheWin:J10a,xiong_tsp}, are used extensively in these works.

Apart from finding applications in network operation tasks, performance limits are also important in their own rights for understanding the efficiency of cooperation among agents. This could help us to determine whether the accuracy improvement introduced by agent cooperation is worth the additional system complexity. Especially, scaling laws of the localization accuracy are particularly desirable for massive networks, since obtaining the performance limits explicitly for these networks would be computationally expansive. However, closed-form expressions of the performance limits are only proposed for some specific network topologies \cite{inf_coupling}, and existing results on the asymptotic localization performance in general large-scale networks are mostly qualitative\cite{value_coop,outage_buehrer,xiong_CL,icc_conf}. Especially, there is a lack of investigation into the localization accuracy scaling laws in multi-hop networks. For these networks, traditional triangulation techniques fail to provide reliable localization services, whereas cooperative network localization becomes particularly useful. The essential difficulty of the asymptotic analysis lies in the information coupling phenomenon \cite{inf_coupling}, because of which the \ac{efim} does not possess a block-diagonal structure, and is intractable to be inverted analytically.

{
\begin{table*}[t]
\renewcommand{\arraystretch}{1.5}
\footnotesize
\centering
\caption{Comparison between the error scaling laws proposed in this paper and those developed in existing contributions.}
\label{tbl:scaling}
\footnotesize
\begin{tabular}{|m{0.18\columnwidth}<{\centering}|m{0.2\columnwidth}<{\centering}|m{0.28\columnwidth}<{\centering}|
m{0.28\columnwidth}<{\centering}|m{0.5\columnwidth}<{\centering}|}
  \hline
   \multicolumn{1}{|m{0.18\columnwidth}|}{\centering \textbf{Original Contribution}} & \textbf{Scheduling} & \textbf{Measurement Type} & \textbf{Anchor Deployment} & \textbf{Error Scaling (as $N_{\rm a}\rightarrow \infty$)} \\
  \hline\hline
  \cite{SheWymWin:J10} &  &  & Constant density &\multicolumn{1}{m{0.5\columnwidth}|}{\centering Converges to a \emph{constant}} \\[2pt]
  \cline{1-1}  \cline{4-5}
  \cite{concentric} & \multirow{6}{0.18\columnwidth}[-3mm]{Fully cooperated (optimal)} &  \multirow{3}{0.23\columnwidth}[-2mm]{Distance only ($N_{\rm t} = 1$)} &  \multirow{2}{0.28\columnwidth}[-1.75mm]{Located in the innermost circle} & \multicolumn{1}{m{0.5\columnwidth}|}{\centering \textit{Quadratic} in the number of hops from the anchors} \\[3pt]
  \cline{1-1}  \cline{5-5}
  \cellcolor{green!10} &  &  &  & \cellcolor{green!10} {\bf\textit{Logarithmic}} in the radial direction; {\bf \textit{Quadratic}} in the tangential direction \\[3pt]
  \cline{4-5}
  \multirow{4}{*}[4mm]{\!\!\cellcolor{green!10} \bf This paper} &  &  & \multirow{2}{*}[-2mm]{Single anchor} & \multicolumn{1}{m{0.5\columnwidth}|}{\centering \cellcolor{green!10} {\bf\textit{Logarithmic}} in the radial direction; {\bf \textit{Unbounded}} in the tangential direction}  \\[6pt]
  \cline{3-3} \cline{5-5}
  \cellcolor{green!10}& & Both distance and angle ($N_{\rm t}\geq 2$)& & \cellcolor{green!10} {\bf \textit{Logarithmic}} in both directions\\
  \cline{3-5}
  \cellcolor{green!10}&  & \multirow{2}{0.28\columnwidth}[-2mm]{Distance only or both distance and angle ($N_{\rm t}=1$ or $N_{\rm t}\geq 2$)}  & \multirow{2}{0.28\columnwidth}[-2mm]{Density $\rightarrow 0$} &\cellcolor{green!10}  {\bf \textit{Logarithmic}} in (anchor density)$^{-1}$ \\[3pt]
  \cline{1-2} \cline{5-5}
  \cite{dvhop,avg_hop_dist,ppp_mhl} & Sequential &  &  & \multicolumn{1}{m{0.5\columnwidth}|}{\centering At least proportional to the \textit{square root} of  (anchor density)$^{-1}$} \\[3pt]
  \hline
\end{tabular}
\end{table*}
}

Recently, we have developed a framework for understanding the scaling laws of relative and absolute synchronization errors in large networks \cite{netsync}. The network synchronization problem share some similarities with the cooperative localization problem. Especially, they both rely on the pairwise measurement between adjacent nodes. However, it is not straightforward to apply the results in \cite{netsync} to the cooperative localization problem, since the analysis method therein relies on the fact that the local parameter (i.e., the delay) in network synchronization is a scalar, while the location parameters are typically vectors.

In this paper, we generalize the framework proposed in \cite{netsync} to aid the asymptotic analysis of network localization. In particular, we develop a random walk interpretation of the \ac{efim} for location parameters, by introducing generalized random walks having matrix-valued ``pseudo-probabilities''. This enables us to depict the asymptotic localization performance quantitatively. For better clarity of illustration, we have summarized the asymptotic error scaling laws proposed in this paper, and contrasted them to the existing ones in Table \ref{tbl:scaling}.

The contributions of this paper are summarized as follows.
\begin{itemize}
\item We propose a decomposition of the \ac{efim}, based on which we propose the concepts of nominal position information (NPI) and efficiency of cooperation (EoC) to characterize the information coupling between an agent and its neighbors.
\item We develop a random walk interpretation of the EoC. Inspired by this interpretation, we describe quantitatively an intuition of ``position information routing'' for network localization.
\item We show that in large lattice and stochastic geometric networks\footnote{In this paper, we consider the theoretical models of lattice and stochastic geometric networks instead of real-world networks, since massive cooperative location-aware networks typically consist of mobile nodes, and hence do not have fixed topologies. These models are also widely used in other research works that analyze large-scale wireless networks \cite{HaeAndBacDouFra:09,DarConBurVer:07,sg_2,WinPinShe:J09,tutorial_sg,sg_uav,sg_vnet1,sg_vnet2,sg_add1,sg_add2,sg_add3}.}, the average position \ac{mse} of agents grows only \emph{logarithmically} with the reciprocal of anchor density in an asymptotic regime, if anchors are distributed uniformly in the network.
\item We show that for the concentric multi-hop network model considered in \cite{concentric} where each node is equipped with a single antenna, the \ac{dpeb} \cite{SheWymWin:J10} grows \emph{logarithmically} on the direction to the centroid of that region, while it increases \emph{quadratically} on the orthogonal direction.
\end{itemize}

The rest of this paper is organized as follows. Section \ref{sec:system_model} formulates the network localization problem, specifies the system model, and describes the general form of \ac{efim}. In Section \ref{sec:quasi_rw} we conduct a random-walk-inspired analysis of \ac{efim}, and present the position information routing interpretation. Section \ref{sec:lattice} develops expressions for the asymptotic behavior of \ac{efim} and \ac{speb} in large lattice and stochastic geometric networks. We further discuss the practical implications of our results in Sec. \ref{sec:discussions}. The analytical results are illustrated in Section \ref{sec:numerical}, and the conclusions are drawn in Section \ref{sec:con}.

\emph{Notations:} Throughout this paper, $\rv{a}$, $\RV{a}$, $\RM{A}$, and $\RS{A}$ represent random variables (scalars), random vectors, random matrices and random sets, respectively; Their realizations, or the corresponding deterministic quantities, are denoted by $a$, $\V{a}$, $\M{A}$, and $\Set{A}$, respectively. The $m$-by-$n$ matrix of zeros (resp. ones) is denoted by $\M{0}_{m\times n}$ (resp. $\M{1}_{m\times n}$). The $m$-dimensional vector of zeros (resp. ones) is denoted by $\V{0}_{m}$ (resp. $\M{1}_{m}$). The $m$-by-$m$ identity matrix is denoted by $\M{I}_{m}$. $\M{Q}_{n}\in\mathbb{R}^{2n\times 2}$ denotes a matrix given by $[\M{I}_2~\M{I}_2~\dotsc~\M{I}_2]^{\rm T}$. These subscripts are omitted if they are clear from the context. $\M{E}_{ij}^{M,N}$ denotes a $M\times N$ matrix with all zeros but a $1$ on the $(i,j)$-th entry, which is simplified as $\M{E}_{ij}^N$ when $M=N$. The notation $[\cdot]_{i,j}$ denotes the $(i,j)$-th $2\times 2$ sub-matrix of its argument; $[\cdot]_{r_1:r_2,c_1:c_2}$ denotes a submatrix obtained by extracting the $(2r_1-1)$-th to the $(2r_2)$-th row and the $(2c_1-1)$-th to the $(2c_2)$-th column of its argument. $[\cdot]^\dagger$ denotes the Moore-Penrose pseudo inverse of its argument, while $[\cdot]^{\rm H}$ denotes the Hermitian transpose of its argument. $\mathrm{tr}\{\cdot\}$ stands for the trace of a square matrix. The Kronecker product between matrices $\M{A}$ and $\M{B}$ is denoted by $\M{A}\otimes\M{B}$. $\|\V{x}\|_p$ denotes the $l_p$ norm, which represents the $l_2$ norm by default when the subscript is omitted. $\V{u}(\varphi)$ denotes a unit vector $[\cos\varphi~\sin\varphi]^{\rm T}$, and $\M{J}_{\rm r}(\varphi):= \V{u}(\varphi)\V{u}(\varphi)^{\rm T}$. The function $\mathbbm{1}\{\cdot\}$ takes the value $1$ if its argument as a logical expression is true, and $0$ otherwise. The notation $\min^{(i)}(\Set{S})$ denotes the $i$-th smallest element in the set $\Set{S}$.

The notation $\mathbb{E}_{\RV{x}}\{\cdot\}$ denotes the expectation with respect to $\RV{x}$, and the subscript is removed when there is no confusion. The functions $f_{\RV{x}}(\V{x})$ and $f_{\RV{x}}(\V{x},\V{\theta})$ denote the \ac{pdf} of $\RV{x}$ and the \ac{pdf} of $\RV{x}$ parameterized by $\V{\theta}$, respectively. We also define the following function
$$
\begin{aligned}
\hnm(\V{z}, a(\V{\theta}_1,\V{\theta}_2,\V{\theta}_3),\V{\theta}_1,\V{\theta}_2)  &:=  \frac{\partial \ln f_{\RV{z}}(\V{z};a(\V{\theta}_1,\V{\theta}_2,\V{\theta}_3)) }{\partial \V{\theta}_1}\\
&\hspace{3mm} \cdot \frac{\partial \ln f_{\RV{z}}(\V{z};a(\V{\theta}_1,\V{\theta}_2,\V{\theta}_3)) }{ \partial \V{\theta}_2\T}.
\end{aligned}
$$
Some frequently-used Bachmann-Landau notations are given as follows.

\vspace{2mm}
\noindent
\begin{tabular}{ll}
\centering
    $a(n)= O(b(n))$ & $ \mathop{\lim\sup}_{n\rightarrow \infty} a(n)(b(n))^{-1} < \infty$ \\
    $a(n)=\Omega(b(n))$ & $\mathop{\lim\inf}_{n\rightarrow \infty} a(n)(b(n))^{-1} > 0$ \\
    $a(n)=\Theta(b(n))$ & $a(n)=O(b(n))$ and $a(n)=\Omega(b(n))$ \\
\end{tabular}

\section{System Model and Evaluation of \ac{efim}}\label{sec:system_model}
Consider a network with $N_{\rm a}$ agents and $N_{\rm b}$ anchors. The indices of agents and anchors constitute sets $\Set{N}_{\rm a}=\{1,2,\dotsc,N_{\rm a}\}$ and $\Set{N}_{\rm b}=\{N_{\rm a}+1,N_{\rm a}+2,\dotsc,N_{\rm a}+N_{\rm b}\}$, respectively. The position of node $i$ is denoted by $\V{p}_i=[p_{{\rm x}i}~p_{{\rm y}i}]^{\rm T}$. The vector containing positions of all nodes is denoted by $\V{p}_{\rm all}=[\V{p}_1^{\rm T}~\V{p}_2^{\rm T}~\dotsc~\V{p}_{N_{\rm a}+N_{\rm b}}^{\rm T}]^{\rm T}$, and especially for agents we denote $\V{p}=[\V{p}_1^{\rm T}~\V{p}_2^{\rm T}~\dotsc~\V{p}_{N_{\rm a}}^{\rm T}]^{\rm T}$. The angle from node $i$ to $j$ is denoted by $\varphi_{ij}$.

In general, all relative position measurements between node $i$ and $j$ are measurements of the displacement vector $\V{d}_{ij}:= \V{p}_i-\V{p}_j$. In this paper, we assume in particular that each node $i$ is equipped with an array of $N_{\rm t}$ antennas.\footnote{Our results can also be applied to the case in which nodes are equipped with different number of antennas after minor modifications.} The received signals at node $j$ from node $i$ can be modeled as
\begin{equation}\label{waveform_original}
\begin{aligned}
\RV{r}_{ij}(t)= N_{\rm t}\sum_{l=1}^{N_{\rm p}}\alpha_{ij}^{(l)}\V{a}_j\V{a}_i^{\rm H}\V{s}_i(t-\tau_{ij}^{(l)})+\RV{z}_{ij}(t)
\end{aligned}
\end{equation}
where $N_{\rm p}$ denotes the number of propagation paths, $\V{s}_i(t)=[s_{i1}(t)~s_{i2}(t)~\dotsc~s_{iN_{\rm t}}(t)]^{\rm T}$ denotes a set of orthonormal transmitting waveforms, $\V{a}_i$ is the unit array steering vector of node $i$, $\alpha_{ij}^{(l)}$ is the amplitude of the received signal of the $l$-th path, and $\RV{z}_{ij}(t)$ denote the noise modeled as white complex Gaussian processes. The propagation delay of the $l$-th path $\tau_{ij}^{(l)}$ is given by $c^{-1}d_{ij}^{(l)}$ where $c$ is the signal propagation speed and $d_{ij}^{(l)}$ represents the length of the $l$-th propagation path. We denote by $\RV{r}$ the concatenated vector of the Karhunen-Loeve expansions of all $\RV{r}_{ij}(t)$. We assume that there is a \ac{los} path corresponding to $l=1$ for each link, meaning that $\tau_{ij}^{(1)}=c^{-1}\|\V{p}_i-\V{p}_j\|$.

We denote the receiver sensitivity of each node $i$ by $p_{{\rm rs},i}$, below which the receiving power is not sufficient for successful signal detection. For simplicity, we say that node $i$ is in the measurement range of node $j$ (and denote this relation as $i\in\Set{N}_j$), if we have $\alpha_{ij}^2\|\V{p}_i-\V{p}_j\|^{-\gamma}\leq p_{{\rm rs},j}$ where $\gamma$ is the path loss exponent and $\alpha_{i,j}$ denotes the signal amplitude of the \ac{los} path; otherwise node $j$ cannot receive measurements from node $i$. The connectivity between nodes can then be described using a graph $\Set{G}_{\rm net}=\{\Set{V}_{\rm net},\Set{E}_{\rm net}\}$, where the vertex set $\Set{V}_{\rm net}=\Set{N}_{\rm a}\cup\Set{N}_{\rm b}$, and $\Set{E}_{\rm net}$ is the set of edges given by
$$
\Set{E}_{\rm net}:=\{(i,j)|i\in\Set{V}_{\rm net},~j\in\Set{V}_{\rm net},~i\leftrightarrow j\}
$$
where $i \leftrightarrow j $ means $i\in\Set{N}_j$ (or $j\in\Set{N}_i$) holds. Naturally, the connectivity between agents constitutes another graph, denoted by $\Set{G}_{\rm agt}$.

Apart from the agents' positions, there are also some unknown channel parameters denoted by $\V{\eta}$. In light of this, we consider $\V{\theta}=[\V{p}^{\rm T}~\V{\eta}^{\rm T}]^{\rm T}$ as the complete vector of parameters. For any unbiased estimator $\hat{\RV{p}}$ of $\V{p}$, the \ac{mse} matrix of $\hat{\RV{p}}$ satisfies
$$
\mathbb{E}\big\{(\V{p}-\hat{\RV{p}})(\V{p}-\hat{\RV{p}})^{\rm T}\big\}\succeq \invefim{\V{p}}
$$
where $\efim{\V{p}}$ is the \ac{efim} for $\V{p}$ with respect to $\V{\theta}$. To further characterize the performance limit of a specific agent $i$, we define the \ac{speb} of agent $i$ as
$$
{\rm sp}\{\V{p}_i\}:= \tr{\invefim{\V{p}_i}}
$$
and the \ac{dpeb} of agent $i$ along the direction of $\V{u}$ as \cite{SheWymWin:J10}
$$
{\rm dp}\{\V{p}_i;\V{u}\}:= \V{u}^{\rm T}\invefim{\V{p}_i}\V{u}.
$$

Using \ac{speb}, we have $\mathbb{E}\{\|\V{p}_i-\hat{\RV{p}}_i\|^2\}\geq {\rm sp}\{\V{p}_i\}$. Using \ac{dpeb}, the position error along the direction of $\V{u}$ can be bounded by $\mathbb{E}\{\|\V{u}^{\rm T}(\V{p}_i-\hat{\RV{p}}_i)\|^2\}\geq {\rm dp}\{\V{p}_i;\V{u}\}$.

According to \cite[Theorem 1]{SheWymWin:J10}, the \ac{efim} for $\V{p}$ can be written as
\begin{equation}\label{vib_laplacian}
\efim{\V{p}} = \M{D}-\M{A}
\end{equation}
where $\M{D}:= {\rm diag}(\M{D}_1,\dotsc,\M{D}_{N_{\rm a}})$ with $\M{D}_i$ given by $\M{D}_i= \sum_{i\leftrightarrow j}\M{J}_{ij}$, and
$$
\M{A}:=-\sum_{i\leftrightarrow j,i\in\Set{N}_{\rm a},j\in\Set{N}_{\rm a}}\M{E}_{ij}^{N_{\rm a}}\otimes \M{J}_{ij}.
$$
Matrix $\M{J}_{ij}$ denotes the contribution of measurement $\RV{r}_{ij}$ to the position information of agent $i$ (as well as $j$) given by
\begin{equation}\label{jd_original}
\M{J}_{ij} =\mathbb{E}\big\{\hnm(\RV{r}_{ij},[\V{d}_{ij}~\V{\kappa}_{ij}],\V{p}_i,\V{p}_j)\big\}
\end{equation}
where $\V{\kappa}_{ij}$ denotes nuisance parameters related to channel condition, transmission power, signal waveform, etc.

We further assume that the power transmitted from each antenna is identical in every single node. Under previous assumptions, it is known that in far-field environments, $\M{J}_{ij}$ can be approximately decomposed as follows \cite[Corollary 1]{array_localization}
\begin{equation}\label{approx_jd}
\M{J}_{ij} \approx N_{\rm t}^2\Big(\widetilde{\xi}_{ij,{\rm r}}\M{J}_{\rm r}(\varphi_{ij})+\widetilde{\xi}_{ij,{\rm t}}\M{J}_{\rm r}\Big(\varphi_{ij}+\frac{\pi}{2}\Big)\Big)
\end{equation}
where $\widetilde{\xi}_{ij,{\rm r}}:= p_i\beta^2\xi_{ij}$, $\widetilde{\xi}_{ij,{\rm t}}:= p_i f_{\rm c}^2\xi_{ij}G_{ij}\|\V{d}_{ij}\|^{-2}$, $p_i$ is the transmission power of node $i$, $\beta$ is the transmitted signal bandwidth, $\xi_{ij}$ is given by $\xi_{ij} = \zeta_{ij}\|\V{d}_{ij}\|^{-\gamma}$ with $\zeta_{ij}$ being the ranging coefficient \cite{power_opt} determined by channel and waveform parameters, and $G_{ij}$ denotes the \ac{saaf} determined by the shapes of the arrays equipped on nodes. When $N_{\rm t}=1$, \eqref{approx_jd} can be simplified as
\begin{equation}\label{simplifed_jd}
\M{J}_{ij}\approx\widetilde{\xi}_{ij,{\rm r}}\M{J}_{\rm r}(\varphi_{ij}).
\end{equation}
In general, $G_{ij}$ is determined by the orientations of node $i$ and $j$. As a special case, $G_{ij}$ is a constant with respect to orientations in the case of \ac{uoa} \cite[Definition 4]{array_localization}. When $N_{\rm t}=1$, we have $G_{ij}=0$, and hence \eqref{approx_jd} reduces to \eqref{simplifed_jd}.

\section{Information Coupling and Random-Walk-Inspired Analysis}\label{sec:quasi_rw}
To obtain the desired \ac{mse} lower bound, we have to take the inverse of the \ac{efim} for $\V{p}$ derived in Section \ref{sec:system_model}, which is a rather difficult task due to the notorious information coupling phenomenon. To alleviate this difficulty, in this section we propose a decomposition of \ac{efim} inspired by the theory of random walk on graphs.

\subsection{Information Coupling and \ac{efim} Decomposition}\label{ssec:inf_coupling}
In a non-cooperative network, it is well-known that the \ac{efim} for agent $i$ can be expressed as \cite{SheWin:J10a}
\begin{equation}\label{non_coop}
\efim{\V{p}_i} = \M{D}_i  = \sum_{j\in\Set{N}_{\rm b}} \M{J}_{ij}
\end{equation}
due to the block-diagonal structure of the \ac{efim} for $\V{p}$. Here, the second equality follows from the fact that agents can only communicate with anchors in non-cooperative networks. Unfortunately, this simple formula does not hold for cooperative networks since the corresponding \ac{efim}s are not block-diagonal.

Intuitively, for any agent $i$ in a cooperative network, the neighboring agents can be viewed as ``weak anchors'' in the sense that they provide less position information, due to their position uncertainty. Hence we may have
$$
\efim{\V{p}_i} =\M{D}_i \M{X}_i \preceq \M{D}_i
$$
where the matrix $\M{X}_i$ depicts the overall efficiency that agent $i$ utilizes the position information obtained from neighboring nodes. Formally speaking, we have the following result.
\begin{theorem}\label{thm:main}
When the inverse of $\efim{\V{p}}$ exists, the \ac{efim} of agent $i$ can be decomposed as
\begin{equation}\label{main_thm}
\efim{\V{p}_i} = \M{D}_i(\M{I}+\M{\Delta}_i)^{-1}~\forall i\in\Set{N}_{\rm a}
\end{equation}
where $\M{\Delta}_i:= \sum_{n=1}^{\infty} \M{T}_{ii}^{(n)}\succeq\M{0}$ with $\M{T}_{ij}^{(n)}$ given by the following recursion
\begin{equation}\label{Kolmogorov}
\M{T}_{ij}^{(n)} = \sum_{k=1}^{N_{\rm a}}\M{T}_{ik}^{(n-1)}\M{T}_{kj}^{(1)}
\end{equation}
where
\begin{equation}\label{modified_tp}
\M{T}_{ij}^{(1)}= \left\{
                             \begin{array}{ll}
                               \M{D}_i^{-1}\M{J}_{ij}, & \hbox{$i\neq j,~i\notin \Set{N}_{\rm b}$;} \\
                               \M{I}_2, & \hbox{$i=j,~i\in\Set{N}_{\rm b}$;}\\
                               \M{0}_{2\times 2}, & \hbox{otherwise.}
                             \end{array}
                           \right.
\end{equation}
\begin{IEEEproof}
Please refer to Appendix \ref{sec:proof_qrw}.
\end{IEEEproof}
\end{theorem}

Considering the similarity between \eqref{non_coop} and the definition of $\M{D}_i$, matrix $\M{D}_i$ can be interpreted as the nominal position information provided by agent $i$'s neighbors when information coupling is ignored, and hence we give the following definition.
\begin{definition}[Nominal Position Information]
The \ac{npi} of agent $i$ is defined as $\M{D}_i$.
\end{definition}

Since \ac{efim} characterizes the effective position information acquired by a certain agent, it can be seen from \eqref{main_thm} that the term $(\M{I}+\M{\Delta}_i)^{-1}$ quantifies the efficiency of cooperation between agent $i$ and its neighbors, and can be viewed as a generalized version of the \ac{eoc} defined in \cite{netsync}. Note that \ac{eoc} satisfies $\M{0}\preceq (\M{I}+\M{\Delta}_i)^{-1}\preceq \M{I}$ due to the non-negativity of $\M{\Delta}_i$, implying that the information coupling among agents will lead to degeneration of cooperation efficiency. As an extreme case, when all neighboring nodes of agent $i$ are anchors, there is no coupling between agent $i$ and its neighbors, and thus we have $(\M{I}+\M{\Delta}_i)^{-1}=\M{I}$ and $\efim{\V{p}_i}=\M{D}_i$.

To simplify further discussions, we will denote
\begin{equation}\label{transition_matrix}
\M{T}:= \M{D}^{-1}\M{A}
\end{equation}
in the rest of this paper.

\begin{remark}
Most of the results in this subsection have their analogue in \cite{netsync}. However, it should be noted that the \ac{eoc} in this paper is a matrix instead of a scalar. Furthermore, the quantity $\M{T}_{ij}^{(n)}$ defined in \eqref{Kolmogorov} corresponds to the $n$-step transition probability in \cite{netsync}, but now it can no longer be viewed formally as a probability, since it is a matrix. Fortunately, as we shall discuss in more detail in Sec. \ref{ssec:random_walk_2d}, these matrix-valued quantities can still be regarded as probabilities in a certain sense, and some classical results in the random walk theory can still be applied.
\end{remark}

\subsection{Random-Walk-Inspired Formalism: The 1-D Toy Example}\label{ssec:random_walk}
\begin{figure}[t]
    \centering
    \vspace{2mm}
    \begin{overpic}[width=.42\textwidth]{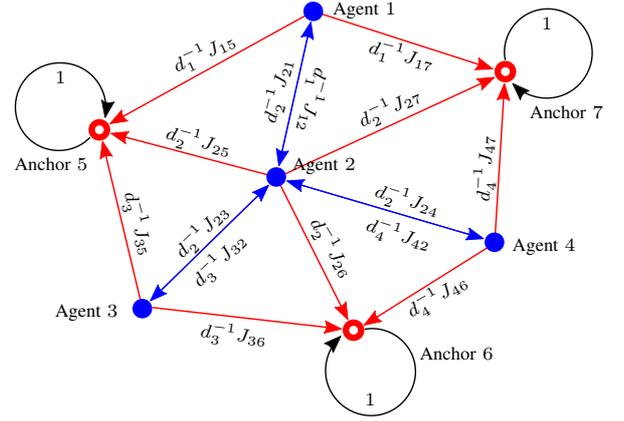}
    \put(55,70){\scriptsize Agent $1$}
    \put(48,43){\scriptsize Agent $2$}
    \put(7,17.5){\scriptsize Agent $3$}
    \put(86,29){\scriptsize Agent $4$}
    \put(0,43){\scriptsize Anchor $5$}
    \put(70,10){\scriptsize Anchor $6$}
    \put(89,52){\scriptsize Anchor $7$}
    \put(7,58){\scriptsize $1$}
    \put(60.5,2){\scriptsize $1$}
    \put(91,67.5){\scriptsize $1$}
    \put(28,60){\scriptsize \begin{rotate}{25}$d_1^{-1}J_{15}$\end{rotate}}
    \put(61,63.5){\scriptsize \begin{rotate}{-15}$d_1^{-1}J_{17}$\end{rotate}}
    \put(52,61){\scriptsize \begin{rotate}{-105}$d_1^{-1}J_{12}$\end{rotate}}
    \put(45,51){\scriptsize \begin{rotate}{75}$d_2^{-1}J_{21}$\end{rotate}}
    \put(26,48.5){\scriptsize \begin{rotate}{-15}$d_2^{-1}J_{25}$\end{rotate}}
    \put(60,50.5){\scriptsize \begin{rotate}{25}$d_2^{-1}J_{27}$\end{rotate}}
    \put(29,27.5){\scriptsize \begin{rotate}{45}$d_2^{-1}J_{23}$\end{rotate}}
    \put(62,38.5){\scriptsize \begin{rotate}{-17.5}$d_2^{-1}J_{24}$\end{rotate}}
    \put(51,34){\scriptsize \begin{rotate}{-60}$d_2^{-1}J_{26}$\end{rotate}}
    \put(32,22.5){\scriptsize \begin{rotate}{45}$d_3^{-1}J_{32}$\end{rotate}}
    \put(19,39){\scriptsize \begin{rotate}{-75}$d_3^{-1}J_{35}$\end{rotate}}
    \put(32,14){\scriptsize \begin{rotate}{-7.5}$d_3^{-1}J_{36}$\end{rotate}}
    \put(60.5,32.5){\scriptsize \begin{rotate}{-17.5}$d_4^{-1}J_{42}$\end{rotate}}
    \put(68.5,17.25){\scriptsize \begin{rotate}{30}$d_4^{-1}J_{46}$\end{rotate}}
    \put(82,37.5){\scriptsize \begin{rotate}{87.5}$d_4^{-1}J_{47}$\end{rotate}}
    \end{overpic}
    \caption{The transition diagram portraying the random walk characterized by \eqref{transition_recurrence} and \eqref{transition_prob}. The quantities next to the edges represent the corresponding transition probability.}
    \label{fig:random_walk_portrayal}
\end{figure}

The \ac{eoc} characterizes the effect of information coupling. In fact, if we can obtain the \ac{eoc} of an agent $i$, we can readily compute its \ac{efim} using \eqref{main_thm} without inverting the whole \ac{efim} for $\V{p}$. However, \eqref{Kolmogorov} and \eqref{modified_tp} are still rather intricate, and hence it is not clear that how \ac{eoc} is related to other characteristics of the network. To facilitate understanding, we first consider the toy example of 1-D cooperative localization, in which all nodes locate along a line. Although this example is much different from real-world scenarios, the technique used in the analysis will be similar in spirit to those applied in the generic case.

In this simplified scenario, quantities such as $\M{D}_i$, $\M{J}_{ij}$ and $\M{\Delta}_i$ are all scalars, and hence we denote them as $d_i$, $J_{ij}$ and $\Delta_i$, respectively. Using the definition \eqref{transition_matrix}, the equations \eqref{main_thm}, \eqref{Kolmogorov} and \eqref{modified_tp} can be rewritten as
\begin{subequations}\label{line_random_walk}
\begin{align}
\efim{p_i}& = \frac{d_i}{1+\Delta_i}~\forall i \in \Set{N}_{\rm a} \\
\M{T}^n&= \M{T}^{n-1} \M{T} \label{transition_recurrence} \\
T_{ij}&=\left\{
                 \begin{array}{ll}
                   d_i^{-1}J_{ij}, & \hbox{$i\neq j,~i\notin \Set{N}_{\rm b}$;} \\
                   1, & \hbox{$i=j,~i\in \Set{N}_{\rm b}$;} \\
                   0, & \hbox{otherwise.}
                 \end{array}
               \right. \label{transition_prob}
\end{align}
\end{subequations}
respectively, where $T_{ij}$ denotes the $(i,j)$-th entry of matrix $\M{T}$. In addition, we will denote the $(i,j)$-th entry of $\M{T}^n$ as $T_{ij}^{(n)}$. Due to the non-negativity of Fisher information, the term $T_{ij}$ is non-negative, and satisfies $\sum_{j\in\Set{N}_i} T_{ij} = 1$ according to the definition of $\M{D}$. In light of this, we may regard matrix $\M{T}$ as the \emph{transition probability matrix} of a random walk, as portrayed in Fig. \ref{fig:random_walk_portrayal}.

From Fig. \ref{fig:random_walk_portrayal} we can see that anchors correspond to the \emph{absorbing states} of the random walk, since the probability of staying at any state representing an anchor is $1$. The quantity $\Delta_i$ can now be expressed as
\begin{equation}\label{returning_probs}
\Delta_i = \sum_{n=1}^\infty [\M{T}^n]_{i,i}
\end{equation}
which can be interpreted as the sum over all $n$-step returning probabilities from node $i$ to itself, where $n=1,2,\dotsc, \infty$. In light of \eqref{returning_probs}, we can rewrite the \ac{eoc} of any agent $i$ as
\begin{equation}\label{green_hitting}
\frac{1}{1+\Delta_i} = 1-F_{ii}
\end{equation}
where $F_{ii}$ is the \emph{hitting probability} from node $i$ to itself \cite[Sec. 1.6]{sto_process}. To elaborate a little further, $F_{ii}$ denotes the probability that the random walk starting from $i$ would ever return to $i$. Formally, we have
$$
F_{ii} := \mathbb{P}\{\rv{x}_t = i,~\exists t\geq 0 |\rv{x}_0=i\}
$$
where $\rv{x}_t$ denotes the state of the random walk at time step $t$. The equation \eqref{green_hitting} is a classical result in the random walk theory \cite[Sec. 1.5]{sto_process}.

Another important fact about one-dimensional random walks is that they are always \emph{recurrent} when there is no absorbing states \cite[Sec. 28]{random_walk_principles}, meaning that we have $F_{ii}=1$ for all $i\in\Set{N}_{\rm a}$ if there is not any anchor in the network. This implies that when there are some anchors, the hitting probability $F_{ii}$ satisfies
\begin{equation}\label{recurrence_alias}
F_{ii} + \sum_{j\in \Set{N}_{\rm b}} F_{\Set{N}_{\rm b}\cup\{i\}}(i,j) = 1
\end{equation}
where $F_{\Set{N}_{\rm b}\cup\{i\}}(i,j)$ denotes the probability that the random walk starting from $i$ would ever reach the set $\Set{N}_{\rm b}\cup\{i\}$ at least once, and that it is at state $j$ when it reaches $\Set{N}_{\rm b}\cup\{i\}$ for the first time. In other words, \eqref{recurrence_alias} can be interpreted as that a random walk starting from $i$ would always reach the set $\Set{N}_{\rm b}\cup \{i\}$ at least once.

\begin{figure}[t]
    \centering
    \psfrag{A}{\footnotesize Agent $i$}
    \psfrag{B}{\footnotesize Agent}
    \psfrag{C}{\footnotesize Anchor}
    \begin{overpic}[width=.4\textwidth]{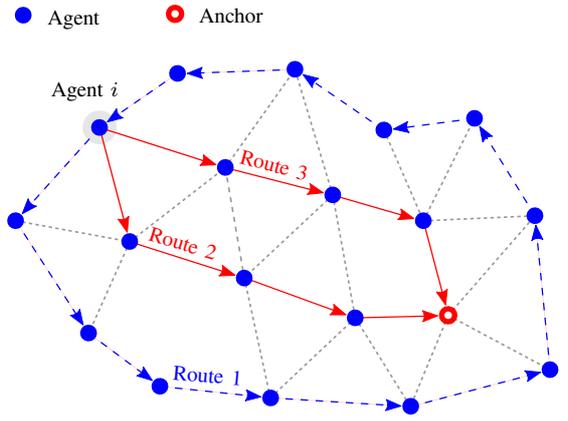}
    \put(30,6.5){\begin{rotate}{-5}\footnotesize \color{blue} Route 1\end{rotate}}
    \put(42,46){\begin{rotate}{-15}\footnotesize \color{red} Route 3\end{rotate}}
    \put(25.5,32){\begin{rotate}{-17}\footnotesize \color{red} Route 2\end{rotate}}
    \end{overpic}
    \caption{Graphical illustration of the ``position information routing'' interpretation. Here the ineffective route (route 1) does not include anchors, and hence does not provide agent $i$ with any position information. However, effective routes (routes 2 and 3) do provide position information.}
    \label{fig:routing_interpretation}
\end{figure}

Combining \eqref{green_hitting} and \eqref{recurrence_alias}, we have
\begin{equation}\label{position_information_routing_eoc}
\frac{1}{1+\Delta_i} = \sum_{j\in\Set{N}_{\rm b}} F_{\Set{N}_{\rm b}\cup\{i\}}(i,j)
\end{equation}
and hence
\begin{equation}\label{alt2_main}
d_i = \underbrace{d_i\sum_{j\in\Set{N}_{\rm b}} F_{\Set{N}_{\rm b}\cup\{i\}}(i,j)}_{\efim{\V{p}_i}} + d_iF_{ii}.
\end{equation}
Intuitively, we may now interpret \ac{eoc} as the ``position information routing'' efficiency of an agent. To elaborate, \eqref{alt2_main} shows that the \ac{npi} of agent $i$ consists of two parts: effective information (i.e., $\efim{\V{p}_i}$) coming from effective routes that start from agent $i$ and arrive at anchors before returning to agent $i$, and ineffective information coming from ineffective routes that returns to agent $i$ before visiting any anchor, as portrayed in Fig. \ref{fig:routing_interpretation}. Since agent $i$ does not have \textit{a priori} position information, it cannot provide itself any information via ineffective routes. By contrast, anchors do provide position information via effective routes.

\subsection{Random-Walk-Inspired Formalism: The 2-D Scenario}\label{ssec:random_walk_2d}
In this subsection, we generalize the discussions in Sec. \ref{ssec:random_walk} to the more practical 2-D scenario, and develop the ``position information routing'' interpretation of \ac{eoc}. To elaborate, the quantities $\M{D}_i$, $\M{J}_{ij}$, $\M{T}_{ij}^{(1)}$ and $\M{\Delta}_i$ have a strong resemblance to their 1-D counterparts, except that they are $2\times 2$ matrices. In particular, the term $\M{T}_{ij}^{(1)}$ satisfies the (slightly modified) normalization property of $\sum_{j\in\Set{N}_{i}} \M{T}_{ij}^{(1)} =\M{I}_2$. However, it does not satisfy the non-negativity property. Although $\M{T}_{ij}^{(1)}$ does not constitute a properly defined probability measure, in the following discussions, we will show that many results in the classical random walk theory, including \eqref{green_hitting}, can be obtained in a similar form by treating $\M{T}_{ij}^{(1)}$ as \emph{pseudo-probability}.

{\linespread{1}
\begin{table}
\renewcommand{\arraystretch}{1.5}

\centering
\caption{Notations related to the pseudo-probabilities.}
\label{tbl:notation}
\begin{tabular}{|m{0.12\columnwidth}<{\centering}|m{0.34\columnwidth}<{\centering}|m{0.39\columnwidth}<{\centering}|}
  \hline
  \textbf{Notation} & \textbf{Definition} & \textbf{Interpretation}  \\
  \hline\hline
 $\Omega_n^{i,j}$  & $\big\{\V{\omega}|\V{\omega}\in\mathbb{Z}^{n+1},\omega_1=i,\omega_{n+1}=j,\omega_k\in\Set{N}_{\rm a}\cup\Set{N}_{\rm b}, ~k=1,2,\dotsc, n+1\big\}$ & The set of all possible sequences of node indices of length $n+1$ that start with $i$ and end with $j$. \\
 \hline
 $\Omega_n^{i,j}(\Set{S})$ & $\{\V{\omega}|\V{\omega}\in\Set{\Omega}_n^{i,j},~\omega_k\notin\Set{S}~\forall 2 \leq k\leq n\}$ & A subset of $\Omega_n^{i,j}$ comprising the sequences that exclude the nodes in the set $\Set{S}$ (except $i$ and $j$). \\
 \hline
 $\M{T}_{\V{\omega}}^{(n)}$ & $\M{T}_{\omega_1\omega_2}^{(1)}\dotsc\M{T}_{\omega_{n}\omega_{n+1}}^{(1)}$ & The pseudo-probability of a random walk in the \emph{original} network travelling from $i$ to $j$ following path $\V{\omega}$ given by $\V{\omega}=[\omega_1~\omega_2~\dotsc~\omega_{n+1}]^{\rm T}$. \\
 \hline
  $\widetilde{\M{T}}_{\V{\omega}}^{(n)}$ & $\widetilde{\M{T}}_{\omega_1\omega_2}^{(1)}\dotsc\widetilde{\M{T}}_{\omega_{n}\omega_{n+1}}^{(1)}$ & The pseudo-probability of a random walk in the \emph{auxiliary} network travelling from $i$ to $j$ following the path $\V{\omega}$.\\
  \hline
\end{tabular}
\end{table}
}

To formalize these ideas, we will use the notations defined in Table \ref{tbl:notation} hereafter. Using these notations, matrix $\M{T}_{ij}^{(n)}$ can be expressed as
\begin{equation}
\M{T}_{ij}^{(n)} = \sum_{\V{\omega}\in\Set{\Omega}_n^{i,j}} \M{T}_{\V{\omega}}^{(n)}.
\end{equation}
Furthermore, we may have the following result by simply imitating \eqref{green_hitting}.
\begin{proposition}\label{prop:hitting}
The \ac{eoc} of agent $i$ $(\M{I}+\M{\Delta}_i)^{-1}$ can be rewritten as (as long as it exists)
\begin{equation}\label{hitting_prob}
(\M{I}+\M{\Delta}_i)^{-1} = \M{I} - \M{F}_{ii}
\end{equation}
where $\M{F}_{ij}$ is defined as $\M{F}_{ij}:= \sum_{n=1}^\infty\M{F}_{ij}^{(n)}$ with
\begin{equation}\label{hitting_relation}
\M{F}_{ij}^{(n)} := \sum_{\V{\omega} \in\Set{\Omega}_n^{i,j}(\{j\})}\M{T}_{\V{\omega}}^{(n)}.
\end{equation}
\begin{IEEEproof}
Please refer to Appendix \ref{ssec:proof_hitting}.
\end{IEEEproof}
\end{proposition}

Apparently, $\M{F}_{ij}$ is an analogue of the hitting probability, and hence will be referred to as the \emph{hitting pseudo-probability} hereafter. Following a similar line of reasoning as in the one-dimensional case, next we will show the recurrence of the ``pseudo-random walk'' in the absence of anchors. To formulate this idea, we define the an auxiliary network of the original network, in which there is no anchor.

\begin{definition}[Auxiliary Network]\label{def:auxiliary}
The auxiliary network is obtained by treating all anchors in the original network as agents, for which the \ac{efim} takes the following form
\begin{equation}\label{efim_auxiliary}
\widetilde{\M{J}}_{\rm e}(\V{p}_{\rm all}) = \widetilde{\M{D}}-\widetilde{\M{A}} \in \mathbb{R}^{2(N_{\rm a}+N_{\rm b})\times2(N_{\rm a}+N_{\rm b})}
\end{equation}
where $\widetilde{\M{D}} := {\rm diag}(\M{D}_1,\dotsc,\M{D}_{N_{\rm a}+N_{\rm b}})$ is an analogue of $\M{D}$ but includes the \ac{npi}s of nodes in $\Set{N}_{\rm b}$ (previously anchors).
\end{definition}

Similar to the matrix $\M{T}$ in the original network, in the auxiliary network we define
\begin{equation}\label{def_t}
\widetilde{\M{T}}=\widetilde{\M{D}}^{-1}\widetilde{\M{A}}.
\end{equation}
With these definitions we are able to show the recurrence of the random walk over auxiliary networks.

\begin{proposition}[Recurrence of Finite Auxiliary Networks]\label{prop:recurrence}
For any network in which there is no anchor (including the auxiliary network) with $N_{\rm a}+N_{\rm b}<\infty$, if the graph $\Set{G}_{\rm net}$ is connected, we have $\M{F}_{ii}=\M{I}_2$ for all $i\in \Set{N}_{\rm a}$.
\begin{IEEEproof}
Please refer to Appendix \ref{ssec:proof_recurrence_finite}.
\end{IEEEproof}
\end{proposition}

Similar to the formula $F_{ii}=1$ in the one-dimensional case, $\M{F}_{ii}=\M{I}_2$ can be interpreted as ``the random walk starting from $i$ returns almost surely'', in the sense of pseudo-probability. Moreover, with the following definition
\begin{equation}\label{hitting_set_2d}
\M{F}_{\Set{S}}(i,j)=\sum_{n=1}^\infty \sum_{\V{\omega}\in\Omega_n^{i,j}(\Set{S})}\widetilde{\M{T}}_{\V{\omega}}^{(n)}
\end{equation}
we may obtain
\begin{equation}\label{recurrence_normalize}
\M{F}_{ii} + \sum_{j\in \Set{N}_{\rm b}} \M{F}_{\widetilde{\Set{R}}_i}(i,j) = \M{I}_2,~\forall i \in\Set{N}_{\rm a}
\end{equation}
using Proposition \ref{prop:hitting} and \ref{prop:recurrence}, where $\widetilde{\Set{R}}_i:= \{i\}\cup \Set{N}_{\rm b}$. Thus the \ac{efim} for $\V{p}_i$ can be written as
\begin{equation}\label{alt_main}
\efim{\V{p}_i}= \sum_{j\in\Set{N}_{\rm b}}\M{D}_i\M{F}_{\widetilde{\Set{R}}_i}(i,j),~\forall i \in\Set{N}_{\rm a}.
\end{equation}

Now we can see that in the 2-D scenario, we may have a similar ``position information routing'' interpretation of \ac{eoc}. Indeed, using \eqref{hitting_set_2d} and the definitions in Table \ref{tbl:notation}, we can derive the following result from \eqref{alt_main}
\begin{equation}\label{alt2_main_2d}
\M{D}_i\!=\!\underbrace{\sum_{j\in\Set{N}_{\rm b}}\sum_{n=1}^{\infty} \sum_{\V{\omega}\in\Set{\Omega}_n^{i,j}(\widetilde{\Set{R}}_i)}\M{D}_i\widetilde{\M{T}}_{\V{\omega}}^{(n)}}_{\efim{\V{p}_i}} + \sum_{n=1}^{\infty} \sum_{\V{\omega}\in\Set{\Omega}_n^{i,i}(\widetilde{\Set{R}}_i)}\M{D}_i\widetilde{\M{T}}_{\V{\omega}}^{(n)}
\end{equation}
which takes a similar form as \eqref{alt2_main}.

\section{Large-Scale Networks}\label{sec:lattice}
In this section, we develop asymptotic expressions for \ac{eoc} (hence \ac{efim}) in certain types of large-scale networks, using the method introduced in Sec. \ref{sec:quasi_rw}. Especially, we are interested in the asymptotic \ac{eoc} of agents as their distances to the nearest anchors increase. Intuitively, \ac{eoc} should be decreasing with the distance to the nearest anchors. This may be interpreted as an \emph{error propagation behavior}, which is a major issue in cooperative networks. To elaborate, when anchors are spatially sparse, the agents being next to anchors are likely to have the best localization accuracy, while those being far from anchors may perform worse since they have to rely entirely on neighboring agents that are themselves subject to position uncertainty.

For simplicity, in this section we consider the case where nodes are equipped with \ac{uoa}s when $N_{\rm t}\geq 2$, and hence the orientations of nodes do not have an impact on the results.\footnote{For \ac{uoa}s, this implies $N_{\rm t}\geq 3$. When nodes are equipped with other types of arrays, if their orientations can be modeled as independently uniform distributed random variables, the results in this section can still serve as looser lower bounds of the expected position \ac{mse} due to Jensen's inequality.}

\subsection{Large Lattice Networks}\label{ssec:lattice}
We first investigate the asymptotic \ac{eoc} in large lattice networks. The notion ``lattice networks'' considered here refers to networks in which agents are located in a connected subset of the space $\mathbb{Z}^2$. They may be regarded as highly simplified models of real-world large-scale wireless networks, but one can typically obtain insightful results therefrom due to their symmetry. In particular, by saying a lattice network is ``large'', we mean a network satisfying the following technical conditions.

\begin{condition}\label{asu:lattice}
We consider a finite lattice network within an expanding rotational symmetric region centered at the origin $\V{0}$. The network tends to an infinite lattice network in the sense that $D_{\rm net}\rightarrow \infty$, where $D_{\rm net}:= \max_{\{j,k\}\subseteq\Set{N}_{\rm a}\cup\Set{N}_{\rm b}}\|\V{d}_{jk}\|$ is the network diameter. The centroid of the network does not change as it expands. Whenever we say the distance between two nodes $\|\V{d}_{ij}\|\rightarrow \infty$, its rate of growth is lower than that of $D_{\rm net}$, i.e., $\|\V{d}_{ij}\|=o(D_{\rm net})$.
\end{condition}

To elaborate on our method of analysis, we take a brief look back on the 1-D toy example. For an infinitely large lattice network, it is known that the $n$-step transition probability $T_{ij}^{(n)}$ tends to be the probability density function of a Gaussian distribution as $n\rightarrow \infty$ \cite[Sec. 7]{random_walk_principles}, taking the following form
\begin{equation}\label{clt}
T_{ij}^{(n)} \sim \frac{1}{\sqrt{2\pi n\sigma^2}} \exp\Big\{-\frac{1}{2n\sigma^2}(p_i-p_j)^2\Big\}
\end{equation}
where $\sigma^2$ is a constant. The desired asymptotic behavior of \ac{eoc} can then be derived by calculating the asymptotic hitting probabilities with the help of \eqref{clt}.

In its essence, \eqref{clt} is a result of the celebrated central limit theorem, which may be derived using the technique of characteristic function \cite[Sec. 7]{random_walk_principles}. In light of this, we may follow a similar line of reasoning in the analysis of the 2-D scenario, while bearing in mind some technical differences between transition probabilities and pseudo-probabilities. To prevent the paper from being crammed with technical details that are not directly related to the main topic, here we will only present the main results, while the technicalities will be deferred to Appendix \ref{sec:prw}.

We first present the simple case where there is only a single anchor $\nu$ located at the origin of the network, namely we have $\V{p}_{\nu}=\V{0}$. Using the notations introduced in Sec. \ref{ssec:random_walk_2d}, we have $\widetilde{\Set{R}}_i=\{\nu,i\}$ for any agent $i$.

\subsubsection{Single Anchor, $N_{\rm t}\geq 2$}
When $N_{\rm t}\geq 2$, each agent is able to obtain both angular and distance information from its neighbors. In this case, we have the following result on the asymptotic behavior of \ac{efim}.

\begin{proposition}[Position Information Path Loss]\label{prop:inf_pathloss}
For lattice networks satisfying Condition \ref{asu:lattice}, when $\nu$ is the only anchor, for the $N_{\rm t}\geq 2$ case, the \ac{speb} of agent $i$ scales as
\begin{equation}\label{single_anchor_limit}
{\rm sp}\{\V{p}_i\} =\Theta(\log \|\V{d}_{i\nu}\|)
\end{equation}
as $\|\V{d}_{i\nu}\|\rightarrow \infty$, if $\|\V{d}_{i\nu}\| = o(D_{\rm net}/2-\|\V{p}_i\|)$.
\begin{IEEEproof}
Please refer to Appendix \ref{ssec:proof_inf_pathloss}.
\end{IEEEproof}
\end{proposition}

Proposition \ref{prop:inf_pathloss} implies that the position information an agent acquired decreases logarithmically with the distance to the anchor in an asymptotic regime, as long as agent $i$ is closer to the anchor than the network edge (as implied by the condition $\|\V{d}_{i\nu}\| = o(D_{\rm net}/2-\|\V{p}_i\|)$). This may be viewed as a law of \emph{position information path loss}. An interesting fact is that the path loss of the receiving signal power does not have an effect on the order of position information path loss. We will revisit this fact in Sec. \ref{ssec:rgg}.

\subsubsection{Single Anchor, $N_{\rm t}=1$}
When $N_{\rm t}=1$, each agent is only able to receive distance information from its neighbors. Next we show that when there is only a single anchor in the network, this will prevent agents from obtaining the position information on the direction perpendicular to that from them to the anchor.
\begin{proposition}[Range-only Position Information Path Loss]\label{prop:rank3}
For lattice networks satisfying Condition \ref{asu:lattice}, when $\nu$ is the only anchor, for the $N_{\rm t}=1$ case, the \ac{efim} of agent $i$ is a rank-1 matrix given by
\begin{equation}\label{efim_rank3}
\efim{\V{p}_i} = \lambda_{i\nu} \M{J}_{\rm r}(\varphi_{i\nu}).
\end{equation}
The term $\lambda_{i\nu}$ satisfies
\begin{equation}\label{radial_rank3}
\lambda_{i\nu}=\Theta\Big((\log\|\V{d}_{i\nu}\|)^{-1}\Big)
\end{equation}
as long as $\|\V{d}_{i\nu}\| = o(D_{\rm net}/2-\|\V{p}_i\|)$.
\begin{IEEEproof}
Please refer to Appendix \ref{ssec:proof_rank3_sketch}.
\end{IEEEproof}
\end{proposition}

The fact that $\efim{\V{p}_i}$ is rank-1 agrees with our intuition that the network suffers from rotational ambiguity when there is only one anchor. In fact, according to the discussions in Appendix \ref{ssec:proof_rank3_sketch}, \eqref{efim_rank3} holds in any network (i.e., not limited to lattice networks) where there is only a single anchor $\nu$, while the asymptotic behavior of $\lambda_{i\nu}$ may vary.

\subsubsection{Uniformly Distributed Anchors}
Next we investigate the scenario where anchors are uniformly distributed on a lattice with constant density $\lambda_{\rm anc}$. Specifically, the positions of anchors take the following form
\begin{equation}\label{uni_anchor}
\V{p}_i = (\lambda_{\rm anc})^{-\frac{1}{2}}[k~l]^{\rm T},~k\in\mathbb{Z},~l\in\mathbb{Z},~\forall i \in \Set{N}_{\rm b}.
\end{equation}
Since anchors are spread out over the entire network area, it may not be insightful to investigate the \ac{speb} of a certain agent, and hence we consider the average \ac{speb} of agents.

To relate the previous results in the single anchor case to the multi-anchor scenario considered here, we first conceive a simple \ac{efim} lower bound. Specifically, for any agent $i$, the projection of its \ac{efim} on an arbitrary direction corresponding to the unit vector $\V{u}$, i.e., $\V{u}^{\rm T}\efim{\V{p}_i}\V{u}$, can be bounded from below by
\begin{equation}\label{more_is_more}
\V{u}^{\rm T}\efim{\V{p}_i}\V{u}\geq \max_{j \in \Set{N}_{\rm b}} \V{u}^{\rm T}\M{J}_{\rm e}^{(j)}(\V{p}_i)\V{u}
\end{equation}
where the notation $\M{J}_{\rm e}^{(j)}(\V{p}_i)$ denotes the \ac{efim} when anchor $j$ is the only anchor in the network. To elaborate, it is clear that introducing new anchors into the network leads to a non-negative contribution to $\efim{\V{p}}$, and hence to $\efim{\V{p}_i},~\forall i \in \Set{N}_{\rm a}$. Therefore, using the definition of positive-semidefiniteness we have
$$
\V{u}^{\rm T}\Big(\efim{\V{p}_i}- \max_{j \in \Set{N}_{\rm b}}\M{J}_{\rm e}^{(j)}(\V{p}_i)\Big)\V{u}\geq 0
$$
yielding \eqref{more_is_more}. With the help of \eqref{more_is_more}, we generalize the results in the single anchor case as follows.
\begin{proposition}\label{prop:avg_speb_uda}
For lattice networks satisfying Condition \ref{asu:lattice}, when anchors are distributed as \eqref{uni_anchor}, the average \ac{speb} of agents in the ``interior area'' $\Set{I}_\epsilon$ of the network scales as
\begin{equation}\label{avg_speb_uda}
\frac{1}{|\Set{I}_\epsilon|} \sum_{\V{p}_i\in \Set{I}_\epsilon} {\rm sp}\{\V{p}_i\} = \Theta(\log \lambda_{\rm anc}^{-1})
\end{equation}
as $\lambda_{\rm anc}\rightarrow 0_+$ for any constant $\epsilon>0$, if $\lambda_{\rm anc}^{-\frac{1}{2}} = o(D_{\rm net})$. The area $\Set{I}_\epsilon$ is defined as $\Set{I}_\epsilon:= \{\V{p}_i|\|\V{p}_i\|<(1-\epsilon) D_{\rm net}/2,i\in \Set{N}_{\rm a}\}$.
\begin{IEEEproof}
Please refer to Appendix \ref{ssec:proof_avg_speb_uda_sketch}.
\end{IEEEproof}
\end{proposition}

Note that Proposition \ref{prop:avg_speb_uda} is valid for both $N_{\rm t}\geq 2$ and $N_{\rm t}=1$. Intuitively, though a single anchor can only provide position information on a specific direction in the $N_{\rm t}=1$ case, uniformly distributed anchors can provide position information from all directions. Therefore, in light of \eqref{more_is_more}, the asymptotic average \ac{speb} can be bounded as \eqref{avg_speb_uda}.

\subsection{Stochastic Geometric Networks}\label{ssec:rgg}
Stochastic geometric networks are networks in which nodes are uniformly distributed within a specific region $\Set{B}\subseteq \mathbb{R}^2$. Nodes in such networks constitute binomial point processes \cite{HaeAndBacDouFra:09,DarConBurVer:07,sg_2,WinPinShe:J09,tutorial_sg}. Compared to lattice networks, they are better models for practical wireless networks, and have been widely applied to model cellular networks, unmanned aerial vehicle networks, and terrestrial vehicular networks \cite{tutorial_sg,sg_uav,sg_vnet1,sg_vnet2}. In particular, agents in networks employing cooperative localization are typically mobile devices, and hence the stochastic geometric modelling is often preferable to its deterministic counterpart. However, stochastic geometric networks no longer possess the translation-invariant property we relied substantially on for proving the results in Sec. \ref{sec:lattice}.

Nevertheless, some recent works indicate that large \acp{rgg} are in some sense ``asymptotically equivalent'' to lattice graphs \cite{rai2004the}. In light of this, in this subsection, we extend some of the results in Section \ref{ssec:lattice} to stochastic geometric networks. For technical reasons, we assume that the networks satisfy the following condition.

\begin{condition}\label{cond:connected}
We consider stochastic geometric networks which remains to be connected as $|\Set{R}_{\rm net}|\rightarrow\infty$ with probability $1-o(N_{\rm a}^{-1})$ as $|\Set{R}_{\rm net}|\rightarrow \infty$.
\end{condition}

\subsubsection{Single anchor}
Consider a stochastic geometric network with a single anchor $\nu$. Since the nodes are randomly located within the network area, it may be difficult to depict the asymptotic \ac{speb} of any single agent. Instead, we will characterize the average \ac{speb} of agents within a certain area. Specifically, we consider a ring-shaped region $\RS{S}_{\nu}(R,r)$ centered at $\RV{p}_{\nu}$ given by
$$
\RS{S}_{\nu}(R,r) := \Big\{\V{x}\Big|R-\frac{r}{2}\le\|\V{x}-\RV{p}_{\nu}\|\leq R+\frac{r}{2},~\V{x}\in\mathbb{R}^2\Big\}
$$
where $r$ is a positive constant. Roughly speaking, $\RS{S}_{\nu}(R,r)$ contains all agents having distance approximately $R$ to the anchor.
\begin{theorem}\label{thm:rgg}
For a stochastic geometric network satisfying Condition \ref{cond:connected}, if $R\rightarrow \infty$ as $|\Set{R}_{\rm net}|\rightarrow \infty$ but $RN_{\rm a}^{-1}\rightarrow 0$, when $N_{\rm t}\geq 2$, the average \ac{speb} of the agents within $\RS{S}_{\nu}(R,r)$ scales as
\begin{equation}\label{na2_rgg}
\frac{1}{|\RS{A}_{\nu}(R,r)|}\sum_{\RV{p}_i\in\RS{S}_{\nu}(R,r)}{\rm sp}\{\RV{p}_i\}=\Theta\big(\log R\big)
\end{equation}
with probability approaching $1$ as $N_{\rm a}\rightarrow \infty$, where $\RS{A}_{\nu}(R,r)$ denote the set of agents located in $\RS{S}_{\nu}(R,r)$. For the $N_{\rm t}=1$ case, with probability approaching $1$ we also have
\begin{equation}\label{na1_rgg}
\frac{1}{|\RS{A}_{\nu}(R,r)|}\sum_{\RV{p}_i\in\RS{S}_{\nu}(R,r)}\big(\RV{u}(\rv{\phi}_{i\nu})^{\rm T}\refim{\RV{p}_i}\RV{u}(\rv{\phi}_{i\nu})\big)^{-1}=\Theta\big(\log R\big).
\end{equation}
\begin{IEEEproof}
Please refer to  Appendix \ref{sec:proof_rgg}.
\end{IEEEproof}
\end{theorem}

In extended networks, it is known that if the agents do not cooperate, the Fisher information an anchor provides for an agent is proportional to the received \ac{snr} \cite{SheWin:J10a}, which typically exhibits a polynomial decay as the distance between these two nodes increases due to the path loss of receiving power. In contrast, Theorem \ref{thm:rgg} indicates that with cooperation among agents, the rate of position information path loss is only \emph{logarithmic}, showing the great superiority of cooperative localization over its non-cooperative counterpart.

\subsubsection{Uniformly Distributed Anchors}
Now we consider the case that anchors constitute a binomial point process in the network region $\Set{R}_{\rm net}$ with intensity $\lambda_{\rm anc}$. As the network region $\Set{R}_{\rm net}$ expands, $\lambda_{\rm anc}\rightarrow 0$ slowly such that $\lambda_{\rm anc}^{-1}=o(|\Set{R}_{\rm net}|)$, namely there remains at least a constant number of anchors in the network. We assume for convenience that the network region $\Set{R}_{\rm net}$ is a circular area with diameter $D_{\rm net}$. For such networks, we have the following result.

\begin{proposition}\label{prop:rgg_uda}
The average \ac{speb} of agents in the ``interior area'' $\Set{I}_{\epsilon}$ of the network scales as
\begin{equation}
\frac{1}{|\Set{I}_{\epsilon}|}\sum_{\V{p}_i\in \Set{I}_{\epsilon}}{\rm sp}\{\V{p}_i\} = O(\log\lambda_{\rm anc}^{-1})
\end{equation}
with probability approaching $1$ as $N_{\rm a}\rightarrow \infty$, for any constant $\epsilon >0$. The area $\Set{I}_\epsilon$ is defined as $\Set{I}_\epsilon:= \{\V{p}_i|\|\V{p}_i\|<(1-\epsilon) D_{\rm net}/2,i\in \Set{N}_{\rm a}\}$.
\begin{IEEEproof}
Please refer to Appendix \ref{sec:proof_rgg_uda}.
\end{IEEEproof}
\end{proposition}

For a given anchor intensity $\lambda_{\rm anc}$, the distance from a ``typical agent'' to the nearest anchor would be at the order of $\lambda_{\rm anc}^{-\frac{1}{2}}$. In light of this, the logarithmic error scaling in Proposition \ref{prop:rgg_uda} follows intuitively from Theorem \ref{thm:rgg}, since the average \ac{speb} of ``typical agents'' would increase logarithmically with their distances to the nearest anchors.

\section{Discussions}\label{sec:discussions}
In this section we discuss the practical implications of our results, and relate them to some existing works.
\begin{figure}[t]
    \centering
    \begin{overpic}[width=.4\textwidth]{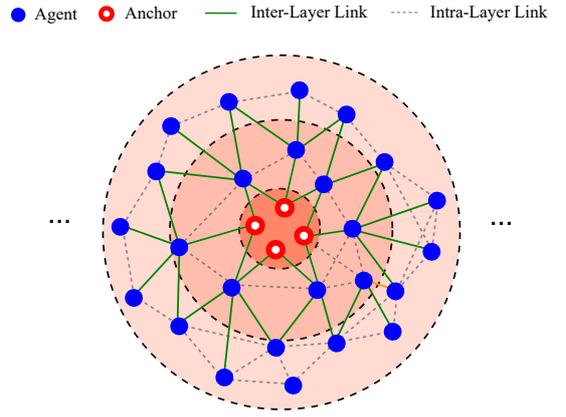}
    \end{overpic}
    \caption{Illustration of the concentric network model considered in \cite{concentric}. Agents in the first layer can communicate with anchors directly, while those in the $n$th layer ($n>1$) are $n$ hops away from anchors.}
    \label{fig:concentric}
\end{figure}
\subsection{Error Propagation in Concentric Multi-hop Wireless Network Model}\label{ssec:concentric}
In the paper \cite{concentric}, the performance limit of cooperative localization in a concentric multi-hop wireless network model (for which $N_{\rm t}=1$) is investigated using methods from stochastic geometry. In this network model, anchors locate in the innermost circular area of the network, while agents locate in the outer ring-shaped area, as illustrated in Fig. \ref{fig:concentric}. It is shown in \cite{concentric} that the expected \ac{speb} of an agent $i$ scales as $\Theta(\|\V{d}_{i{\rm \bar{A}}}\|^2)$, where $\V{d}_{i{\rm \bar{A}}}$ is the distance from $\V{p}_i$ to the network center.

Using the methods proposed in this paper, we can obtain further insights into the asymptotic localization performance in this network. More precisely, from \eqref{more_is_more} and Theorem \ref{thm:rgg}, we see that the expected \ac{dpeb} on the direction from $\V{p}_i$ to the network center scales as $\Theta(\log \|\V{d}_{i{\rm \bar{A}}}\|)$, which is a far lower increasing rate than $\Theta(\|\V{d}_{i{\rm \bar{A}}}\|^2)$ in \cite{concentric}. This implies that the \ac{dpeb} on the orthogonal direction scales as $\Theta(\|\V{d}_{i{\rm \bar{A}}}\|^2)$.  Intuitively, since anchors are confined within a small region, the network is subject to large rotational uncertainty, which leads to the $\Theta(\|\V{d}_{i{\rm \bar{A}}}\|^2)$ error scaling.

In summary, for concentric networks where $N_{\rm t}=1$, when our main interest is to estimate the distance from agent $i$ to the network center, we can expect an error scaling of $\Theta(\log\|\V{d}_{i{\rm \bar{A}}}\|)$ instead of $\Theta(\|\V{d}_{i{\rm \bar{A}}}\|^2)$.

\subsection{Trade-off Between Communication Complexity and Localization Accuracy}\label{ssec:complexity_vs_accuracy}
Typical distributed cooperative localization algorithms include Bayesian message-passing-based methods \cite{proc_henk} and distributed convex optimization-based methods \cite{geert_ml}. These methods may attain near-optimal localization accuracy, but they generally require iterative information exchange among agents, leading to considerable delay and communication overhead.

By contrast, sequential multi-hop localization methods, as exemplified by DV-hop-based methods \cite{dvhop}, work in a layer-by-layer manner and do not require iterations. Basically, DV-hop-based methods consist of a distance estimation stage and a position estimation stage. In the first stage, the distances between each pair of agent and anchor that are $n$ hops from each other are approximated by the sum of all ``raw'' distance estimates in each hop. The ``raw'' distance estimates can be obtained by ranging techniques or from range-free metrics such as average hop-distance \cite{avg_hop_dist} and number of common neighbors \cite{ppp_mhl}. In the second stage, the positions of all agents are estimated locally based on the anchor-agent distance estimates via standard non-cooperative localization algorithms, e.g., weighted least squares.

Naturally, it is of great interest whether the relative simplicity of DV-hop-based methods comes with the cost of accuracy degradation. For simplicity, we assume that the initial distance estimates for each hop have errors with constant variance, and are mutually independent. Thus for any agent $i$, the \ac{mse}s of the final distance estimates are at least proportional to the number of hops between $i$ and the nearest anchors. This implies that for a given anchor density $\lambda_{\rm anc}$, the average localization \ac{mse} is at least linear in $\lambda_{\rm anc}^{-\frac{1}{2}}$. By contrast, from Proposition \ref{prop:rgg_uda} we see that for the same $\lambda_{\rm anc}$, the optimal localization \ac{mse} is at most logarithmic in $\lambda_{\rm anc}^{-1}$.

In summary, in large networks where anchors are sparse, the localization accuracy of DV-hop-based methods are far from optimal, in the sense that their average localization \ac{mse} scales as $\Omega(\lambda_{\rm anc}^{-\frac{1}{2}})$, as opposed to the optimal scaling $O(\log \lambda_{\rm anc}^{-1})$. This can be regarded as an accuracy penalty due to their low communication overhead. For a more in-depth analysis on the trade-off between communication complexity and localization accuracy, one may have to resort to rate-distortion methods \cite{cover_eit}.

\subsection{Reducing the Complexity of Network Operation Techniques in Massive Cooperative Localization Networks}
Network operation techniques, including power allocation and node placement methods, have been introduced to cooperative localization networks to improve the localization performance \cite{DaiSheWin:J15a,netOp_TIT}. For massive networks, the pseudo-probability analysis in Appendix \ref{sec:prw} may be used to derive low-complexity optimization algorithms for network operation tasks.

In particular, for a stochastic geometric network, a general network operation problem may be formulated as follows
\begin{equation}\label{general_opt}
\min_{\V{x}} ~ C(\V{x}):=\mathbb{E}_{\RV{p}}\{\tr{\RM{J}_{\rm e}^{-1}(\RV{p})}\}
\end{equation}
where $\V{x}$ denotes the parameters to be optimized, which corresponds to the power allocation coefficients in power allocation problems, or the positions of anchors in anchor placement tasks. In practice, one may approximate the expectation operation in \eqref{general_opt} by Monte Carlo sampling, and obtain the following objective function
\begin{equation}\label{obj_opt_approx}
\widetilde{C}(\V{x}) = \frac{1}{N_{\rm s}} \sum_{n=1}^{N_{\rm s}} \tr{\M{J}_{\rm e}^{-1}\big(\V{p}^{(n)}\big)}
\end{equation}
where $\V{p}^{(n)}$ denotes the $n$-th sample of the random vector $\RV{p}$, and $N_{\rm s}$ represents the number of samples. The minimization of $\widetilde{C}(\V{x})$ can then be solved by means of gradient-based methods, or gradient-free methods such as the Nelder-Mead algorithm \cite{nm_algo}. In either case, we assume that the optimization algorithm requires $K$ evaluations of the objective function $\widetilde{C}(\V{x})$.\footnote{In gradient-based methods, the gradient can be approximately computed using a fixed number of objective function evaluations.} A naive method is to directly compute the objective function \eqref{obj_opt_approx}, which involves inversion of the $2N_{\rm a}\times 2N_{\rm a}$ matrices $\M{J}_{\rm e}^{-1}\big(\V{p}^{(n)}\big)$. In this case, the computational complexity of the entire optimization algorithm is thus $O(KN_{\rm s}N_{\rm a}^3)$, which could be prohibitive for massive networks having large $N_{\rm a}$.

Alternatively, we may rearrange the objective function as
\begin{equation}
\begin{aligned}
\widetilde{C}(\V{x}) &= \frac{1}{N_{\rm s}}\sum_{n=1}^{N_{\rm s}} \sum_{i=1}^{N_{\rm a}} \tr{\M{J}_{\rm e}^{-1}\big(\V{p}_i^{(n)}\big)} \\
&=\frac{1}{N_{\rm s}}\sum_{n=1}^{N_{\rm s}} \sum_{i=1}^{N_{\rm a}}\tr{\M{D}_i^{(n)}(\M{I}-\M{F}_{ii}^{(n)})}
\end{aligned}
\end{equation}
where the notations having superscript $(n)$ denote the corresponding quantities in the $n$-th network sample. In the following discussion, we will omit the superscript when there is no confusion. Under this formulation, the complexity is given by $O(KN_{\rm s}N_{\rm a}F)$, where $F$ denotes the complexity of computing the matrix $\M{F}_{ii}$. According to Theorem \ref{thm:potential} in Appendix \ref{ssec:potential}, the complexity of computing $\M{F}_{ii}$ is dominated by computing the $(2N_{\rm b}+1)\times (2N_{\rm b}+1)$ matrix $\M{P}_{\widetilde{\Set{R}}_i}^{-1}$. Note that the matrix $\M{P}_{\widetilde{\Set{R}}_i}$ can be partitioned as
$$
\M{P}_{\widetilde{\Set{R}}_i} = \left[
                                  \begin{array}{cc}
                                    {[\M{P}_{\widetilde{\Set{R}}_i}]}_{1,1} & {[\M{P}_{\widetilde{\Set{R}}_i}]}_{1,2:N_{\rm b}+1} \\
                                    {[\M{P}_{\widetilde{\Set{R}}_i}]}_{2:N_{\rm b}+1,1} & \M{P}_{\Set{N}_{\rm b}} \\
                                  \end{array}
                                \right].
$$
The complexity of computing $\M{P}_{\widetilde{\Set{R}}_i}^{-1}$ can be reduced by exploiting the partitioned structure. Specifically, the inverse of the matrix $\M{P}_{\Set{N}_{\rm b}}$ can be computed in advance, and reused when computing $\M{P}_{\widetilde{\Set{R}}_i}$ for different $i$. Finally, for massive networks, the $2\times 2$ blocks in $\M{P}_{\widetilde{\Set{R}}_i}$ may be approximated by the asymptotic formula in Proposition \ref{prop:logarithmic}. We now see that the complexity of the entire optimization algorithm is given by $O[KN_{\rm s}(N_{\rm a}N_{\rm b}^2+N_{\rm b}^3)]$, which is substantially lower than the $O(KN_{\rm s}N_{\rm a}^3)$ complexity of the naive approach when $N_{\rm b}\ll N_{\rm a}$.

\section{Numerical Results}\label{sec:numerical}
In this section we illustrate some results we obtained with numerical examples.

\subsection{Illustration of the Proposed Scaling Laws}
In this subsection, we provide numerical results demonstrating the scaling laws of localization error discussed in Sec. \ref{sec:lattice}. Unless stated otherwise, the parameters take the following values by default in all simulations. The path loss exponent $\gamma=3$, signal bandwidth $\beta=10$\,MHz, the ranging coefficient (defined after \eqref{approx_jd}) $\zeta_{ij}=1$ for all $i\leftrightarrow j$, the carrier frequency $f_{\rm c} = 2$\,GHz, and the \ac{snr} measured at $1$m from the transmitting antenna is $40$\,dB. All nodes locate in a disk with diameter $D_{\rm net}$, and the node density is $2.5\times 10^{-3}$\,m$^{-2}$. For the $N_{\rm t}\geq 2$ case, we consider the scenario where $N_{\rm t}=3$ and antennas are arranged in a uniform circular array with diameter $0.3$m. Receivers on the nodes are able to detect signals when ${\rm SNR}\geq -15$\,dB.

\begin{figure}[t]
    \centering
    \psfrag{Distance to Anchor [m]}[][]{\fitsize Distance to Anchor [m]}
    \psfrag{Rnet=500, 1000, 2000mmmmmmm}{\hspace{0mm}\fitsizesm $D_{\rm net}\!=\!500,1000,2000$\thinspace m}
    \includegraphics[width=.445\textwidth]{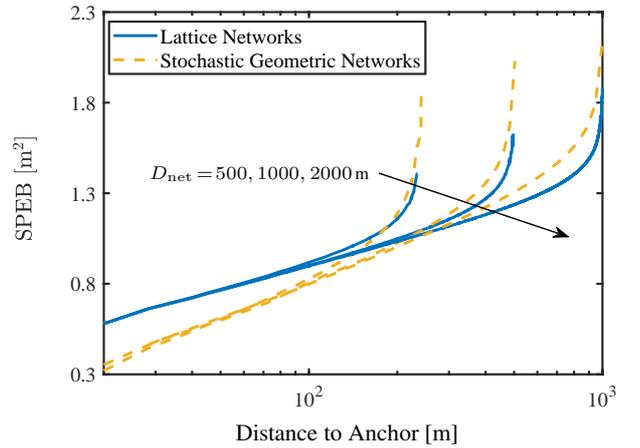}
    \caption{\ac{speb} as a function of the distance to the single anchor in lattice and stochastic geometric networks under the $N_{\rm t}\geq 2$ case for varying network diameter $D_{\rm net}$.}
    \label{fig:inf_pathloss}
\end{figure}

First we investigate the law of position information path loss in both lattice and stochastic geometric networks under the $N_{\rm t}\geq 2$ case. Fig. \ref{fig:inf_pathloss} illustrates the \ac{speb} of agents in the single anchor scenario. A single anchor locates at the network centroid $(0{\rm m},0{\rm m})$, and the network diameter $D_{\rm net}$ varies from $500$m to $2000$m. For stochastic geometric networks, the results are averaged over $1000$ network snapshots. As implied by Proposition \ref{prop:inf_pathloss} and Theorem \ref{thm:rgg}, in both types of networks, the \ac{speb} of agents grows logarithmically as the distance from the anchor increases, except that there is sudden increment in \ac{speb} at the network edge since the agents at network edge can only communicate with fewer neighbors.

\begin{figure}[t]
    \centering
    \includegraphics[width=.348\textwidth]{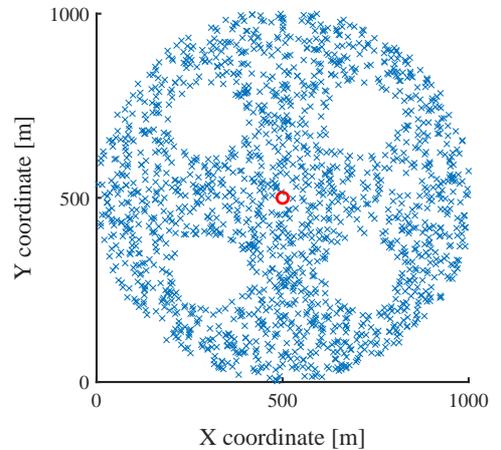}
    \caption{Illustration of a stochastic geometric network containing four circular ``holes'', each with radius $r_{\rm h}=100$m. The crosses represent agents, while a single anchor marked by a circle locating at the center of the network region.}
    \label{fig:four_holes}
\end{figure}

\begin{figure}[t]
    \centering
    \includegraphics[width=.445\textwidth]{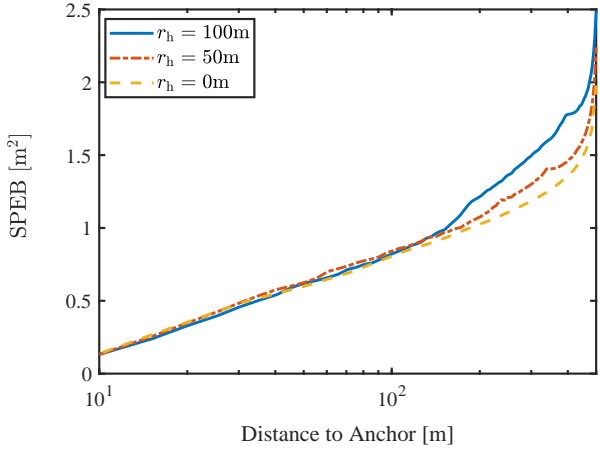}
    \caption{\ac{speb} vs. the distance to the single anchor in stochastic geometric networks with $D_{\rm net}=1000$m under the $N_{\rm t}\geq 2$ case containing four circular ``holes'' with various radius $r_{\rm h}$.}
    \label{fig:rgg_holes}
\end{figure}

To further illustrate the network boundary effect, we consider the case where there are ``holes'' in the network region that do not contain any node. This could be related to practical scenarios where the network region contains large obstacles. In particular, we consider a stochastic geometric network with $D_{\rm net}=1000$m containing four circular holes centered at (300m, 300m), (300m, 700m), (700m, 300m), and (700m, 700m), respectively. Each hole has radius $r_{\rm h}$. A single anchor locates at the network centroid (500m, 500m). An instance of the network is shown in Fig. \ref{fig:four_holes}. Under previous assumptions, we portray the relationship between the \ac{speb} of agents and their distances to the anchor is shown in Fig. \ref{fig:rgg_holes}. We observe that when the distance to the anchor is larger than a threshold (around 150m), the \ac{speb} of agents in networks containing holes starts to deviate from that of the network without holes. Especially, the deviation is exacerbated when $r_{\rm h}$ is larger, since the boundary effect caused by the holes becomes stronger as $r_{\rm h}$ increases.

\begin{figure}[t]
    \centering
    \psfrag{Inverse Anchor Density}[][]{\fitsize Inverse Anchor Density $\lambda_{\rm anc}^{-1}$ [m$^2$]}
    \psfrag{xxxxxxxxxx}{\footnotesize $N_{\rm b}=1,2,3,4$}
     \psfrag{Distance to Centroid [m]}[][]{\fitsize Distance to $\bar{\V{p}}_{\rm A}$ [m]}
    \includegraphics[width=.445\textwidth]{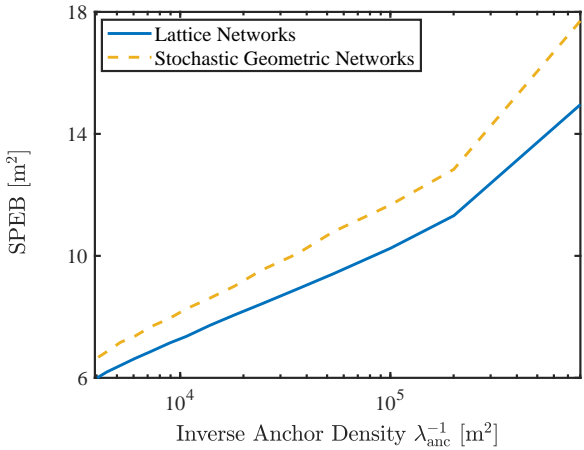}
    \caption{Network average \ac{speb} as a function of the reciprocal of anchor density $\lambda_{\rm anc}^{-1}$ in lattice and stochastic geometric networks where anchors are uniformly distributed under the $N_{\rm t}=1$ case.}
    \label{fig:nt2_multi}
\end{figure}
\begin{figure}[t]
    \centering
    \psfrag{Distance to Anchor [m]}[][]{\fitsize Distance to Anchor [m]}
    \psfrag{Dnet500mxx}{\footnotesize $D_{\rm net}\!=\!500$m}
    \psfrag{Dnet2000mxx}{\footnotesize $D_{\rm net}\!=\!2000$m}
    \psfrag{ple4xxxxx}[r][r]{\footnotesize $\gamma \!=\!3.5$}
    \psfrag{ple375xxxxx}[r][r]{\!\footnotesize $\gamma \!=\!3.25$}
    \psfrag{ple35xxxx}{~~\footnotesize $\gamma \!=\!3$}
    \includegraphics[width=.445\textwidth]{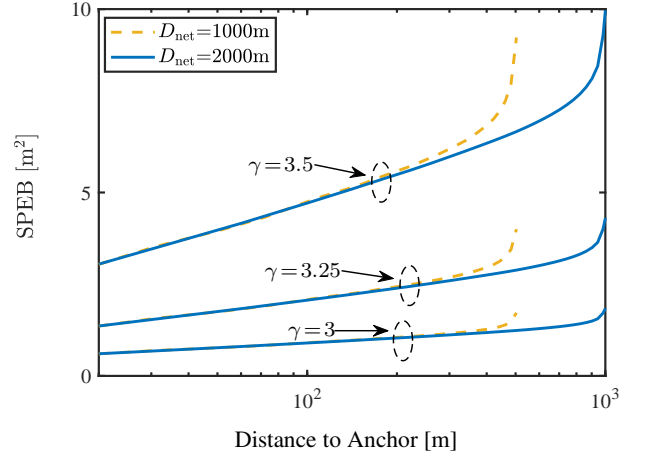}
    \caption{\ac{speb} as a function of the distance to the single anchor in lattice networks under the $N_{\rm t}=2$ case for varying path loss exponent $\gamma$.}
    \label{fig:inf_pathloss_ple}
\end{figure}
\begin{figure}[t]
    \centering
    \psfrag{Distance to Centroid [m]}[][]{\fitsize Distance to $\bar{\V{p}}_{\rm A}$ [m]}
    \psfrag{Rnetabccdd2000}{\fitsizesm $D_{\rm net}\!=\!2000$m}
        \psfrag{Rnetabccdd1000}{\fitsizesm $D_{\rm net}\!=\!1000$m}
        \psfrag{Rnetabccdd500}{\fitsizesm $D_{\rm net}\!=\!500$m}
    \includegraphics[width=.48\textwidth]{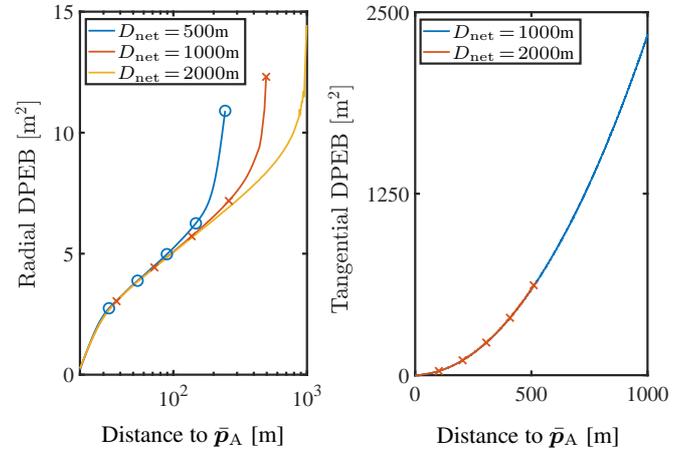}
   \caption{The radial and tangential \ac{dpeb} in lattice networks with varying network diameter $D_{\rm net}$ under the $N_{\rm t}=1$ case. Left: radial \ac{dpeb}; right: tangential \ac{dpeb}.}
    \label{fig:range_only}
\end{figure}

Next we illustrate the asymptotic behavior of network average \ac{speb} in the uniformly distributed anchor scenario. We consider the $N_{\rm t}=1$ case in both lattice and stochastic geometric networks. In lattice networks, anchors are distributed according to \eqref{uni_anchor}, while in stochastic geometric networks they form a binomial point process with the same intensity $\lambda_{\rm anc}$. For such networks with diameter $D_{\rm net}=2000$m, the average \ac{speb} of all agents in the networks are shown in Fig. \ref{fig:nt2_multi}. Results for stochastic geometric networks are averaged over $1000$ network snapshots. It can be seen that the average \ac{speb} grows logarithmically as the reciprocal of density $\lambda_{\rm anc}^{-1}$ increases, as indicated by Propositions \ref{prop:avg_speb_uda} and \ref{prop:rgg_uda}. The abrupt increment of average \ac{speb} when $\lambda_{\rm anc}^{-1}$ is large is due to the network edge effect, which is more significant when there are fewer anchors in the network.

We then demonstrate the impact of path loss exponent $\gamma$ on the \ac{speb} of agents in lattice networks. Here we focus on the single anchor scenario, and two different values of network diameter are considered: $D_{\rm net}=1000$m and $D_{\rm net}=2000$m. The path loss exponent $\gamma$ varies from $3$ to $3.5$, and the \ac{speb} as a function of the distance to the anchor is plotted in Fig. \ref{fig:inf_pathloss_ple}. Similar to Fig. \ref{fig:inf_pathloss}, all \ac{speb}s grows logarithmically as the distance to network centroid increases, except for those agents located at the network edge. However, the slope of these curves are different. As $\gamma$ increases, agents receive less information from their neighbors, and hence the \ac{speb} increases faster.

Finally, we illustrate the localization performance under the $N_{\rm t}=1$ case in the concentric network model discussed in Sec. \ref{ssec:concentric}. We consider lattice networks in which around the network centroid, four anchors are located at $(\pm 20{\rm m},0{\rm m})$ and $(0{\rm m},\pm 20{\rm m})$, respectively. The \ac{dpeb}s of agents in networks with $D_{\rm net}$ varying from $500$m to $2000$m are illustrated in Fig. \ref{fig:range_only}. Here, radial \ac{dpeb} refers to the \ac{dpeb} on the direction from an agent to the network centroid, while tangential \ac{dpeb} refers to the \ac{dpeb} on the orthogonal direction. It can be seen that the radial \ac{dpeb} grows in a similar way as the \ac{speb} in the $N_{\rm t}\geq 2$ case illustrated in Fig. \ref{fig:inf_pathloss}. In addition, the tangential \ac{speb} increases quadratically as the distance to the network centroid increases. These results are consistent with Proposition \ref{prop:inf_pathloss} and the discussions in Sec. \ref{ssec:concentric}.

\begin{figure}
\centering
\subfloat[][Lattice Network]{
    \begin{minipage}[t]{0.47\linewidth}
        \centering
        \includegraphics[width=.97\textwidth]{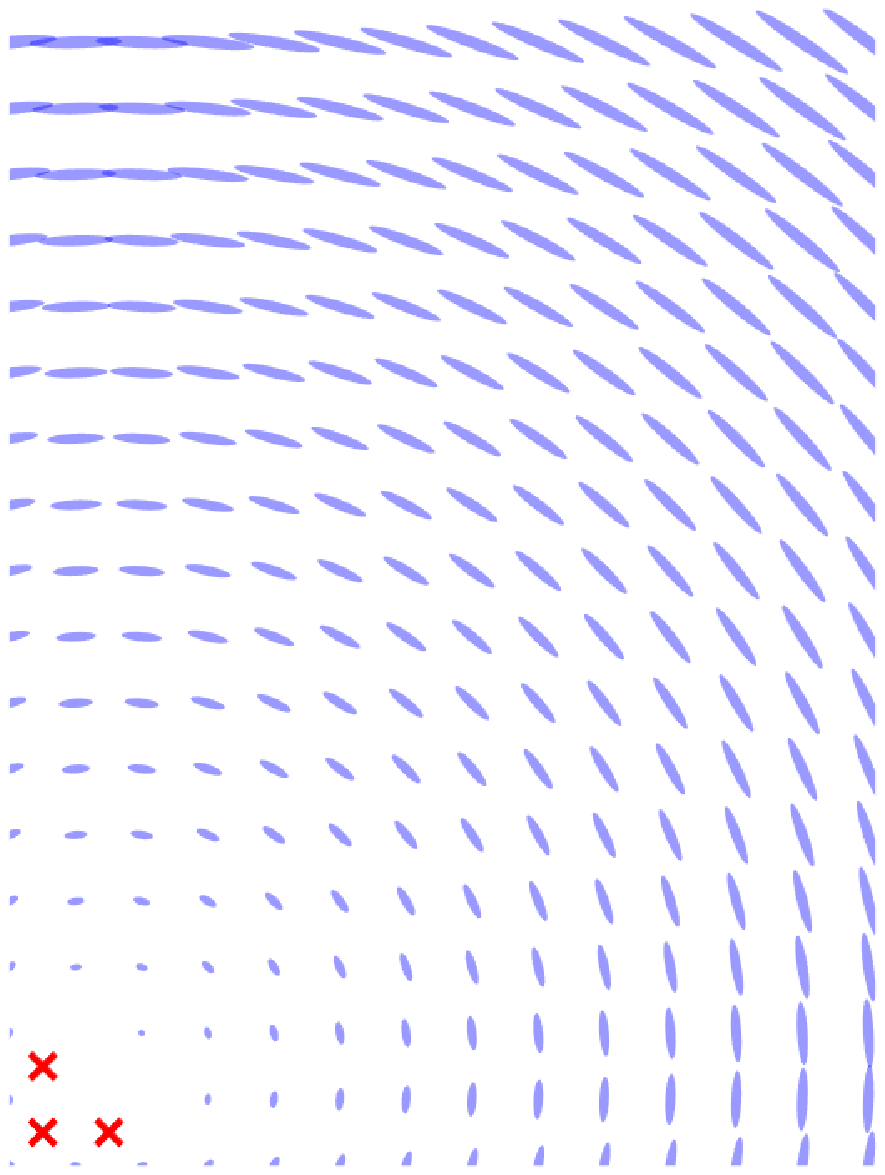}
        \label{fig:dpebs:a}
    \end{minipage}
}
\subfloat[][Stochastic Geometric Network]{
    \begin{minipage}[t]{0.47\linewidth}
        \centering
        \includegraphics[width=.97\textwidth]{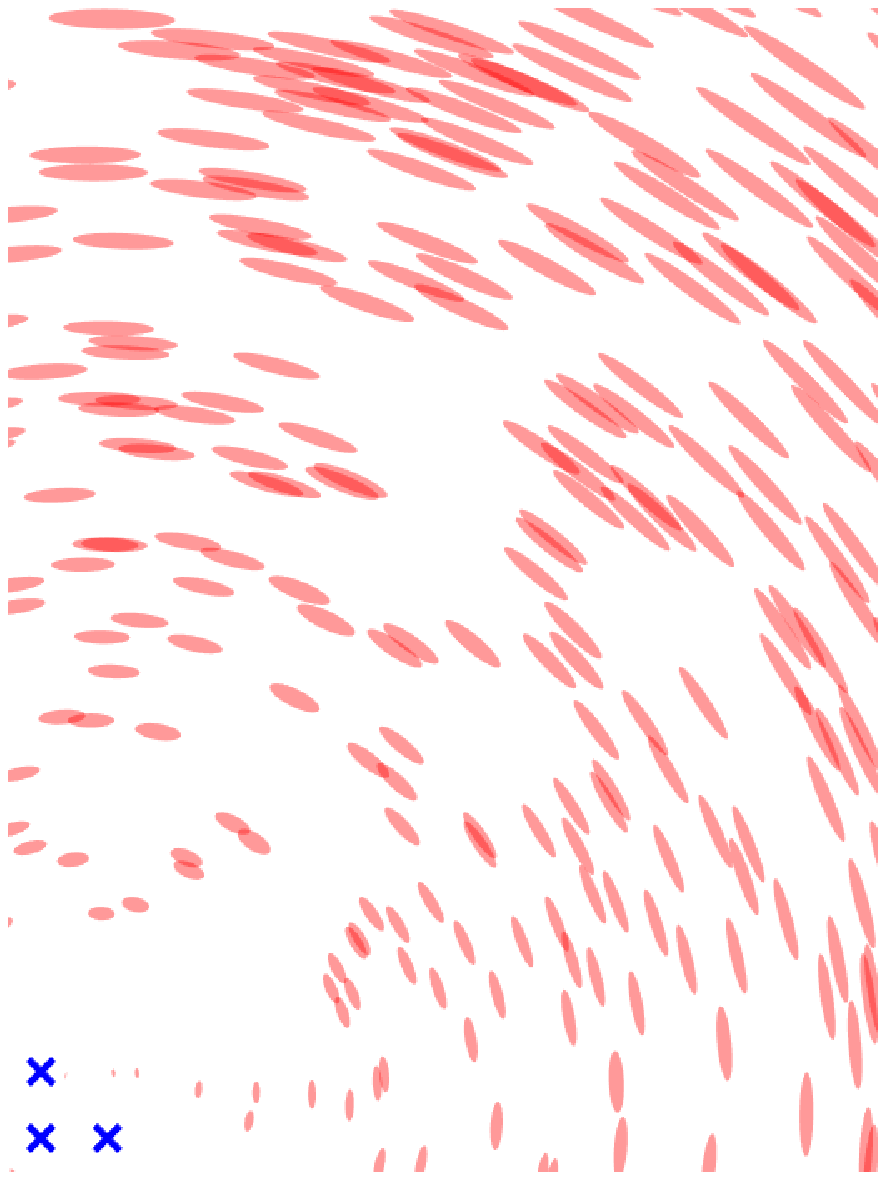}
        \label{fig:dpebs:b}
    \end{minipage}
}
\vspace{2.25mm}
    \caption{Error ellipses of agents in lattice and stochastic geometric networks in the constant anchor region scenario when $N_{\rm t}=1$. Positions of anchors are marked with ``x''.}
    \label{fig:dpebs}
\end{figure}

To better illustrate the behavior of \ac{dpeb} when $N_{\rm t}=1$, Fig. \ref{fig:dpebs} shows the error ellipses in both lattice and stochastic geometric networks. For better demonstration, only part of the networks is shown in the figure. The error ellipse of agent $i$ is defined as the set of point $\V{x}$ satisfying
$$
(\V{x}-\V{p}_i)^{\rm T}\efim{\V{p}_i}(\V{x}-\V{p}_i)=1.
$$
Here the network diameter $D_{\rm net}=500$m. Three anchors locate at a corner of the scene, marked by ``x'' in Fig. \ref{fig:dpebs}.  From the figure it is clear that the tangential \ac{dpeb} contribute most to the \ac{speb} of agents, in either lattice networks or stochastic geometric networks. Furthermore, the length of major axes of the ellipses increase linearly with the distance to the anchors, as implied by Proposition \ref{prop:inf_pathloss} and the discussions in Sec. \ref{ssec:concentric}.

\subsection{Comparison with Actual Algorithms}\label{ssec:comparison_algo}
In this subsection, we compare the proposed error scaling laws with actual algorithms. In particular, we consider two algorithms, including a centralized algorithm (which is expected to be near-optimal) and a sequential algorithm.

For the simulations presented in this subsection, we consider the practical scenario of a wireless sensor network employing pairwise \ac{rss} and \ac{aoa} measurements. Specifically, we assume that the network resides in a disk with diameter $D_{\rm net}=400$m, in which a single anchor locates at the network center. Agents are randomly distributed in the network region with density $3\times 10^{-3}$m$^{-2}$. The \ac{rss} measurements are obtained according to the channel model in the ``park'' scenario described in \cite{pl_model}, given by
\begin{equation}
\rv{y}_{ij}^{\rm RSS} = P_{\rm T} - L_0 - 10\gamma \log_{10}(\|\V{d}_{ij}\|d_0^{-1}) + \rv{n}_{\rm shadow}
\end{equation}
where $\V{d}_{ij}$ is the displacement vector between node $i$ and $j$, $P_{\rm T}=5$dBm denotes the transmitting power, $L_0 = 31$dB denotes the path loss at the reference distance $d_0=1$m, $\gamma=3.69$ represents the path loss exponent \cite{pl_model}, and $\rv{n}_{\rm shadow}$ models the shadowing effect, which follows a zero-mean Gaussian distribution with standard deviation $\sigma_{\rm RSS}=1.42$dB \cite{pl_model}. The bearing measurement errors are modelled as zero-mean von Mises distributed random variables with standard deviation $\sigma_{\rm AOA}=5$ degree \cite{locating_the_nodes}. Each node can acquire pairwise measurements with neighbors within range $R_{\rm max}=43$m, corresponding to a receiver sensitivity of $-100$dBm \cite{pl_model}.

For the centralized algorithm, we consider an iterative approximate maximum likelihood (AML) estimator, which approximates the likelihood function corresponding to the each pairwise measurement as 2-D Gaussian distributions, with an iterative refinement on the estimates of measurement variances. Spefically, in the $\ell$-th iteration, the likelihood function for a pair of nodes $i$ and $j$ is approximated as follows
\begin{equation}\label{likelihood_approx}
\begin{aligned}
&\rv{g}^{(\ell)}_{\rv{\alpha}_{ij},\rv{r}_{ij}}(\alpha_{ij},r_{ij};\V{p}_i,\V{p}_j)\\
&\hspace{2mm}\propto \exp \Big\{\!-\!\frac{1}{2} (\V{p}_i\!-\!\V{p}_j\!-\!\hat{\V{d}}_{ij})^{\rm T}\M{U}_{ij}\RM{\Lambda}_{ij}^{(\ell)}\M{U}_{ij}^{\rm T} (\V{p}_i\!-\!\V{p}_j\!-\!\hat{\V{d}}_{ij})\Big\}
\end{aligned}
\end{equation}
where $\hat{\V{d}}_{ij} = \rv{r}_{ij}[\cos\alpha_{ij}~\sin\alpha_{ij}]^{\rm T}$ is an estimate of the displacement vector $\V{d}_{ij}$ between node $i$ and $j$  based on the range and bearing measurements, the unitary matrix $\M{U}_{ij}$ is given by
$$
\M{U}_{ij} = \left[
               \begin{array}{cc}
                 \cos\alpha_{ij} & -\sin\alpha_{ij} \\
                 \sin\alpha_{ij} & \cos\alpha_{ij} \\
               \end{array}
             \right]
$$
and the diagonal matrix $\RM{\Lambda}_{ij}^{(\ell)}$ denotes the estimate of the measurement variances at the $\ell$-th iteration, given by ${\rm diag}\big((\hat{\rv{\sigma}}_{{\rm r},ij}^{(\ell)})^{-2},(\hat{\rv{\sigma}}^{(\ell)}_{{\rm t},ij})^{-2}\big)$. The estimates $\hat{\rv{\sigma}}_{{\rm r},ij}^{(\ell)}$ and $\hat{\rv{\sigma}}_{{\rm t},ij}^{(\ell)}$ are given by
\begin{equation}\label{estimates_aml}
\begin{aligned}
\hat{\rv{\sigma}}_{{\rm r},ij}^{(\ell)}&= \frac{\ln 10\sigma_{\rm RSS}}{10\gamma}\cdot\Big\|\hat{\RV{p}}_{i,{\rm AML}}^{(\ell-1)}-\hat{\RV{p}}_{j,{\rm AML}}^{(\ell-1)}\Big\|\\
\hat{\rv{\sigma}}_{{\rm t},ij}^{(\ell)}&= \sigma_{\rm AOA}\cdot \Big\|\hat{\RV{p}}_{i,{\rm AML}}^{(\ell-1)}-\hat{\RV{p}}_{j,{\rm AML}}^{(\ell-1)}\Big\|
\end{aligned}
\end{equation}
where $\hat{\RV{p}}_{i,{\rm AML}}^{(\ell-1)}$ denotes the position estimate of node $i$ in the $(\ell-1)$-th iteration. For $\ell=1$, the estimates $\hat{\rv{\sigma}}_{{\rm r},ij}^{(1)}$ and $\hat{\rv{\sigma}}_{{\rm t},ij}^{(1)}$ are obtained by replacing $\Big\|\hat{\RV{p}}_{i,{\rm AML}}^{(\ell-1)}-\hat{\RV{p}}_{j,{\rm AML}}^{(\ell-1)}\Big\|$ in \eqref{estimates_aml} with the range estimate $\rv{r}_{ij}$. The bearing estimate $\rv{\alpha}_{ij}$ is obtained directly using the \ac{aoa} measurement, while the range estimate $\rv{r}_{ij}$ is computed as
$$
\rv{r}_{ij}= 10^{(P_{\rm T}- L_0-\rv{y}_{ij}^{\rm RSS})(10\gamma)^{-1}}\cdot d_0.
$$

After some manipulations, one could rearrange the likelihood function of $\V{p}$ into the following standard form of Gaussian Markov random fields
$$
\rv{g}_{\RV{\alpha},\RV{r}}(\V{\alpha},\V{r};\V{p})\propto \exp\Big\{-\frac{1}{2}\V{p}^{\rm T}\RM{P}^{(\ell)}\V{p}+\big(\RV{h}^{(\ell)}\big)^{\rm T}\V{p}\Big\}
$$
where $\RM{P}^{(\ell)}$ denotes the precision matrix\footnote{We are slightly abusing the notations here. Please do not confuse the precision matrix with the potential kernel discussed in Appendix \ref{ssec:potential}.}, $\RV{h}^{(\ell)}$ represents the linear coefficients, while $\V{\alpha}$ and $\V{r}$ represent the vectors containing all bearing and range measurements, respectively. The closed-form AML estimate at the $\ell$-th iteration can thus be obtained as
\begin{equation}\label{linml_solution}
\hat{\RV{p}}_{\rm AML}^{(\ell)} = \big(\RM{P}^{(\ell)}\big)^{-1}\RV{h}^{(\ell)}.
\end{equation}
A straightforward computation of the matrix $\big(\RM{P}^{(\ell)}\big)^{-1}$ would require centralized operations, which may be unrealistic for large networks. Fortunately, some distributed algorithms such as Gaussian belief propagation are shown to be capable of obtaining the solution in \eqref{linml_solution} \cite{gabp_su}, and hence may be viewed as distributed implementations of the AML estimator.

As for the sequential algorithm, we consider a sequential and non-iterative implementation of the aforementioned AML estimator. To elaborate, the agents that can directly obtain measurements with the anchor, which will be referred to as the ``1-hop agents'', estimate their positions in the first round. In the subsequent rounds of the algorithm, e.g. the $n$-th round, the $n$-hop agents estimate their positions based on their measurements with the $(n-1)$-hop agents that are treated as ``virtual anchors''. More precisely, the sequential estimator for agent $i$ is
\begin{equation}
\hat{\RV{p}}_{i,{\rm Seq}}=\RM{P}_i^{-1}\RV{h}_i
\end{equation}
where $\RM{P}_i\in \mathbb{R}^{2\times 2}$ and $\RV{h}_i\in\mathbb{R}^2$ are constructed using the measurements between agent $i$ and its neighboring virtual anchors. To avoid iteration, the estimate of the measurement variances of each pair of nodes $i$ and $j$ is fixed at its initial value $\RM{\Lambda}_{ij}^{(1)}$.

\begin{figure}[t]
    \centering
    \includegraphics[width=.445\textwidth]{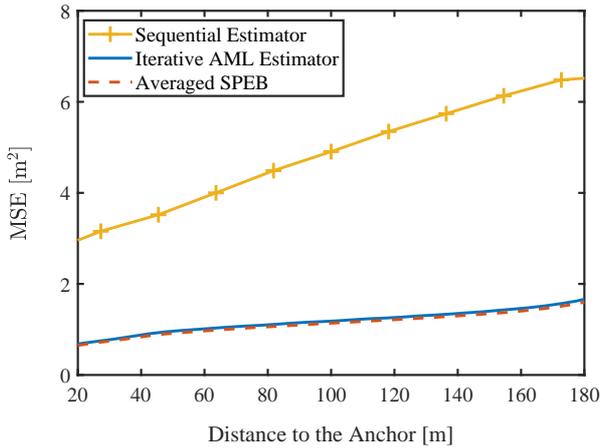}
    \caption{The \ac{mse} of the iterative AML estimator and the sequential estimator, as functions of the distance to the single anchor located at the network centroid. The averaged \ac{speb} is also plotted as a benchmark.}
    \label{fig:bound_vs_sequential}
\end{figure}

The localization \ac{mse} of an agent $i$ based on the iterative AML estimator, as well as that of the sequential estimator, is illustrated in Fig. \ref{fig:bound_vs_sequential} as a function of the distance between $\V{p}_i$ and the anchor. The results are averaged over $1000$ network snapshots, and the averaged \ac{speb} given by
$$
\frac{1}{N_{\rm snapshot}\cdot N_{\rm a}}\sum_{n=1}^{N_{\rm snapshot}} \sum_{i=1}^{N_{\rm a}} {\rm sp}\{\V{p}_{i,n}\}
$$
is also presented as a benchmark. Here, $N_{\rm snapshot} =1000$ is the number of snapshots, and $\V{p}_{i,n}$ denotes the true position of agent $i$ in the $n$-th snapshot. We may observe from the figure that the sequential estimator exhibits approximately a linear error scaling with respect to the distance to the network centroid (i.e. the anchor), which is far from the average \ac{speb}. This corroborates our discussion in Sec. \ref{ssec:complexity_vs_accuracy}.

\begin{figure}[t]
    \centering
    \includegraphics[width=.445\textwidth]{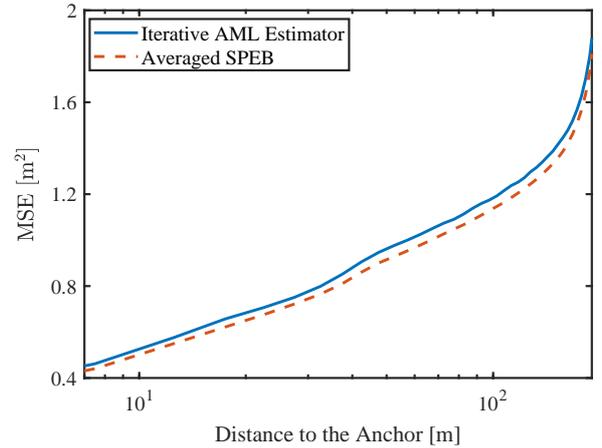}
    \caption{Comparison between the \ac{mse} of the iterative AML estimator and the corresponding averaged \ac{speb}, as functions of the distance to the anchor.}
    \label{fig:bound_vs_linML}
\end{figure}

To observe the performance of the iterative AML estimator more closely, we provide a zoomed-in view of the curves in Fig. \ref{fig:bound_vs_linML}. As it can be seen from the figure, the AML estimator exhibits a similar asymptotic behavior to the averaged \ac{speb} in the sense that it grows logarithmically as the distance to the network centroid increases. The residual error, which appears to be constant (with respect to the distance to the network centroid), can be interpreted as a performance penalty due to the approximation.

To conclude, these simulation results suggest that the lower bounds of localization error that we proposed are likely to be attainable by actual algorithms. Especially, from the localization performance of the iterative AML estimator we could observe the logarithmic scaling law that we have predicted in the theoretical analysis. The sequential estimator is not capable of attaining the bounds, but it exhibits an error scaling behavior as expected in Sec. \ref{ssec:complexity_vs_accuracy}.

\section{Conclusion}\label{sec:con}
This paper provides a new approach to understanding the effect of cooperation among agents in cooperative network localization. By introducing the \ac{npi}-\ac{eoc} decomposition of \ac{efim}, we develop a random-walk-inspired formalism for the treatment of \ac{eoc}. We provide an interpretation of this analysis, showing that the \ac{eoc} among nodes can be regarded as the efficiency of position information routing from anchors to agents. Moreover, following this line of reasoning we show that in large lattice and stochastic geometric networks, the \ac{speb} of an agent scales logarithmically as the reciprocal of anchor density increases, and interpret this result as a consequence of position information path loss. The scaling laws of localization performance proposed in this paper provide a new set of insights into the network localization problem from a network-level perspective, and can serve as guidelines to network deployment and operation tasks in large-scale networks.

\appendices

\section{Proof of Theorem \ref{thm:main}}\label{sec:proof_qrw}
\begin{IEEEproof}
Denoting $\M{T}=\M{D}^{-1}\M{A}$, from \eqref{vib_laplacian} matrix $\efim{\V{p}}$ can be rewritten as
\begin{equation}
\efim{\V{p}} = \M{D}(\M{I}-\M{T}).
\end{equation}
Expanding $(\M{I}-\M{T})^{-1}$ as a matrix power series, the inverse \ac{efim} can be expressed as
\begin{equation}
\invefim{\V{p}} = \bigg(\M{I}+\sum_{n=1}^\infty \M{T}^n\bigg)\M{D}^{-1}.
\end{equation}
The power series converges as long as the inverse of $\efim{\V{p}_i}$ exists. Hence the \ac{efim} for agent $i$ takes the following form
\begin{equation}\label{pre_main}
\efim{\V{p}_i} = \M{D}_i \bigg(\M{I}+\sum_{n=1}^\infty \subblk{\M{T}^n}{i}{i}\bigg)^{-1}.
\end{equation}

Comparing \eqref{main_thm} and \eqref{pre_main}, it now suffices to showing that
\begin{equation}\label{delta_pre1}
\sum_{n=1}^\infty \subblk{\M{T}^n}{i}{i} = \sum_{n=1}^\infty \M{T}_{ii}^{(n)}
\end{equation}
for all $i,j\in\Set{N}_{\rm a}$ and that $\M{\Delta}_i\succeq \M{0}$. Note that the matrix $\M{T}$ can be expressed as
\begin{equation}
\M{T} = \sum_{i\notin \Set{N}_{\rm b}}\sum_{j\leftrightarrow i} \Big(\M{E}_{ij}^{N_{\rm a}}\otimes(\M{D}_i^{-1}\M{J}_{ij})\Big).
\end{equation}
We can construct another matrix $\underline{\M{T}}\in\mathbb{R}^{2(N_{\rm a}+N_{\rm b})\times 2(N_{\rm a}+N_{\rm b})}$ using $\M{T}_{ij}^{(1)}$ as
\begin{equation}
\underline{\M{T}} = \sum_{i\in\Set{N}_{\rm a}\cup\Set{N}_{\rm b}}\sum_{j\in\Set{N}_{\rm a}\cup\Set{N}_{\rm b}} \M{E}_{ij}^{N_{\rm a}+N_{\rm b}}\otimes \M{T}_{ij}^{(1)}
\end{equation}
which can be partitioned as
\begin{equation}\label{partition_t}
\underline{\M{T}} = \left[
                      \begin{array}{cc}
                       \M{T} & \sum_{i\in\Set{N}_{\rm a}}\sum_{j\in\Set{N}_{\rm b}}\M{E}_{ij}^{N_{\rm a},N_{\rm b}}\otimes \M{T}_{ij}^{(1)} \\
                        \M{0}_{2N_{\rm b}\times 2N_{\rm a}} & \M{I}_{2N_{\rm b}} \\
                      \end{array}
                    \right].
\end{equation}
By this construction we have $[\underline{\M{T}}^n]_{i,j} = \M{T}_{ij}^{(n)}$. Furthermore, from \eqref{partition_t} we have $[\underline{\M{T}}^{n}]_{1:N_{\rm a},1:N_{\rm a}}=\M{T}^{n}$, which implies \eqref{delta_pre1}.

Now note that $\invefim{\V{p}_i}=(\M{I}+\M{\Delta}_i)\M{D}_i^{-1}$ is a real symmetric matrix, and that $\invefim{\V{p}_i}\succeq \M{D}_i^{-1}$. It then follows that $\M{\Delta}_i\succeq \M{0}$, which completes the proof.
\end{IEEEproof}

\section{Proof of Results in Section \ref{ssec:random_walk_2d}}
\subsection{Proposition \ref{prop:hitting}}\label{ssec:proof_hitting}
\begin{IEEEproof}
Expressing $\M{T}_{ij}^{(n)}$ in terms of $\Set{\Omega}_n^{i,j}$, we have
\begin{equation}
\begin{aligned}
\M{T}_{ij}^{(n)} &= \sum_{\V{\omega}\in\Set{\Omega}_n^{i,j}}\M{T}_{\V{\omega}}^{(n)}\\
&=\sum_{m=1}^n\sum_{\V{\alpha}\in\Set{\Omega}_m^{i,j}(\{j\})}\M{T}_{\V{\alpha}}^{(m)}\sum_{\V{\beta}\in\Set{\Omega}_{n-m}^{j,j}}\M{T}_{\V{\beta}}^{(n-m)}\\
&=\sum_{m=1}^n\M{F}_{ij}^{(m)}\M{T}_{jj}^{(n-m)}.
\end{aligned}
\end{equation}
Therefore $\M{\Delta}_i$ can be written as
\begin{equation}
\begin{aligned}
\M{\Delta}_i &=\sum_{m=1}^\infty\M{F}_{ii}^{(m)}\sum_{n=m}^\infty\M{T}_{ii}^{(n-m)}=\M{F}_{ii}(\M{I}+\M{\Delta}_i).
\end{aligned}
\end{equation}
Hence we have $\M{I}+\M{\Delta}_i=\M{I}+\M{F}_{ii}(\M{I}+\M{\Delta}_i)$ implying \eqref{hitting_prob}.
\end{IEEEproof}

\subsection{Proposition \ref{prop:recurrence}}\label{ssec:proof_recurrence_finite}
\begin{IEEEproof}
To prove $\M{F}_{ii}=\M{I}_2$, it is equivalent to show that both eigenvalues of $\sum_{n=0}^m \widetilde{\M{T}}_{ii}^{(n)}$ tend to infinity as $m\rightarrow \infty$. A sufficient condition is that $\lim_{n\rightarrow\infty}\widetilde{\M{T}}_{ii}^{(n)}$ is full-rank. From \eqref{efim_auxiliary} and \eqref{def_t} we see that
\begin{equation}
\M{I}-\widetilde{\M{T}} = \widetilde{\M{D}}^{-1} \widetilde{\M{J}}_{\rm e}(\V{p}_{\rm all})
\end{equation}
implying that the eigenspace of $\widetilde{\M{T}}$ corresponding to eigenvalue $1$ is the null space of $\widetilde{\M{D}}^{-1} \widetilde{\M{J}}_{\rm e}(\V{p}_{\rm all})$. Thus, as long as matrix $\widetilde{\M{D}}^{-1} \widetilde{\M{J}}_{\rm e}(\V{p}_{\rm all})$ is rank-deficient, the matrix limit $\widetilde{\M{T}}^{\infty}$ exists, with its eigenvalues being either $1$ or $0$. In light of this, the column space of $\widetilde{\M{T}}^{\infty}$ is the null space of $\M{D}^{-1}\widetilde{\M{J}}_{\rm e}(\V{p}_{\rm all})$. Since there is no anchor in the auxiliary network, matrix $\M{D}^{-1}\widetilde{\M{J}}_{\rm e}(\V{p}_{\rm all})$ is indeed rank-deficient with deficiency of at least $2$ (corresponding to two-dimensional translation \cite{relative}). Therefore, as long as the network is finite, the null space of $\M{D}^{-1}\widetilde{\M{J}}_{\rm e}(\V{p}_{\rm all})$ is at least two-dimensional.

More precisely, when only the range information is available (corresponding to $N_{\rm t}=1$), the null space of $\widetilde{\M{J}}_{\rm e}(\V{p}_{\rm all})$ is known to be $3$-dimensional, spanned by the vectors \cite{relative}
\begin{equation}\label{nullspace}
\begin{aligned}
&\V{v}_1 \propto [1~0~\dotsc~1~0]^{\rm T}, ~\V{v}_2 \propto [0~1~\dotsc~0~1]^{\rm T}, \\
&\V{v}_3 \propto [\V{u}(\phi_1+\pi/2)^{\rm T}~\dotsc~\V{u}( \phi_{N_{\rm a}+N_{\rm b}}+\pi/2)^{\rm T}]^{\rm T}
\end{aligned}
\end{equation}
where $\phi_i$ is the angle from node $i$ to the network centroid. Furthermore, the null space of $\widetilde{\M{J}}_{\rm e}(\V{p}_{\rm all})$ is spanned by $\V{v}_1$ and $\V{v}_2$ when both range and bearing information is available. In either case, $\lim_{n\rightarrow\infty}\widetilde{\M{T}}_{ii}^{(n)} = \M{D}_i^{-1}$ can be seen to be the product of two full-rank matrices hence is full-rank, and the proof is completed.
\end{IEEEproof}

\section{The Pseudo-Random Walk Formalism}\label{sec:prw}
\subsection{Additional Notations}\label{ssec:extra_notation}
To facilitate the analysis, we define
\begin{equation}
\M{H}_{\Set{S}}(i,j)=\left\{
                        \begin{array}{ll}
                          \M{F}_{\Set{S}}(i,j), & \hbox{$i\notin\Set{S}$;} \\
                          \M{I}_2, & \hbox{$i\in\Set{S},~i=j$;} \\
                          \M{0}_{2\times 2}, & \hbox{$i\in\Set{S},~i\neq j$.}
                        \end{array}
                      \right.
\end{equation}
In addition, the quantity $\M{F}_{\Set{S}}(i,j)$ will also be used extensively in the analysis, and hence we reproduce its definition here as
\begin{equation}
\M{F}_{\Set{S}}(i,j)=\sum_{n=1}^\infty \sum_{\V{\omega}\in\Omega_n^{i,j}(\Set{S})}\widetilde{\M{T}}_{\V{\omega}}^{(n)}.
\end{equation}
We may also rearrange $\M{H}_{\Set{S}}(i,j)$ and $\M{F}_{\Set{S}}(i,j)$ in a compact matrix form as
\begin{equation}
\begin{aligned}
\M{H}_{\Set{S}} &= \sum_{i\in\Set{N}_{\rm a}\cup\Set{N}_{\rm b}} \sum_{j\in\Set{S}} \M{E}_{ij}^{N_{\rm a}+N_{\rm b}}\otimes \M{H}_{\Set{S}}(i,j) \\
\M{F}_{\Set{S}} &=\sum_{i\in\Set{N}_{\rm a}\cup\Set{N}_{\rm b}} \sum_{j\in\Set{S}} \M{E}_{ij}^{N_{\rm a}+N_{\rm b}}\otimes \M{F}_{\Set{S}}(i,j) \,.
\end{aligned}
\end{equation}

\setcounter{tempEqCounter}{\value{equation}}
\setcounter{equation}{64}
\begin{figure*}[t]
\begin{equation}\label{potential}
\begin{aligned}
\M{F}_{\widetilde{\Set{R}}_i}(k,l) = \left\{
                                     \begin{array}{ll}
                                      \big[\M{P}_{\widetilde{\Set{R}}_i}^{-1}\big]_{m,n} -\Big(\sum_{j\in\widetilde{\Set{R}}_i}\big[\M{P}_{\widetilde{\Set{R}}_i}^{-1}\big]_{\min^{(j)}(\widetilde{\Set{R}}_i),m}\Big)
                                      \M{Z}_{\widetilde{\Set{R}}_i}^{-1}\Big(\sum_{j\in\widetilde{\Set{R}}_i}\M{D}_n\big[\M{P}_{\widetilde{\Set{R}}_i}^{-1}\big]_{n,\min^{(j)}(\widetilde{\Set{R}}_i)}\Big),
 & \hbox{$m\neq n$;} \\
                                       \M{I}_2+\big[\M{P}_{\widetilde{\Set{R}}_i}^{-1}\big]_{n,n} -\Big(\sum_{j\in\widetilde{\Set{R}}_i}\big[\M{P}_{\widetilde{\Set{R}}_i}^{-1}\big]_{\min^{(j)}(\widetilde{\Set{R}}_i),n}\Big)
                                      \M{Z}_{\widetilde{\Set{R}}_i}^{-1}\Big(\sum_{j\in\widetilde{\Set{R}}_i}\M{D}_n\big[\M{P}_{\widetilde{\Set{R}}_i}^{-1}\big]_{n,\min^{(j)}(\widetilde{\Set{R}}_i)}\Big)
 & \hbox{$m=n$.}
                                     \end{array}
                                   \right.
\end{aligned}
\end{equation}
\hrulefill
\end{figure*}
\setcounter{equation}{\value{tempEqCounter}}

\subsection{Potential Kernel}\label{ssec:potential}
For the asymptotic analysis task considered in this paper, the central problem is to characterize the position information an agent $i$ received from an anchor $j$ as a function of their relative displacement $\V{d}_{ij}$. Intuitively, the amount of position information should decrease with $\|\V{d}_{ij}\|$, exhibiting a ``position information path loss'' behavior. According to previous analysis, this problem amounts to expressing the pseudo-probabilities $\M{F}_{\widetilde{\Set{R}}_i}(i,j)$ as functions of $\V{d}_{ij}$. In classical random walk theory, such a task relies on the construction of \textit{potential kernels} \cite{hitting_prob}. In this subsection, we construct analogous quantities.

First we note that
\begin{equation}\label{singular_fim_auxiliary}
\widetilde{\M{J}}_{\rm e}(\V{p}_{\rm all})=\widetilde{\M{D}}(\M{I}-\widetilde{\M{T}})
\end{equation}
where matrix $\M{I}-\widetilde{\M{T}}$ is an analogue of the \textit{random walk normalized graph Laplacian} \cite{chung1997spectral}. It is clear that $\M{I}-\widetilde{\M{T}}$ is non-invertible, but we may calculate its \emph{generalized inverse}.
\begin{lemma}\label{lem:potential}
We have
\begin{equation}\label{pi_g_relation}
(\M{I}-\widetilde{\M{T}}^\infty)\M{H}_{\Set{S}} = \lim_{n\rightarrow \infty}\{\M{P}_n(\M{F}_{\Set{S}} - \M{I}_{\Set{S}})\}
\end{equation}
where $\M{P}_n =\widetilde{\M{G}}_n-\M{G}_n$, $\M{G}_n=\sum_{k=0}^n \widetilde{\M{T}}^k$, and
\begin{equation}\label{tgn_def}
\widetilde{\M{G}}_n = \sum_{\{i,j\}\subseteq\Set{N}_{\rm a}\cup\Set{N}_{\rm b}}\M{E}_{ij}^{N_{\rm a}+N_{\rm b}} \otimes \big(\subblk{\M{G}_n}{i}{i}\M{D}_i^{-1}\M{D}_j\big).
\end{equation}
\begin{IEEEproof}
Using the additional notations in Sec. \ref{ssec:extra_notation}, and after some manipulations, one can obtain
\begin{equation}\label{first_iteration}
(\widetilde{\M{T}}-\M{I})\M{H}_{\Set{S}} = \M{F}_{\Set{S}} - \M{I}_{\Set{S}}
\end{equation}
where $\M{I}_{\Set{S}} = \sum_{i\in\Set{S}} \M{E}_{ii}^{N_{\rm a}+N_{\rm b}} \otimes \M{I}_2$. Multiplying $\M{G}_n$ from the left at both sides of \eqref{first_iteration} and taking limit $n\rightarrow \infty$, we have
\begin{equation}\label{limit_iteration}
\M{M}_{\Set{S}}-\M{H}_{\Set{S}} = -\lim_{n\rightarrow \infty}\big\{\M{G}_n(\M{F}_{\Set{S}} - \M{I}_{\Set{S}})\big\}
\end{equation}
where $\M{M}_{\Set{S}} = \widetilde{\M{T}}^\infty\M{H}_{\Set{S}}$. Note that \eqref{first_iteration} implies $\widetilde{\M{T}}^\infty(\M{F}_{\Set{S}}-\M{I}_{\Set{S}})=\M{0}$, and thus the matrix limit in \eqref{limit_iteration} does exist. For the $N_{\rm t}\geq 2$ case, because the relative measurements in the network is invariant under two-dimensional translation, we have
\begin{equation}\label{right_eigenspace}
\begin{aligned}
\M{Q}^{\rm T}\widetilde{\M{J}}_{\rm e}(\V{p}_{\rm all}) = \M{0},~\widetilde{\M{J}}_{\rm e}(\V{p}_{\rm all})\M{Q} = \M{0}
\end{aligned}
\end{equation}
where $\widetilde{\M{J}}_{\rm e}(\V{p}_{\rm all})$ satisfies \eqref{singular_fim_auxiliary}. Now, in light of \eqref{right_eigenspace}, we can construct a matrix $\widetilde{\M{G}}_n$ as in \eqref{tgn_def} satisfying $\widetilde{\M{G}}_n(\M{F}_{\Set{S}}-\M{I}_{\Set{S}})=\M{0}$. Therefore with $\M{P}_n := \widetilde{\M{G}}_n-\M{G}_n$ we have
\begin{equation}\label{iteration_limit2}
\M{H}_{\Set{S}}-\M{M}_{\Set{S}} = \lim_{n\rightarrow \infty}\big\{\M{P}_n(\M{F}_{\Set{S}} - \M{I}_{\Set{S}})\big\}.
\end{equation}
Thus the proof is completed.
\end{IEEEproof}
\end{lemma}

In Lemma \ref{lem:potential}, we are essentially trying to construct a \emph{generalized inverse} for $\M{I}-\widetilde{\M{T}}$. To elaborate, if $\M{P}:= \lim_{n\rightarrow \infty}\M{P}_n$ exists, $-\M{P}$ can serve as a valid generalized inverse. Then we can solve a system of linear equations for the desired quantity $\M{F}_{\Set{S}}$. Indeed, the limit on the right hand side of \eqref{pi_g_relation} does exist as long as $\widetilde{\M{T}}^{\infty}$ exists. However, a subtle fact is that this does not imply the existence of $\lim_{n\rightarrow \infty}\M{P}_n$. Fortunately, this is true when $N_{\rm t}\geq 2$, for which we have the following result.

\setcounter{equation}{65}
\begin{theorem}\label{thm:potential}
When $N_{\rm t}\geq 2$, for $k=\min^{(m)}(\widetilde{\Set{R}}_i)$ and $l=\min^{(n)}(\widetilde{\Set{R}}_i)$, the term $\M{F}_{\widetilde{\Set{R}}_i}(k,l)$ can be expressed as \eqref{potential}, where matrix $\M{P}_{\widetilde{\Set{R}}_i}$ is given by $\big[\M{P}_{\widetilde{\Set{R}}_i}\big]_{k,l} = \subblk{\M{P}}{m}{n}$, $\M{Z}_{\widetilde{\Set{R}}_i}=\sum_{\{a,b\}\subseteq\widetilde{\Set{R}}_i}\M{D}_a\big[\M{P}_{\widetilde{\Set{R}}_i}^{-1}\big]_{\min^{(a)}(\widetilde{\Set{R}}_i),\min^{(b)}(\widetilde{\Set{R}}_i)}$ is a normalization constant, and
\begin{equation}\label{potential_kernel_def}
\M{P} \!=\! \sum_{\{a,b\}\subseteq \Set{N}_{\rm a}\cup\Set{N}_{\rm b}}\M{E}_{ab}^{N_{\rm a}+N_{\rm b}}\otimes \bigg\{\sum_{p=0}^\infty \Big(\widetilde{\M{T}}_{aa}^{(p)}\M{D}_a^{-1}\M{D}_b\!-\!\widetilde{\M{T}}_{ab}^{(p)}\Big)\bigg\}.
\end{equation}
\begin{IEEEproof}
We shall first show that the matrix limit $\M{P}=\lim_{n\rightarrow\infty}\M{P}_n$ exists for the case $N_{\rm t}\geq 2$, and hence
\begin{equation}\label{rank2_limit}
\M{H}_{\Set{S}} - \M{M}_{\Set{S}} =\M{P}(\M{F}_{\Set{S}} - \M{I}_{\Set{S}}).
\end{equation}
First we use the definition of $\M{G}_n$ and obtain
\begin{equation}\label{series_p}
\subblk{\M{P}\M{D}^{-1}}{i}{j} = \sum_{n=0}^\infty \Big(\widetilde{\M{T}}_{ii}^{(n)}\M{D}_i^{-1}-\widetilde{\M{T}}_{ij}^{(n)}\M{D}_j^{-1}\Big).
\end{equation}
Matrix $\M{P}$ exists if the series in \eqref{series_p} converges for every pair $(i,j)$. It is clear from \eqref{right_eigenspace} that
$$
\lim_{n\rightarrow \infty}\big\|\widetilde{\M{T}}_{ii}^{(n)}\M{D}_i^{-1}-\widetilde{\M{T}}_{ij}^{(n)}\M{D}_j^{-1}\big\|_2=0
$$
holds for all $i$ and $j$. Moreover, it can be shown that $\big\|\widetilde{\M{T}}_{ii}^{(n)}\M{D}_i^{-1}-\widetilde{\M{T}}_{ij}^{(n)}\M{D}_j^{-1}\big\|_2$ exhibits a linear rate of convergence as $n\rightarrow \infty$, since $\|\widetilde{\M{T}}^n-\widetilde{\M{T}}^\infty\|_2$ converges to zero linearly as $n\rightarrow \infty$. Therefore, we can conclude that matrix $\M{P}$ does exist.

Now, restricting the rows and columns of the matrices in \eqref{rank2_limit} to the elements in the index set $\Set{S}$, we can construct $2|\Set{S}|\times 2|\Set{S}|$ matrices $\M{F}_{\Set{S},\Set{S}},\M{H}_{\Set{S},\Set{S}},\M{P}_{\Set{S}},\M{M}_{\Set{S},\Set{S}}$ and $\M{D}_{\Set{S}}$ as
\begin{equation}\label{restricted_matrices}
\begin{aligned}
\subblk{\M{P}_{\Set{S}}}{i}{j} &\!=\! \subblk{\M{P}}{k}{l},~\M{H}_{\Set{S},\Set{S}} \!=\! \M{I}_{2|\Set{S}|},~\subblk{\M{M}_{\Set{S},\Set{S}}}{i}{j} \!=\! \subblk{\M{M}_{\Set{S}}}{k}{l},\\
\subblk{\M{F}_{\Set{S},\Set{S}}}{i}{j} &\!=\! \subblk{\M{F}_{\Set{S}}}{k}{l},~\subblk{\M{D}_{\Set{S}}}{i}{j} \!=\! \subblk{\M{D}}{k}{l}
\end{aligned}
\end{equation}
where $k=\min^{(i)}(\Set{S})$ and $l=\min^{(j)}(\Set{S})$. Thus using \eqref{rank2_limit} we obtain
\begin{equation}\label{restricted_limit}
\M{F}_{\Set{S},\Set{S}} - \M{I} = \M{P}_{\Set{S}}^{-1}(\M{I}-\M{M}_{\Set{S},\Set{S}}).
\end{equation}
From \eqref{restricted_limit} we also have $\M{Q}^{\rm T}\M{D}_{\Set{S}}\M{P}_{\Set{S}}^{-1}(\M{I}-\M{M}_{\Set{S},\Set{S}})=\M{0}$, and hence
\begin{equation}\label{ms}
\M{M}_{\Set{S},\Set{S}} =\M{Q}\big(\M{Q}^{\rm T}\M{D}_{\Set{S}}\M{P}_{\Set{S}}^{-1}\M{Q}\big)^{-1}\M{Q}^{\rm T}\M{D}_{\Set{S}}\M{P}_{\Set{S}}^{-1}.
\end{equation}
Substituting \eqref{ms} into \eqref{restricted_limit} we obtain
\begin{equation}\label{pi_ps}
\M{F}_{\Set{S},\Set{S}} \!=\! \M{I}\!+\!\M{P}_{\Set{S}}^{-1} \!-\! \M{P}_{\Set{S}}^{-1} \M{Q}\big(\M{Q}^{\rm T}\M{D}_{\Set{S}}\M{P}_{\Set{S}}^{-1}\M{Q}\big)^{-1}\M{Q}^{\rm T}\M{D}_{\Set{S}}\M{P}_{\Set{S}}^{-1}
\end{equation}
yielding \eqref{potential} after some manipulations.
\end{IEEEproof}
\end{theorem}

The matrix $[\M{P}_{\widetilde{\Set{R}}_i}]_{k,l}$ resembles the potential between $m$ and $n$ in classical random walk theory in the sense that, as will be illustrated in Section \ref{sec:lattice}, it is a function of solely $\V{d}_{mn}$ in infinite lattice networks. It is also important in finite networks, since $\M{P}_{\widetilde{\Set{R}}_i}\in\mathbb{R}^{2(N_{\rm b}+1)\times 2(N_{\rm b}+1)}$ and $\{\M{D}_i\}_{i\in N_{\rm a}\cup N_{\rm b}}$ retain all the necessary information to describe the information coupling between agent $i$ and other nodes. Typically, the number of anchors $N_{\rm b}$ is far less than the number of agents $N_{\rm a}$, and thus the dimensionality of $\M{P}_{\widetilde{\Set{R}}_i}$ is far lower than that of $\efim{\V{p}}$. Therefore, it is much simpler to derive \ac{eoc} using $\M{P}_{\widetilde{\Set{R}}_i}$ instead of manipulating $\efim{\V{p}}$ directly.

\subsection{Infinite Lattice Networks: Pseudo-Characteristic Function}\label{ssec:harmonic}
We now give the asymptotic expressions of \ac{eoc} in infinitely large lattice networks. Thanks to the symmetry of these networks, we are able to perform harmonic analysis on infinite lattice networks since the underlying graph $\Set{G}_{\rm net}$ is invariant with respect to translation. Since we have assumed that nodes are equipped with \ac{uoa}s when $N_{\rm t}\geq 2$ which implies that $\M{J}_{ij}$ only depends on $\V{d}_{ij}$, the singular \ac{efim} $\widetilde{\M{J}}_{\rm e}(\V{p}_{\rm all})$ built upon $\Set{G}_{\rm net}$ should also be translation-invariant. In this regard, we can define \emph{pseudo-characteristic functions} on auxiliary networks.
\begin{definition}[Pseudo-Characteristic Function]\label{def:CF}
The pseudo-characteristic function of an auxiliary network associated with an infinite lattice network is defined as
\begin{equation}\label{CF}
\M{\Phi}(\V{\vartheta}) =\sum_{\V{x}\in\mathbb{Z}^2}\widetilde{\M{T}}_1(\V{0},\V{x})e^{\jmath\V{x}^{\rm T}\V{\vartheta}}
\end{equation}
where $\widetilde{\M{T}}_k(\V{0},\V{x})=\widetilde{\M{T}}_{ij}^{(k)}$ for $\V{p}_i=\V{0}$ and $\V{p}_j=\V{x}$, and $\jmath$ is the imaginary unit $\sqrt{-1}$. We define the notation $\widetilde{\M{T}}_k(\V{0},\V{x})$ to emphasize that it is a function of  not only the indices of agents, but also their positions.
\end{definition}

To ensure that Theorem \ref{thm:potential} is applicable, we first show the recurrence of the random walk on the auxiliary network of infinite lattice networks, using the pseudo-characteristic function.
\begin{proposition}\label{prop:recurrence_infinite}
For any auxiliary network (or original network without anchor) of infinite lattice networks, we have $\M{F}_{ii}=\M{I}_2$ for all $i\in \Set{N}_{\rm a}$.
\begin{IEEEproof}
Using the pseudo-characteristic function we have
$$
\sum_{n=0}^\infty \widetilde{\M{T}}_n(\V{0},\V{0}) = \frac{1}{2\pi} \int_{\Set{C}} (\M{I}-\M{\Phi}(\V{\vartheta}))^{-1} {\rm d} \V{\vartheta}
$$
We only need to show that the integral does not converge. We consider a decomposition as follows
\begin{equation}\label{decomp_recurrence}
\begin{aligned}
\frac{1}{(2\pi)^2} \int_{\Set{C}} (\M{I}-\M{\Phi}(\V{\vartheta}))^{-1} {\rm d} \V{\vartheta} &=\frac{1}{2\pi^2}\int_{\Set{C}} \|\V{\vartheta}\|^{-2} \M{\Sigma}_2^{-1}(\V{\vartheta}){\rm d} \V{\vartheta} \\
&\hspace{3mm}+\frac{1}{(2\pi)^2}\int_{\Set{C}} \widetilde{\M{\Psi}}(\V{\vartheta}) {\rm d}\V{\vartheta}
\end{aligned}
\end{equation}
where $\widetilde{\M{\Psi}}:= (\M{I}-\M{\Phi}(\V{\vartheta}))^{-1}-2\|\V{\vartheta}\|^{-2}\M{\Sigma}_2^{-1}$. For the first term on the right hand side of \eqref{decomp_recurrence}, note that $c_1\M{I}\preceq \M{\Sigma}_2^{-1}(\V{\vartheta})\preceq c_2\M{I}$ holds for all $\V{\vartheta}$ where $c_1=(\max_{\V{\vartheta}}\lambda_{\max}(\M{\Sigma}_2(\V{\vartheta})))^{-1}$ and $c_2=(\min_{\V{\vartheta}}\lambda_{\min}(\M{\Sigma}_2(\V{\vartheta})))^{-1}$ are positive constants, and that
$$
\int_{\Set{C}}\|\V{\vartheta}\|^{-2}{\rm d}\V{\vartheta}\\=\int_0^{2\pi}{\rm d}\phi \int_0^\pi r^{-1} {\rm d}r
$$
does not converge. Therefore, to prove this proposition, it is now sufficient to show that $\widetilde{\M{\Psi}}(\V{\vartheta})$ is integrable. Taking the spectral norm of $\widetilde{\M{\Psi}}$, we have
\begin{equation}\label{expand_potential}
\begin{aligned}
\|\widetilde{\M{\Psi}}(\V{\vartheta})\|_2&= \Big\|(\M{I}-\M{\Phi}(\V{\vartheta}))^{-1}-\Big(\frac{\|\V{\vartheta}\|^2}{2}\M{\Sigma}_2(\V{\vartheta})\Big)^{-1}\Big\|_2\\
&\leq \|(\M{I}-\M{\Phi}(\V{\vartheta}))^{-1}\|_2\cdot\Big\|\frac{2}{\|\V{\vartheta}\|^2}\M{\Sigma}_2^{-1}(\V{\vartheta})\Big\|_2\\
&\hspace{3mm}\cdot\Big\|(\M{I}-\M{\Phi}(\V{\vartheta}))-\frac{\|\V{\vartheta}\|^2}{2}\M{\Sigma}_2(\V{\vartheta})\Big\|_2.
\end{aligned}
\end{equation}

\setcounter{tempEqCounter}{\value{equation}}
\setcounter{equation}{78}
\begin{figure*}[t]
\begin{equation}\label{simplified_potential}
\begin{aligned}
\widetilde{\M{F}}_{\widetilde{\Set{R}}_i}(k,l)= \left\{
                                     \begin{array}{ll}
                                       \begin{aligned}
                                       \big[\M{P}_{\widetilde{\Set{R}}_i}^{-1}\big]_{m,n}-\big[\M{P}_{\widetilde{\Set{R}}_i}^{-1}\big]^{(m)}\left(\M{Q}^{\rm T}\M{P}_{\widetilde{\Set{R}}_i}^{-1}\M{Q}\right)^{-1}\Big(\big[\M{P}_{\widetilde{\Set{R}}_i}^{-1}\big]^{(n)}\Big)^{\rm T},
\end{aligned} & \hbox{$m\neq n$;} \\
                                        \begin{aligned}
                                       \M{I}_2+\big[\M{P}_{\widetilde{\Set{R}}_i}^{-1}\big]_{n,n}-\big[\M{P}_{\widetilde{\Set{R}}_i}^{-1}\big]^{(n)}\left(\M{Q}^{\rm T}\M{P}_{\widetilde{\Set{R}}_i}^{-1}\M{Q}\right)^{-1}\Big(\big[\M{P}_{\widetilde{\Set{R}}_i}^{-1}\big]^{(n)}\Big)^{\rm T}
\end{aligned} & \hbox{$m=n$.}
                                     \end{array}
                                   \right.
\end{aligned}
\end{equation}
\hrulefill
\end{figure*}
\setcounter{equation}{\value{tempEqCounter}}

To bound the term $\|(\M{I}-\M{\Phi}(\V{\vartheta}))^{-1}\|_2$, note that $\widetilde{\M{T}}_1(\V{0},\V{x})$ is Hermitian in infinite lattice networks, and thus using \eqref{def_potential} one can verify that $\M{\Phi}(\V{\vartheta})$ is also Hermitian for any $\V{\vartheta}$. Hence $\|(\M{I}-\M{\Phi}(\V{\vartheta}))^{-1}\|_2$ is in fact $[\lambda_{\min}(\M{I}-\M{\Phi}(\V{\vartheta}))]^{-1}$. From the symmetry of $\widetilde{\M{T}}_1(\V{0},\V{x})$, i.e., $\widetilde{\M{T}}_1(\V{0},\V{x})=\widetilde{\M{T}}_1(\V{0},-\V{x})$, we have
$$
\M{\Phi}(\V{\vartheta}) = \Re\{\M{\Phi}(\V{\vartheta})\} =\sum_{\V{x}\in\mathbb{Z}^2}\widetilde{\M{T}}_1(\V{0},\V{x})\cos (\V{x}^{\rm T}\V{\vartheta}).
$$
Since $\widetilde{\M{T}}_1(\V{0},\V{x})$ is positive semi-definite for all $\V{x}$ and positive definite for some $\V{x}$, we can bound matrix $\M{I}-\M{\Phi}(\V{\vartheta})$ from below as
\begin{equation}\label{bound_phi1}
\begin{aligned}
\M{I}-\M{\Phi}(\V{\vartheta})
&\succeq 2\pi^{-2}\sum_{\|\V{x}\|\leq K}(\V{x}^{\rm T}\V{\vartheta})^2\widetilde{\M{T}}_1(\V{0},\V{x}) \\
&=\|\V{\vartheta}\|^2\M{\Phi}_{\rm c}
\end{aligned}
\end{equation}
for all $\V{\vartheta}$ satisfying $\|\V{\vartheta}\|\leq \pi K^{-1}$, where $\M{\Phi}_{\rm c}$ is a constant positive definite matrix with respect to $\V{\vartheta}$. For the case $\|\V{\vartheta}\| >\pi K^{-1}$, we also have
\begin{equation}\label{bound_phi2}
\M{I}-\M{\Phi}(\V{\vartheta})\succeq K^2\pi^{-2}\|\V{\vartheta}\|^2\M{\Phi}_{\min},
\end{equation}
where $\M{\Phi}_{\min}:= \min_{\V{\vartheta}\in\{\V{\vartheta}|\|\V{\vartheta}\|> \pi K^{-1}\}} \{\M{I}-\M{\Phi}(\V{\vartheta})\}\succ \M{0}_{2\times 2}$. The minimum is in taken the sense of L\"{o}wner partial ordering \cite[Sec. 2.4]{convex_opt}.

Combining \eqref{bound_phi1} and \eqref{bound_phi2}, we see that $\|(\M{I}-\M{\Phi}(\V{\vartheta}))^{-1}\|_2\leq c_3\|\V{\vartheta}\|^{-2}$ for some constant $c_3>0$, and hence for some constant $c_4>0$, \eqref{expand_potential} can be further simplified as
$$
\|\widetilde{\M{\Psi}}(\V{\vartheta})\|_2 \leq c_4\|\V{\vartheta}\|^{-4}\Big\|(\M{I}-\M{\Phi}(\V{\vartheta}))-\frac{\|\V{\vartheta}\|^2}{2}\M{\Sigma}_2(\V{\vartheta})\Big\|_2.
$$
Denote $\widetilde{\M{E}}_{\V{x}}\{f(\V{x})\}=\sum_{\V{x}\in\mathbb{Z}^2}f(\V{x})\widetilde{\M{T}}_1(\V{0},\V{x})$, and note that $\|\V{\vartheta}\|\M{\Sigma}_2(\V{\vartheta})=\widetilde{\M{E}}_{\V{x}}\{(\V{x}^{\rm T}\V{\vartheta})^2\}$ and $\widetilde{\M{E}}_{\V{x}}\{\V{x}\}=\M{0}_{2\times 2}$, using Taylor expansion we obtain
\begin{equation}\label{convergence_sigma2}
\begin{aligned}
&\Big\|(\M{I}-\M{\Phi}(\V{\vartheta}))-\frac{\|\V{\vartheta}\|^2}{2}\M{\Sigma}_2(\V{\vartheta})\Big\|_2\\
&\hspace{3mm}\leq \Big\|\widetilde{\M{E}}_{\V{x}}\Big\{\Big|e^{\jmath\V{x}^{\rm T}\V{\vartheta}}-1-\jmath\V{x}^{\rm T}\V{\vartheta}-\frac{1}{2}(\jmath\V{x}^{\rm T}\V{\vartheta})^2\Big|\Big\}\Big\|_2 \\
&\hspace{3mm}\leq c_5 \|\V{\vartheta}\|^{2+\delta}\|\M{\Sigma}_{2+\delta}(\V{\vartheta})\|_2
\end{aligned}
\end{equation}
for some constant $c_5>0$ and $\delta>0$, where $\M{\Sigma}_{a}(\V{\vartheta}):= \widetilde{\M{E}}_{\V{x}}\{|\V{x}^{\rm T}(\V{\vartheta}/\|\V{\vartheta}\|)|^a\}$. From the system model in Section \ref{sec:system_model} we see that $\M{\Sigma}_a$ always exists for $a>0$ and is positive definite with finite norm. Furthermore, it does not depend on the norm $\|\V{\vartheta}\|$. Hence the proof is completed.
\end{IEEEproof}
\end{proposition}

By application of the pseudo-characteristic function, we can rewrite Theorem \ref{thm:potential} in a simpler form in the case of infinite lattice networks. To achieve this, first note that in infinite lattice networks the matrix $\M{D}$ is proportional to the identity matrix $\M{I}$, and thus we have the following corollary.
\begin{corollary}\label{coro:simplified_potential}
In infinite lattice networks, for $k=\min^{(m)}(\widetilde{\Set{R}}_i)$ and $l=\min^{(n)}(\widetilde{\Set{R}}_i)$, if $N_{\rm t}\geq 2$, $\M{F}_{\widetilde{\Set{R}}_i}(k,l)$ can be expressed as \eqref{simplified_potential}, where $\M{P}_{\widetilde{\Set{R}}_i}^{(m)}:=\sum_{j=1}^{N_{\rm b}+1}\big[\M{P}_{\widetilde{\Set{R}}_i}^{-1}\big]_{i,j}$.
\begin{IEEEproof}
In infinite lattice networks, for any node $i$, the neighboring nodes are symmetrically distributed around $\V{p}_i$. Using the \ac{uoa} assumption we have $\M{D}_i=\M{D}_j\propto\M{I}_2$ holds for all nodes $i$ and $j$, and this corollary follows directly from Theorem \ref{thm:potential}.
\end{IEEEproof}
\end{corollary}

From Corollary \ref{coro:simplified_potential} it is clear that $\M{F}_{\widetilde{\Set{R}}_i}(k,l)$ can be expressed solely in terms of the inverse of the potential kernel, i.e., $\M{P}_{\widetilde{\Set{R}}_i}^{-1}$, in infinite lattice networks. Matrix $\M{P}$ can now be rewritten as
$$
\M{P} = \lim_{n\rightarrow\infty} \big\{\M{G}_n(\V{0},\V{0})\otimes (\V{1}\V{1}^{\rm T})-\M{G}_n\big\}
$$
where $\M{G}_n(\V{x},\V{y}):= \sum_{k=0}^n\widetilde{\M{T}}_k(\V{x},\V{y})$ and $\M{G}_n:=\sum_{k=0}^n\widetilde{\M{T}}_{\V{q}}^k$, and each of the $2\times 2$ blocks of $\M{P}$ can be expressed in terms of the following matrix limit
\setcounter{equation}{79}
\begin{equation}\label{def_potential}
\M{P}(\V{x},\V{y}):= \lim_{n\rightarrow\infty}\sum_{k=0}^n \big(\widetilde{\M{T}}_k(\V{0},\V{0})-\widetilde{\M{T}}_k(\V{x},\V{y})\big).
\end{equation}
Due to the translation-invariance of infinite lattice networks, $\M{P}(\V{x},\V{y})$ is in fact a function of $\V{x}-\V{y}$. Hence $[\M{P}_{\widetilde{\Set{R}}_i}]_{k,l}$ is a function of solely $\V{d}_{mn}$, where $k=\min^{(m)}(\widetilde{\Set{R}}_i)$ and $l=\min^{(n)}(\widetilde{\Set{R}}_i)$. We will show this in the following proposition using the pseudo-characteristic function.
\begin{proposition}\label{prop:existence_potential}
Matrix limit $\M{P}(\V{x},\V{y})$ can be expressed in terms of the pseudo-characteristic function $\M{\Phi}(\V{\vartheta})$ as
\begin{equation}
\M{P}(\V{x},\V{y})=\frac{1}{(2\pi)^2}\int_{\Set{C}} \left(\M{I}-\M{\Phi}(\V{\vartheta})\right)^{-1}\big(1-e^{\jmath(\V{x}-\V{y})^{\rm T}\V{\vartheta}}\big) {\rm d}\V{\vartheta}
\end{equation}
where $\Set{C}:= \{\V{\vartheta}|\|\V{\vartheta}\|\leq \pi\}$.
\begin{IEEEproof}
Using the definition of characteristic function, for $\M{\Phi}^2(\V{\vartheta})$ we have
$$
\M{\Phi}^2(\V{\vartheta}) = \sum_{\V{x}\in\mathbb{Z}^2}\sum_{\V{z}\in\mathbb{Z}^2}\widetilde{\M{T}}_1(\V{0},\V{x})\widetilde{\M{T}}_1(\V{0},\V{z})
e^{\jmath(\V{x}+\V{z})^{\rm T}\V{\vartheta}}.
$$
But in infinite lattice networks, we have $\widetilde{\M{T}}_1(\V{0},\V{z})=\widetilde{\M{T}}_1(\V{x},\V{x+z})$, and thus
$$
\M{\Phi}^2(\V{\vartheta}) = \sum_{\V{y}\in\mathbb{Z}^2}\widetilde{\M{T}}_2(\V{0},\V{y})
e^{\jmath\V{y}^{\rm T}\V{\vartheta}}
$$
where $\V{y}=\V{x}+\V{z}$. Repeating this iteration and taking Fourier transform from both sides, we obtain
$$
\widetilde{\M{T}}_n(\V{x},\V{y}) = \frac{1}{(2\pi)^2}\int_{\Set{C}} \M{\Phi}^n(\V{\vartheta})e^{\jmath\V{\vartheta}^{\rm T}(\V{x}-\V{y})} {\rm d}\V{\vartheta}.
$$
Therefore, from the definition of $\M{P}(\V{x},\V{y})$ \eqref{def_potential}, we see that it can be expressed as
\begin{equation}
\begin{aligned}
\M{P}(\V{x},\V{y})&=\sum_{k=0}^\infty \big(\widetilde{\M{T}}_k(\V{0},\V{0})-\widetilde{\M{T}}_k(\V{x},\V{y})\big) \\
&=\frac{1}{(2\pi)^2}\int_{\Set{C}} \big(1-e^{\jmath\V{\vartheta}^{\rm T}(\V{x}-\V{y})}\big) \sum_{k=0}^\infty\M{\Phi}^k(\V{\vartheta}){\rm d}\V{\vartheta} \\
&=\frac{1}{(2\pi)^2}\int_{\Set{C}} (\M{I}-\M{\Phi}(\V{\vartheta}))^{-1} \big(1-e^{\jmath\V{\vartheta}^{\rm T}(\V{x}-\V{y})}\big) {\rm d}\V{\vartheta}
\end{aligned}
\end{equation}
hence the proof is completed.
\end{IEEEproof}
\end{proposition}

With Proposition \ref{prop:existence_potential}, the asymptotic behavior of $\M{P}(\V{x},\V{y})$ can be described as follows.
\begin{proposition}[Asymptotic Behavior of $\M{P}(\V{x},\V{y})$]\label{prop:logarithmic}
As $\|\V{x}-\V{y}\|\rightarrow \infty$, we have
\begin{equation}\label{asymptotic_potential}
\lambda_m(\M{P}(\V{x},\V{y})) = \Theta\Big(\log\frac{\|\V{x}-\V{y}\|}{ \max_{\V{\vartheta}}\lambda_{\max}(\M{\Sigma}_2(\V{\vartheta}))}\Big),~\forall m\in\{1,2\}
\end{equation}
where $\M{\Sigma}_2(\V{\vartheta}):= \|\V{\vartheta}\|^{-2}\sum_{\V{x}\in\mathbb{Z}^2}(\V{x}^{\rm T}\V{\vartheta})^2\widetilde{\M{T}}_1(\V{0},\V{x})$. The term $\max_{\V{\vartheta}}\lambda_{\max}(\M{\Sigma}_2(\V{\vartheta}))$ is a constant with respect to $\|\V{d}_{i\nu}\|$ but increases with the path loss exponent $\gamma$.
\begin{IEEEproof}
We first decompose $\M{P}(\V{x},\V{y})$ into
\begin{equation}\label{decomp_potential}
\begin{aligned}
\M{P}(\V{x},\V{y})&=\frac{1}{(2\pi)^2}\int_{\Set{C}}\big(1-e^{\jmath(\V{x}-\V{y})^{\rm T}\V{\vartheta}}\big)(\M{I}-\M{\Phi}(\V{\vartheta}))^{-1}{\rm d}\V{\vartheta}\\
&=\frac{1}{2\pi^2}\int_{\Set{C}}\frac{1-\cos((\V{x}-\V{y})^{\rm T}\V{\vartheta})}{\|\V{\vartheta}\|^2}\M{\Sigma}_2^{-1}(\V{\vartheta}){\rm d}\V{\vartheta}\\
&\hspace{3mm}+\frac{1}{(2\pi)^2}\int_{\Set{C}}\big(1-e^{\jmath(\V{x}-\V{y})^{\rm T}\V{\vartheta}}\big)\widetilde{\M{\Psi}}(\V{\vartheta}){\rm d}\V{\vartheta}.
\end{aligned}
\end{equation}
For the first term on the right hand side of \eqref{decomp_potential}, we have
$$
\begin{aligned}
&\int_{\Set{C}}\big(1-\cos((\V{x}-\V{y})^{\rm T}\V{\vartheta})\big)\|\V{\vartheta}\|^{-2}{\rm d}\V{\vartheta}\\
&\hspace{3mm}\sim\int_0^{2\pi}{\rm d}\phi \int_0^\pi\big(1-\cos(\|\V{x}-\V{y}\|r\sin \phi)\big)r^{-1} {\rm d}r \\
&\hspace{3mm}\sim \ln \|\V{x}-\V{y}\|
\end{aligned}
$$
where $\phi$ is the angle between $\V{x}-\V{y}$ and $\V{\vartheta}$. The second term on the right hand side of \eqref{decomp_potential} can be simplified as
$$
\frac{1}{(2\pi)^2}\int_{\Set{C}}\big(1-e^{\jmath(\V{x}-\V{y})^{\rm T}\V{\vartheta}}\big)\widetilde{\M{\Psi}}(\V{\vartheta}){\rm d}\V{\vartheta} = \frac{1}{(2\pi)^2}\int_{\Set{C}}\widetilde{\M{\Psi}}(\V{\vartheta}){\rm d}\V{\vartheta}
$$
due to the Riemann-Lebesgue Lemma in Fourier analysis. Therefore, since $\widetilde{\M{\Psi}}(\V{\vartheta})$ is integrable, the proof is completed.
\end{IEEEproof}
\end{proposition}

In Appendix \ref{sec:proofs_many}, Proposition \ref{prop:logarithmic} will be used to investigate the asymptotic characteristics of \ac{efim} when there is a single anchor in the network.

\section{Proofs of Results in Section \ref{ssec:lattice}}\label{sec:proofs_many}
\subsection{Proof of Proposition \ref{prop:inf_pathloss}}\label{ssec:proof_inf_pathloss}
\begin{IEEEproof}
We use Theorem \ref{thm:potential} to simplify the analysis. Since there is only one anchor, $\M{P}_{\widetilde{\Set{R}}_i}$ is a $4\times 4$ matrix given by
\begin{equation}\label{single_anchor}
\M{P}_{\widetilde{\Set{R}}_i} = \left[
                                          \begin{array}{cc}
                                            \M{0}_{2\times 2} & \big[\M{P}_{\widetilde{\Set{R}}_i}\big]_{1,2} \\
                                            \big[\M{P}_{\widetilde{\Set{R}}_i}\big]_{2,1} & \M{0}_{2\times 2} \\
                                          \end{array}
                                        \right].
\end{equation}
Since $\|\V{d}_{i\nu}\|=o(D_{\rm net})$, when the network is sufficiently large, we have $\M{D}_i=\M{D}_{\nu}\propto \M{I}_2$. Hence from \eqref{potential} and \eqref{single_anchor} we have
$$
\M{F}_{\widetilde{\Set{R}}_i}(i,\nu) = \frac{1}{2}\subblk{\M{P}_{\widetilde{\Set{R}}_i}}{2}{1}^{-1}.
$$
According to \eqref{alt_main}, the \ac{efim} for agent $i$ can be rewritten as
\begin{equation}
{\rm sp}\{\V{p}_i\} = \tr{2\Big(\M{D}_i \subblk{\M{P}_{\widetilde{\Set{R}}_i}}{2}{1}^{-1}\Big)^{-1}} \propto \tr{\subblk{\M{P}_{\widetilde{\Set{R}}_i}}{2}{1}}.
\end{equation}
From Proposition \ref{prop:logarithmic} we see that
$$
\tr{\sum_{n=0}^\infty \Big(\widetilde{\M{T}}_{\nu\nu}^{(n),{\rm inf}}-\widetilde{\M{T}}_{\nu i}^{(n),{\rm inf}}\Big)}=\Theta(\log\|\V{p}_\nu-\V{p}_i\|)
$$
where $\widetilde{\M{T}}_{ij}^{(n),{\rm inf}}$ denotes $\widetilde{\M{T}}_{ij}^{(n)}$ in an infinite lattice network. Now it suffices to show that
$$
\begin{aligned}
\lim_{D_{\rm net}\rightarrow \infty} \subblk{\M{P}_{\widetilde{\Set{R}}_i}}{2}{1} &= \lim_{D_{\rm net}\rightarrow \infty} \sum_{n=0}^\infty \Big(\widetilde{\M{T}}_{\nu\nu}^{(n)}-\widetilde{\M{T}}_{\nu i}^{(n)}\Big) \\
&=\sum_{n=0}^\infty \Big(\widetilde{\M{T}}_{\nu\nu}^{(n),{\rm inf}}-\widetilde{\M{T}}_{\nu i}^{(n),{\rm inf}}\Big).
\end{aligned}
$$
Note that we have $\widetilde{\M{T}}_{ij}^{(n),{\rm inf}}=\widetilde{\M{T}}_{ij}^{(n)}$ as long as the random walk cannot reach the network edge from node $i$ in $n$ steps. Thus
$$
\begin{aligned}
&\lim_{D_{\rm net}\rightarrow \infty} \sum_{n=0}^\infty \Big(\widetilde{\M{T}}_{\nu\nu}^{(n)}-\widetilde{\M{T}}_{\nu i}^{(n)}\Big)-\sum_{n=0}^\infty \Big(\widetilde{\M{T}}_{\nu\nu}^{(n),{\rm inf}}-\widetilde{\M{T}}_{\nu i}^{(n),{\rm inf}}\Big)\\
&\hspace{3mm} = \lim_{m\rightarrow \infty}\sum_{n=m}^{\infty}  \Big( \big( \widetilde{\M{T}}_{\nu\nu}^{(n)}-\widetilde{\M{T}}_{\nu\nu}^{(n),{\rm inf}}\big) - \big(\widetilde{\M{T}}_{\nu i}^{(n)}-\widetilde{\M{T}}_{\nu i}^{(n),{\rm inf}}\big)\Big)
\end{aligned}
$$
where $m=\lceil R_{\max}^{-1}(D_{\rm net}/2-\max\{\|\V{p}_i\|,\|\V{p}_\nu\|\}) \rceil$ and $R_{\max}$ is the maximum communication range between nodes. Since $\|\V{d}_{i\nu}\| = o(D_{\rm net}/2-\|\V{p}_i\|)$, as the network expands, the distance between $i$ and $\nu$ is negligible compared with their distance from the network edge, implying that $m\rightarrow \infty$ as $D_{\rm net}\rightarrow \infty$. Thus the proof is completed.
\end{IEEEproof}
\subsection{Proof of Proposition \ref{prop:rank3}}\label{ssec:proof_rank3_sketch}
\begin{IEEEproof}
When $N_{\rm t}=1$, since only information on the relative distances can be obtained, it is known that the null space of $\widetilde{\M{J}}_{\rm e}(\V{p}_{\rm all})$ is 3-dimensional, spanned by \cite{relative}
\begin{equation}\label{null_space_structure}
\begin{aligned}
&\V{v}_1 \propto [1~0~\dotsc~1~0]^{\rm T}, ~\V{v}_2 \propto [0~1~\dotsc~0~1]^{\rm T}, \\
&\V{v}_3 \propto [\V{u}(\phi_1+\pi/2)^{\rm T}~\dotsc~\V{u}( \phi_{N_{\rm a}+1}+\pi/2)^{\rm T}]^{\rm T}
\end{aligned}
\end{equation}
where $\phi_i$ is the angle from node $i$ to the network centroid. Next we show that these vectors also span the row space of the matrix $\widetilde{\M{T}}^{\infty}\widetilde{\M{D}}^{-1}$. Multiplying $\widetilde{\M{T}}^\infty$ from the right at both sides of \eqref{singular_fim_auxiliary}, we have
\begin{equation}\label{null_space_j}
\begin{aligned}
\widetilde{\M{J}}_{\rm e}(\V{p}_{\rm all})\widetilde{\M{T}}^{\infty}&=\widetilde{\M{D}}(\M{I}-\widetilde{\M{T}})\widetilde{\M{T}}^{\infty}\\
&=\M{0}.
\end{aligned}
\end{equation}
Using the definition of $\widetilde{\M{T}}$
$$
\widetilde{\M{T}} = \widetilde{\M{D}}^{-1}\widetilde{\M{A}}
$$
we obtain
\begin{equation}\label{transpose_t}
\widetilde{\M{T}}^{\rm T} = \M{D}\widetilde{\M{T}}\widetilde{\M{D}}^{-1}.
\end{equation}
Combining \eqref{null_space_j} and \eqref{transpose_t}, we have
\begin{equation}
\M{D}\widetilde{\M{T}}^{\infty}\widetilde{\M{D}}^{-1}\widetilde{\M{J}}_{\rm e}(\V{p}_{\rm all}) = \M{0}
\end{equation}
implying that the rows of $\widetilde{\M{T}}^{\infty}\widetilde{\M{D}}^{-1}$ can be written as linear combinations of the vectors given in \eqref{null_space_structure}. Therefore, from \eqref{null_space_structure} we see that $\widetilde{\M{T}}_{ii}^{(\infty)}\M{D}_i^{-1}-\widetilde{\M{T}}_{i\nu}^{(\infty)}\M{D}_{\nu}^{-1}$ is rank-1, thus the following quantity
\begin{equation}\label{rank_1_quantity}
\widetilde{\M{T}}_{ii}^{(\infty)}\M{D}_i^{-1}\M{D}_{\nu}-\widetilde{\M{T}}_{i\nu}^{(\infty)} = (\widetilde{\M{T}}_{ii}^{(\infty)}\M{D}_i^{-1}-\widetilde{\M{T}}_{i\nu}^{(\infty)}\M{D}_{\nu}^{-1})\M{D}_{\nu}
\end{equation}
is also rank-1, and its row space is spanned by $\V{u}_{i,\nu}$ which is a unit vector on the direction of $(\V{u}(\phi_i+\pi/2)-\V{u}(\phi_\nu+\pi/2))\M{D}_\nu$.

In the proof of Lemma \ref{lem:potential} we have shown that
\begin{equation}\label{limit_iteration_restricted}
\M{H}_{\widetilde{\Set{R}}_i}-\M{M}_{\widetilde{\Set{R}}_i} = \lim_{n\rightarrow \infty}\Big\{\M{P}_n(\M{F}_{\widetilde{\Set{R}}_i} - \M{I}_{\widetilde{\Set{R}}_i})\Big\}.
\end{equation}
Under the case where $|\widetilde{\Set{R}}_i|=2$, we have
$$
[\M{P}_n]_{\widetilde{\Set{R}}_i} = \left[
                                          \begin{array}{cc}
                                            \M{0}_{2\times 2} & \big[\M{P}_n\big]_{i,\nu} \\
                                            \big[\M{P}_n\big]_{\nu,i} & \M{0}_{2\times 2} \\
                                          \end{array}
                                        \right]
$$
where $[\M{P}_n]_{\widetilde{\Set{R}}_i}$ denote the matrix obtained by restricting $\M{P}_n$ to the set $\widetilde{\Set{R}}_i$, and
\begin{equation}\label{pn_inu}
\big[\M{P}_n\big]_{i,\nu} = \sum_{k=1}^n\big(\widetilde{\M{T}}_{ii}^{(k)}\M{D}_i^{-1}\M{D}_\nu-\widetilde{\M{T}}_{i\nu}^{(k)}\big)
\end{equation}
and vice versa for $\big[\M{P}_n\big]_{\nu,i}$. Since the left hand side of \eqref{limit_iteration_restricted} does exist, from \eqref{rank_1_quantity}, \eqref{limit_iteration_restricted} and \eqref{pn_inu} we have
$$
\M{X}\M{D}_{\widetilde{\Set{R}}_i}(\M{F}_{\widetilde{\Set{R}}_i,\widetilde{\Set{R}}_i}-\M{I}) = \M{0}
$$
where
$$
\M{X}=\left[
  \begin{array}{cc}
    \M{0} & \widetilde{\M{T}}_{ii}^{(\infty)}\M{D}_i^{-1}-\widetilde{\M{T}}_{i\nu}^{(\infty)}\M{D}_{\nu}^{-1} \\
    \widetilde{\M{T}}_{\nu\nu}^{(\infty)}\M{D}_\nu^{-1}-\widetilde{\M{T}}_{\nu i}^{(\infty)}\M{D}_i^{-1} & \M{0} \\
  \end{array}
\right].
$$
Since $\M{F}_{\widetilde{\Set{R}}_i,\widetilde{\Set{R}}_i}-\M{I}$ is a $4\times 4$ matrix while $\M{X}\M{D}_{\widetilde{\Set{R}}_i}$ is rank-3, $\M{F}_{\widetilde{\Set{R}}_i,\widetilde{\Set{R}}_i}-\M{I}$ has to take the following form
\begin{equation}\label{pi_rank3}
\M{F}_{\widetilde{\Set{R}}_i,\widetilde{\Set{R}}_i}-\M{I} = -m\M{D}_{\widetilde{\Set{R}}_i}^{-1} (\V{v}_{i,\nu}\otimes [1~-1]^{\rm T})(\V{v}_{i,\nu}\otimes [1~-1]^{\rm T})^{\rm T}
\end{equation}
where $m$ is a positive scalar and $\V{v}_{i,\nu}:=\V{u}(\phi_i)-\V{u}(\phi_\nu)$. Therefore, the \ac{efim} for agent $i$ is given by
$$
\begin{aligned}
\efim{\V{p}_i} &= \M{D}_i\M{F}(i,\nu) =m \M{J}_{\rm r}(\varphi_{i\nu}).
\end{aligned}
$$

To investigate the asymptotic behavior of $m$, note that from \eqref{limit_iteration_restricted} we have
\begin{equation}
\M{I}-\M{M}_{\widetilde{\Set{R}}_i,\widetilde{\Set{R}}_i} = \M{P}_{\widetilde{\Set{R}}_i}^{(\rm r)}(\M{F}_{\widetilde{\Set{R}}_i,\widetilde{\Set{R}}_i} - \M{I})
\end{equation}
where $\M{P}_{\widetilde{\Set{R}}_i}^{(\rm r)}\!:=\!\lim_{n\rightarrow \infty}[\M{P}_n]_{\widetilde{\Set{R}}_i}
\M{U}$, $\M{U}=[\V{u}_{i,\nu}^{\rm T}~\V{u}_{i,\nu}^{\rm T}]^{\rm T}[\V{u}_{i,\nu}^{\rm T}~\V{u}_{i,\nu}^{\rm T}]$. As the network expands, $\M{P}_{\widetilde{\Set{R}}_i}^{(\rm r)}$ will tends to $\M{P}_{\widetilde{\Set{R}}_i}\M{U}$, where $\M{P}_{\widetilde{\Set{R}}_i}$ is the matrix $\M{P}$ in infinite lattice networks restricted to the set $\widetilde{\Set{R}}_i$. Hence, these two matrices have the same asymptotic behaviors, implying that $m$ scales as $m=\Theta(\log\|\V{d}_{i\nu}\|)$ as long as both $i$ and $\nu$ are far from the network edge as discussed in Sec. \ref{ssec:proof_inf_pathloss}. Thus the proof is completed.
\end{IEEEproof}
\subsection{Proof of Proposition \ref{prop:avg_speb_uda}}\label{ssec:proof_avg_speb_uda_sketch}
\begin{IEEEproof}
We first prove the $O(\log \lambda_{\rm anc}^{-1})$ scaling, and then proceed to the $\Omega(\log \lambda_{\rm anc}^{-1})$ scaling.
\subsubsection{Upper Bound}
The average \ac{speb} in the $N_{\rm t}=1$ case provides a natural upper bound for that in the $N_{\rm t}\geq 2$ case. Hence we only consider the $N_{\rm t}=1$ case here.

Since anchors are distributed according to \eqref{uni_anchor}, for any agent $i$, in each of the following two areas
$$
\Set{A}_{\rm h}(i):= \{\V{x}| \cos^2\varphi_{\V{x},\V{p}_i} \geq c_{\rm h} \},~
\Set{A}_{\rm v}(i):= \{\V{x}| \sin^2 \varphi_{\V{x},\V{p}_i} \geq c_{\rm v} \}
$$
the nearest anchor is within range $O(\lambda_{\rm anc}^{-\frac{1}{2}})$ of agent $i$, where $c_{\rm h}$ and $c_{\rm v}$ are constants in $(1/2,1)$, and $\varphi_{\V{x},\V{p}_i}$ is the angle from $\V{x}$ to $\V{p}_i$. Therefore, for any agent $i$ that is far from the network edge, i.e., satisfying $\lambda_{\rm anc}^{-\frac{1}{2}}=o(D_{\rm net}/2-\|\V{p}_i\|)$, we have ${\rm sp}\{\V{p}_i\}=O(\log \lambda_{\rm anc}^{-1})$ using Propositions \ref{prop:hitting} and \ref{prop:rank3}.

\subsubsection{Lower Bound}
Similar to the upper bound, it suffices to consider the $N_{\rm t}\geq 2$ case here. Since anchors are distributed according to \eqref{uni_anchor}, agents in the following area
$$
\begin{aligned}
\Set{A}_{\rm sq}:= &\Big\{\V{x}=[x_1~x_2]^{\rm T}\Big|x_1\in \Big(\Big(k+\frac{1}{4}\Big)\lambda_{\rm anc}^{-\frac{1}{2}},\Big(k+\frac{3}{4}\Big)\lambda_{\rm anc}^{-\frac{1}{2}}\Big), \\
&\hspace{3mm}x_2\in \Big(\Big(l+\frac{1}{4}\Big)\lambda_{\rm anc}^{-\frac{1}{2}},\Big(l+\frac{3}{4}\Big)\lambda_{\rm anc}^{-\frac{1}{2}}\Big),k\in\mathbb{Z},l\in\mathbb{Z}\Big\}
\end{aligned}
$$
are at least $1/4 \lambda_{\rm anc}^{-\frac{1}{2}}$ away from anchors. We can construct a lower bound for the \ac{speb} of any agent $i$ in $\Set{A}_{\rm sq}$, by considering the case where all nodes that are at least $1/4 \lambda_{\rm anc}^{-\frac{1}{2}}$ away from agent $i$ are anchors. This implies that random walks started from $i$ in the original network do not return if they once arrive a place at least $1/4 \lambda_{\rm anc}^{-\frac{1}{2}}$ away from $\V{p}_i$. Hence we have
\begin{subequations}
\begin{align}
\M{I}-\M{F}_{ii} &\succeq  \sum_{n=0}^m \widetilde{\M{T}}_n(\V{0},\V{0}) =\frac{1}{(2\pi)^2}\int_{\Set{C}} \sum_{n=0}^m \M{\Phi}^n(\V{\vartheta}){\rm d}\V{\vartheta} \\
&\sim \frac{1}{(2\pi)^2}\int_{\Set{C}} \|\V{\vartheta}\|^{-2}(\M{I}-\M{\Phi}^m(\V{\vartheta}))\M{\Sigma}_2^{-1}(\V{\vartheta}){\rm d}\V{\vartheta} \label{lb_uda_surrounding}
\end{align}
\end{subequations}
where we have applied techniques used in the proof of Proposition \ref{prop:logarithmic}, and $m=\lfloor (2R_{\max})^{-1} \lambda_{\rm anc}^{-\frac{1}{2}}\rfloor$. Note that \eqref{convergence_sigma2} implies that
$$
\begin{aligned}
\lim_{m\rightarrow \infty} \M{\Phi}^m(\V{\vartheta}m^{-\frac{1}{2}})&=\lim_{m\rightarrow \infty} \Big(\M{I}-\frac{\|\V{\vartheta}\|^2}{2m}\M{\Sigma}_2(\V{\vartheta})\Big)^m \\
&=\exp\Big\{-\frac{\|\V{\vartheta}\|^2}{2}\M{\Sigma}_2(\V{\vartheta})\Big\}.
\end{aligned}
$$
Thus with the substitution $\V{\alpha} = \sqrt{m}\V{\vartheta}$ we can rewrite the right hand side of \eqref{lb_uda_surrounding} as
$$
\frac{1}{(2\pi)^2}\int_{\sqrt{n}\Set{C}}\|\V{\alpha}\|^{-2}(\M{I}-e^{-\frac{\|\V{\alpha}\|^2}{2}\M{\Sigma}_2(\V{\alpha})})
\M{\Sigma}_2^{-1}(\V{\alpha}){\rm d}\V{\alpha}.
$$
Using again the arguments in the proof of Proposition \ref{prop:logarithmic}, we have
\begin{equation}\label{lb_uda_scaling}
\Big\|\int_{\Set{C}} \|\V{\vartheta}\|^{-2}(\M{I}-\M{\Phi}^m(\V{\vartheta}))\M{\Sigma}_2^{-1}(\V{\vartheta}){\rm d}\V{\vartheta}\Big\|_2\sim \int_0^{\sqrt{m}\pi} \frac{1-e^{-\frac{r^2}{2}}}{r}{\rm d}r
\end{equation}
as $\lambda_{\rm anc}\rightarrow 0_+$. Note that the right hand side of the last line of \eqref{lb_uda_scaling} scales as $\Theta(\log \lambda_{\rm anc}^{-1})$. Since the number of agents in the area $\Set{A}_{\rm sq}$ is proportional to the number of total agents, the proof is completed.
\end{IEEEproof}

\section{Proof of Theorem \ref{thm:rgg}}\label{sec:proof_rgg}
\begin{IEEEproof}
For the single anchor case, we have $\widetilde{\Set{R}}_i = \{i,\nu\}$. For any agent $i$ ($i<\nu$ by definition), consider the equation \eqref{pi_ps} in Sec. \ref{ssec:potential}
$$
\M{F}_{\Set{S},\Set{S}} = \M{I}+\M{P}_{\Set{S}}^{-1} - \M{P}_{\Set{S}}^{-1}\M{Q}(\M{Q}^{\rm T}\M{D}_{\Set{S}}\M{P}_{\Set{S}}^{-1}\M{Q})^{-1}\M{Q}^{\rm T}\M{D}_{\Set S}\M{P}_{\Set S}^{-1}
$$
which can be rewritten as follows for stochastic geometric networks and $\Set{S}=\widetilde{\Set{R}}_i$
\begin{equation}\label{rgg_piri}
\RM{F}_{\widetilde{\Set{R}}_i,\widetilde{\Set{R}}_i} \!=\! \M{I}\!+\!\RM{P}_{\widetilde{\Set{R}}_i}^{-1}-\RM{P}_{\widetilde{\Set{R}}_i}^{-1}\M{Q}(\M{Q}^{\rm T}\RM{D}_{\widetilde{\Set{R}}_i}\RM{P}_{\widetilde{\Set{R}}_i}^{-1}\M{Q})^{-1}\M{Q}^{\rm T}\RM{D}_{\widetilde{\Set{R}}_i}\RM{P}_{\widetilde{\Set{R}}_i}^{-1}.
\end{equation}
Note that we have $\RM{F}_{\widetilde{\Set{R}}_i}(i,\nu) = [\RM{F}_{\widetilde{\Set{R}}_i,\widetilde{\Set{R}}_i}]_{1,2}$, and
$$
\RM{P}_{\widetilde{\Set{R}}_i} = \left[
                                          \begin{array}{cc}
                                            \M{0}_{2\times 2} & \big[\RM{P}_{\widetilde{\Set{R}}_i}\big]_{1,2} \\
                                            \big[\RM{P}_{\widetilde{\Set{R}}_i}\big]_{2,1} & \M{0}_{2\times 2} \\
                                          \end{array}
                                        \right]
$$
The submatrices on the diagonal of $\RM{P}_{\widetilde{\Set{R}}_i}$ are $\V{0}_{2 \times 2}$, as can be seen from the definition of matrix $\M{P}$
$$
\M{P} = \sum_{\{i,j\}\subseteq \Set{N}_{\rm a}\cup\Set{N}_{\rm b}}\M{E}_{ij}^{N_{\rm a}+N_{\rm b}}\otimes \bigg\{\sum_{n=0}^\infty \Big(\widetilde{\M{T}}_{ii}^{(n)}\M{D}_i^{-1}\M{D}_j\!-\!\widetilde{\M{T}}_{ij}^{(n)}\Big)\bigg\}
$$
in which for diagonal blocks (i.e., $i=j$), we have $[\M{P}]_{i,i} = \sum_{n=0}^{\infty} \Big(\widetilde{\M{T}}_{ii}^{(n)}\M{D}_i^{-1}\M{D}_i-\widetilde{\M{T}}_{ii}^{(n)}\Big) = \V{0}_{2\times 2}$. To simplify the notation, we denote $\big[\RM{P}_{\widetilde{\Set{R}}_i}^{-1}\big]_{1,2}$ by $\RM{K}_{i,\nu}$ and $\big[\RM{P}_{\widetilde{\Set{R}}_i}^{-1}\big]_{2,1}$ by $\RM{K}_{\nu,i}$, whereas denote $\big[\RM{P}_{\widetilde{\Set{R}}_i}\big]_{1,2}$ by $\RM{P}_{i,\nu}$ and $\big[\RM{P}_{\widetilde{\Set{R}}_i}\big]_{2,1}$ by $\RM{P}_{\nu,i}$.
From \eqref{potential_kernel_def} we have $\RM{P}_{i,\nu}=\RM{K}_{\nu,i}$. Thus
\begin{equation}
\RM{P}_{\widetilde{\Set{R}}_i} = \left[
                                    \begin{array}{cc}
                                      \M{0}_{2\times 2} & \RM{P}_{i,\nu} \\
                                      \RM{P}_{\nu,i} & \M{0}_{2\times 2} \\
                                    \end{array}
                                  \right],~~
\RM{P}_{\widetilde{\Set{R}}_i}^{-1} = \left[
                                    \begin{array}{cc}
                                      \M{0}_{2\times 2} & \RM{K}_{i,\nu} \\
                                      \RM{K}_{\nu,i} & \M{0}_{2\times 2} \\
                                    \end{array}
                                  \right].
\end{equation}

The terms $\RM{P}_{\widetilde{\Set{R}}_i}^{-1}\M{Q}$ and $\M{Q}^{\rm T}\RM{D}_{\widetilde{\Set{R}}_i}$ can be written as
\begin{equation}\label{pqqd}
\RM{P}_{\widetilde{\Set{R}}_i}^{-1}\M{Q} = \left[
                                         \begin{array}{c}
                                           \RM{K}_{i,\nu} \\
                                           \RM{K}_{\nu,i} \\
                                         \end{array}
                                       \right],~~~\M{Q}^{\rm T}\RM{D}_{\widetilde{\Set{R}}_i} = [\RM{D}_i~\RM{D}_{\nu}]
\end{equation}
Substituting \eqref{pqqd} into \eqref{rgg_piri}, we have
$$
\begin{aligned}
\RM{F}_{\widetilde{\Set{R}}_i,\widetilde{\Set{R}}_i} &=\M{I} + \left[
                                                   \begin{array}{cc}
                                                     \M{0}_{2\times 2} & \RM{K}_{i,\nu} \\
                                                     \RM{K}_{\nu,i} & \M{0}_{2\times 2} \\
                                                   \end{array}
                                                 \right] - \left[
                                         \begin{array}{c}
                                           \RM{K}_{i,\nu} \\
                                           \RM{K}_{\nu,i} \\
                                         \end{array}
                                       \right]\\
                                       &\hspace{5mm}(\RM{D}_i\RM{K}_{i,\nu}+\RM{D}_\nu\RM{K}_{\nu,i})^{-1} [\RM{D}_i~\RM{D}_{\nu}] \RM{P}_{\widetilde{\Set{R}}_i}^{-1} \\
&= \left[
                                                   \begin{array}{cc}
                                                     \M{I}_2 & \RM{K}_{i,\nu} \\
                                                     \RM{K}_{\nu,i} & \M{I}_2 \\
                                                   \end{array}
                                                 \right] - \left[
                                         \begin{array}{c}
                                           \RM{K}_{i,\nu} \\
                                           \RM{K}_{\nu,i} \\
                                         \end{array}
                                       \right]\\
                                       &\hspace{5mm}(\RM{D}_i\RM{K}_{i,\nu}+\RM{D}_\nu\RM{K}_{\nu,i})^{-1} [\RM{D}_\nu \RM{K}_{\nu,i}~\RM{D}_i \RM{K}_{i,\nu}]
\end{aligned}
$$
For the desired term $\RM{F}_{\widetilde{\Set{R}}_i}(i,\nu)$, we have
$$
\begin{aligned}
\RM{F}_{\widetilde{\Set{R}}_i}(i,\nu)
&= \RM{K}_{i,\nu}\Big(\M{I} - (\RM{D}_i\RM{K}_{i,\nu}+\RM{D}_\nu\RM{K}_{\nu,i})^{-1}\RM{D}_i \RM{K}_{i,\nu}\Big)\\
&=\RM{K}_{i,\nu}\Big(\RM{D}_i\RM{K}_{i,\nu}+\RM{D}_\nu\RM{K}_{\nu,i}\Big)^{-1}\RM{D}_\nu\RM{K}_{\nu,i}
\end{aligned}
$$
Note that the expression of $\RM{P}_{\widetilde{\Set{R}}_i}^{-1}$ above implies $\RM{P}_{i,\nu} = \RM{K}_{\nu,i}$, the inverse of $\RM{F}_{\widetilde{\Set{R}}_i}(i,\nu)$ takes the following form
$$
\begin{aligned}
\RM{F}_{\widetilde{\Set{R}}_i}^{-1}(i,\nu) &= \Big(\RM{K}_{i,\nu}\Big(\RM{D}_i\RM{K}_{i,\nu}+\RM{D}_\nu\RM{K}_{\nu,i}\Big)^{-1}\RM{D}_\nu\RM{K}_{\nu,i}\Big)^{-1}\\
&=\RM{P}_{i,\nu}\RM{D}_\nu^{-1}\Big(\RM{D}_i\RM{K}_{i,\nu}+\RM{D}_\nu\RM{K}_{\nu,i}\Big)\RM{P}_{\nu,i}\\
&=\Big(\M{I}+\RM{P}_{i,\nu}\RM{D}_{\nu}^{-1}\RM{D}_i\RM{K}_{i,\nu}\Big)\RM{P}_{\nu,i} \\
&=\RM{P}_{\nu,i}\Big(\M{I}+\RM{K}_{i,\nu}\RM{P}_{i,\nu}\RM{D}_{\nu}^{-1}\RM{D}_i\Big).
\end{aligned}
$$

Now let us consider the convergence from $\RM{F}_{\widetilde{\Set{R}}_i}^{-1}(i,\nu)$ to $2\RM{P}_{\nu,i}$ using the following inequality
\begin{equation}\label{ineq_conv1}
\begin{aligned}
&\Big\|\RM{P}_{\nu,i}(\RM{K}_{i,\nu}\RM{P}_{i,\nu}\RV{D}_\nu^{-1}\RM{D}_i-\M{I})\Big\|\\
&\hspace{3mm}\leq\|\RM{P}_{\nu,i}\|\Big(\|\M{I}\!-\!\RM{D}_\nu^{-1}\RM{D}_i\|\!+\!\|\RM{K}_{i,\nu}\|\|\RM{P}_{i,\nu}\!-\!\RM{P}_{\nu,i}\|\big\|\RM{D}_\nu^{-1}\RM{D}_i\big\|\Big)
\end{aligned}
\end{equation}
Here, $\|\cdot\|$ can be any submultiplicative matrix norm. From \eqref{ineq_conv1} it is clear that we have to show the convergence from $\RM{D}_i$ to $\RM{D}_\nu$ and that from $\RM{P}_{i,\nu}$ to $\RM{P}_{\nu,i}$. Apparently, this requires us to quantify the similarity between \ac{rgg} and lattice graphs. To this end, we first introduce the concept of \textit{minimax lattice matching error} \cite{shor1991minimax}.
\begin{definition}[Minimax Lattice Matching Error]
Consider a set $\Set{S}_{\rm L}$ of $N$ lattice points arranged as a $\sqrt{N}\times \sqrt{N}$ array within a square in $\mathbb{R}^2$ with area $A$. For any set $\Set{S}_{\rm R}$ of $N$ uniformly random distributed points, the \emph{minimax lattice matching error} $\rv{\epsilon}$ is the minimum length such that there exists a one-to-one mapping from $\Set{S}_{\rm R}$ to $\Set{S}_{\rm L}$ in the square, for which the distance between every matched pair does not exceed $\rv{\epsilon}$.
\end{definition}

Intuitively, if the minimax lattice matching error $\rv{\epsilon}$ tends to zero, the stochastic geometric network would have an almost identical structure as that of lattice networks, and hence the corresponding average \ac{speb} may also exhibit similar asymptotic behaviors. According to \cite{shor1991minimax}, for any stochastic geometric networks containing $N\rightarrow \infty$ nodes, if the network keeps connected with probability approaching $1$ as its area $|\Set{R}_{\rm net}|\rightarrow \infty$, we have $\rv{\epsilon}\rightarrow 0$ with probability at least $1-N^{-1}$.

Next, we proceed to analyze the relation between $\widetilde{\RM{T}}_{i\nu}^{(1)}$ in the stochastic geometric network and $\widetilde{\M{T}}_{j\mu}^{(1)}$ in the minimax matched lattice network, where $j$ and $\mu$ correspond to the matched nodes of $i$ and $\nu$, respectively. Note that
$$
\frac{\|\widetilde{\RM{T}}_{i\nu}^{(1)}-\widetilde{\M{T}}_{j\mu}^{(1)}\|}{\|\widetilde{\RM{T}}_{i\nu}^{(1)}\|}
\leq \|\M{I}-\M{D}_j^{-1}\M{J}_{j\mu}\RM{J}_{i\nu}^{-1}\RM{D}_i\|
$$
which implies that $\widetilde{\RM{T}}_{i\nu}^{(1)}$ converges to $\widetilde{\M{T}}_{j\mu}^{(1)}$ if $\M{D}_j^{-1}\M{J}_{j\mu}\RM{J}_{i\nu}^{-1}\RM{D}_i$ converges to $\M{I}$.

We first show that the eigenvalues of $\M{J}_{j\mu}\RM{J}_{i\nu}^{-1}$ converge to $1$ for all matched pairs $(i,j)$. Since $\M{J}_{j\mu}\RM{J}_{i\nu}^{-1}$ is a $2\times 2$ matrix, using \eqref{approx_jd} we can derive its eigenvalues by directly solving the eigenvalues equations, and obtain
\begin{equation}\label{eigenvalues1}
\begin{aligned}
\lambda_1 &\!=\! \frac{\widetilde{\rv{\xi}}_{i\nu,{\rm r}}}{\widetilde{\xi}_{j\mu,{\rm r}}}\!\cdot \!\frac{1}{2\rv{d}^2}\Big\{(\rv{d}^2d_{\rm L}^2+1)\sin^2\rv{\theta} \!+\!(\rv{d}^2\!+\!d_{\rm L}^2)\cos^2\rv{\theta}\\
&\hspace{2mm}\!+\!\sqrt{\!\big[(\rv{d}^2d_{\rm L}^2\!\!+\!\!1)\sin^2\rv{\theta} \!+\!(\rv{d}^2\!+\!d_{\rm L}^2)\cos^2\rv{\theta}\big]^2\!\!-\!\!4\rv{d}^2d_{\rm L}^2}\Big\}
\end{aligned}
\end{equation}
\begin{equation}\label{eigenvalues2}
\begin{aligned}
\lambda_2 &\!=\!\frac{\widetilde{\rv{\xi}}_{i\nu,{\rm r}}}{\widetilde{\xi}_{j\mu,{\rm r}}}\!\cdot \!\frac{1}{2\rv{d}^2}\Big\{(\rv{d}^2d_{\rm L}^2\!+\!1)\sin^2\rv{\theta} \!+\!(\rv{d}^2+d_{\rm L}^2)\cos^2\rv{\theta}\\
&\hspace{2mm}\!-\!\!\sqrt{\!\big[(\rv{d}^2d_{\rm L}^2\!\!+\!\!1)\sin^2\rv{\theta}\!+\!(\rv{d}^2\!+\!d_{\rm L}^2)\cos^2\rv{\theta}\big]^2\!\!-\!\!4\rv{d}^2d_{\rm L}^2}\Big\}
\end{aligned}
\end{equation}
where $\rv{d}$ is the distance between node $i$ and $\nu$, $d_{\rm L}$ is the distance between $j$ and $\mu$, and $\rv{\theta}$ is the angle between $\RV{d}_{i\nu}$ and $\V{d}_{j\mu}$. Denoting the minimax lattice matching error by $\rv{\epsilon}$, using the triangle inequality we see that $|\rv{d}-d_{\rm L}|\leq 2\rv{\epsilon}$. Thus for all pairs $(i,\nu)$ satisfying $\rv{\epsilon}=o(\rv{d})$, we have $d_{\rm L}=\rv{d}+o(\rv{d})$. From \eqref{eigenvalues1} and \eqref{eigenvalues2} we can now see that both $\lambda_1$ and $\lambda_2$ tend to $1$ with probability\footnote{For simplicity, we shall omit the statement of ``with probability $1-1/N_{\rm a}$'' in the rest of this proof.} exceeding $1-1/N_{\rm a}$. As for the eigenvectors, we observe that
\begin{equation}\label{eigenvectors}
\M{J}_{j\mu} = \M{U}_{j\mu}\M{\Lambda}_{j\mu}\M{U}_{j\mu}^{\rm T},~~\RM{J}_{i\nu} = \RM{U}_{i\nu}\RM{\Lambda}_{i\nu}\RM{U}_{i\nu}^{\rm T}
\end{equation}
where $\M{\Lambda}_{j\mu}$ and $\RM{\Lambda}_{i\nu}$ tend to $\M{I}_2$ according to previous discussions. Thus it now suffices to show that $\|\M{I}-\M{U}_{j\mu}^{\rm T}\RM{U}_{i\nu}\|\rightarrow 0$. This is equivalent to showing that $\cos (\varphi_{j\mu})\cos(\rv{\varphi}_{i\nu})+\sin (\varphi_{j\mu})\sin(\rv{\varphi}_{i\nu})\rightarrow 1$, which is further equivalent to showing that $\rv{\theta}\rightarrow 0$. The latter is straightforward for all pairs satisfying $\rv{\epsilon}=o(\rv{d})$, and hence we have $\|\M{I}-\M{D}_j^{-1}\RM{D}_i\|\rightarrow 0$.

Since $\RM{D}_i=\sum_{k\in\RS{N}_i}\RM{J}_{\RV{d}_{ik}}$, we see that the eigenvalues of $\M{D}_j^{-1}\RM{D}_i$ also tends to $1$ for all matched pairs $(i,j)$ as the maximum communication range satisfies $\rv{\epsilon} = o(R_{\max})$. Following a similar line of reasoning as \eqref{eigenvectors}, we can also show that the eigenvectors of $\M{D}_j^{-1}$ and $\RM{D}_i$ tend to be identical. Therefore, for pairs $(i,\nu)$ satisfying $\rv{\epsilon}=o(\rv{d})$, $\widetilde{\RM{T}}_{i\nu}^{(1)}$ converges to $\widetilde{\M{T}}_{j\mu}^{(1)}$. As the network expands, such pairs will be dominantly many and hence we have the convergence from $\widetilde{\RM{T}}_{i\nu}^{(n)}$ to $\widetilde{\M{T}}_{j\mu}^{(n)}$ for any finite $n$. Consequently, we can now conclude that $\|\RM{P}_{i,\nu}-\M{P}_{j,\mu}\|$ tends to zero.

An important corollary of the previous analysis is that both $\|\RM{P}_{i,\nu}-\RM{P}_{\nu,i}\|$ and $\|\M{I}-\RM{D}_i\RM{D}_{\nu}^{-1}\|$ tend to zero. In light of this, from \eqref{ineq_conv1} we can now work on the simpler term of $2\RM{P}_{i,\nu}$. To finish the proof, it now suffices to show that
$$
\frac{{\rm tr}\{\sum_{\RV{p}_i\in\RS{S}_\nu}\big(\RM{P}_{i,\nu}\RM{D}_i^{-1}-\M{P}_{j,\mu}\M{D}_j^{-1}\big)
\}}{{\rm tr}\{\sum_{\RV{p}_i\in\RS{S}_\nu}\RM{P}_{i,\nu}\RM{D}_i^{-1}\}}\rightarrow 0.
$$
For each $4$-tuple $(i,j,\nu,\mu)$, we have
$$
\begin{aligned}
&\|\RM{P}_{i,\nu}\RM{D}_i^{-1}-\M{P}_{j,\mu}\M{D}_j^{-1}\|\cdot \|\RM{P}_{i,\nu}\RM{D}_i^{-1}\|^{-1} \\
&\hspace{3mm}\leq \|\M{I}-\RM{D}_i\M{D}_j^{-1}\|+\|\RM{D}_i\|\|\M{D}_j^{-1}\|\|\RM{K}_{\nu,i}\|\|\RM{P}_{i,\nu}-\M{P}_{j,\mu}\|
\end{aligned}
$$
which indeed tends to zero since $\|\RM{K}_{\nu,i}\|\rightarrow 0$.

Similar arguments can also be applied to the $N_{\rm t}=1$ case, except that one has to work with the Moore-Penrose pseudo-inverse of $\RM{P}_{\widetilde{\Set{R}}_i}^{(\rm r)}$. With similar manipulations we can obtain \eqref{na1_rgg}, thus the proof is completed.
\end{IEEEproof}

\section{Proof of Proposition \ref{prop:rgg_uda}}\label{sec:proof_rgg_uda}
\begin{IEEEproof}
For any agent $i$ in the interior area of the network, we consider the areas $\RS{A}_{\rm h}(i)$ and $\RS{A}_{\rm v}(i)$ defined as (we use upright letters since they are now random sets)
\begin{equation}\label{areas_upper}
\begin{aligned}
\Set{A}_{\rm h}(i)&:= \{\V{x}| \cos^2\varphi_{\V{x},\V{p}_i} \geq c_{\rm h} \}\\
\Set{A}_{\rm v}(i)&:= \{\V{x}| \sin^2 \varphi_{\V{x},\V{p}_i} \geq c_{\rm v} \}.
\end{aligned}
\end{equation}
Since anchors form a binomial point process, the probability that there is no anchor within range $R$ from $\RV{p}_i$ in $\RS{A}_{\rm h}(i)$ is given by
\begin{equation}
p_{0,{\rm h}}(R) = (|\Set{R}_{\rm net}|-2\cos^{-1}(\sqrt{c_{\rm h}})\pi R^2)^{N_{\rm b}}|\Set{R}_{\rm net}|^{-N_{\rm b}}.
\end{equation}
Choosing $R=k\lambda_{\rm anc}^{-\frac{1}{2}}$, we have
\begin{equation}
p_{0,{\rm h}}(k\lambda_{\rm anc}^{-\frac{1}{2}}) = \Big(1-\frac{2k^2(\pi\cos^{-1}(\sqrt{c_{\rm h}}))}{|\Set{R}_{\rm net}|\lambda_{\rm anc}}\Big)^{|\Set{R}_{\rm net}|\lambda_{\rm anc}}
\end{equation}
and $\lim_{|\Set{R}_{\rm net}|\rightarrow \infty}p_{0,{\rm h}}(k\lambda_{\rm anc}^{\frac{1}{2}}) = e^{-2k\pi\cos^{-1}(\sqrt{c_{\rm h}})}$. Therefore, for $R=\Omega(\lambda_{\rm anc}^{-\frac{1}{2}+\epsilon})$ where $\epsilon>0$ can be arbitrarily small, we have $p_{0,{\rm h}}(R)\rightarrow 0$ as the network expands. Similar arguments also applies to $\RS{A}_{\rm v}(i)$. By application of union bound, we see that with probability approaching $1$, there are anchors in both $\RS{A}_{\rm h}(i)$ and $\RS{A}_{\rm v}(i)$ which are within range $O(\lambda_{\rm anc}^{-1})$ of $\RV{p}_i$. Thus using Theorem \ref{thm:rgg}, we can conclude that the average \ac{speb} of agents in the interior area of the network scales as $O(\log \lambda_{\rm anc}^{-1})$.
\end{IEEEproof}

\bibliographystyle{IEEEtran}
\bibliography{IEEEabrv,Coop-Net-Loc,BiblioCV,WGroup}
\end{document}